\documentclass[
reprint,
a4paper,
australian,
amsmath,
amssymb,
aps,
prd,
twocolumn,
superscriptaddress,
preprintnumbers
]{revtex4-2}
\usepackage{braket}
\usepackage{cancel}
\usepackage{lipsum}
\usepackage{tikz, pgf}
\usetikzlibrary{positioning}
\usepackage{graphicx}
\usepackage{amssymb}
\usepackage{amsmath}
\usepackage{color,xcolor}
\definecolor{darkgreen}{rgb}{0.1,0.4,0.0}
\usepackage{hyperref}
\usepackage{times}
\usepackage{mathrsfs}
\usepackage[utf8]{inputenc}
\usepackage[capitalise]{cleveref}
\usepackage[normalem]{ulem}
\crefrangelabelformat{equation}{(#3#1#4--#5#2#6)}
\graphicspath{{./figs/}}
\newcommand{\rcite}[1]{\cite{#1}}
\newcommand{\rref}[1]{\rcite{#1}}
\newcommand{\refref}[1]{Ref.~\rcite{#1}}
\newcommand{\eref}[1]{Eq.~(\ref{#1})}
\newcommand{\sref}[1]{Sec.~\ref{#1}}
\newcommand{\tref}[1]{Table~\ref{#1}}
\newcommand{\fref}[1]{Fig.~\ref{#1}}

\newcommand{\dkt}{d^3 k\,}
\newcommand{\dq}{d q\,}
\newcommand{\dqt}{d^3 q\,}
\renewcommand{\vec}{\mathbf}
\renewcommand{\bar}[1]{\mkern 1.8mu\overline{\mkern-1.8mu#1\mkern-1.8mu}\mkern 1.8mu}
\newcommand{\lqcd}{lattice QCD~}
\newcommand{\CSSM}{Special Research Centre for the Subatomic Structure
  of Matter (CSSM),\\Department of Physics, University of
  Adelaide, Adelaide, South Australia 5005, Australia}
\newcommand{\UCAS}{School of Physical Sciences, University of Chinese Academy of \\ Sciences (UCAS), Beijing 100049, China}
\begin{document}
\preprint{ADP-21-17/T1164}
\title{{Regularisation in Nonperturbative Extensions of Effective Field Theory}}
\author{Curtis D. Abell}
\email[Corresponding author: ]{curtis.abell@adelaide.edu.au}
\affiliation{\CSSM}
\author{Derek B. Leinweber}
\affiliation{\CSSM}
\author{Anthony W. Thomas}
\affiliation{\CSSM}
\author{Jia-Jun Wu}
\affiliation{\UCAS}
\begin{abstract}
The process of renormalisation in nonperturbative Hamiltonian Effective Field Theory (HEFT) is
examined in the $\Delta$-resonance scattering channel.  As an extension of effective field theory incorporating the L\"uscher formalism,
HEFT provides a bridge between the infinite-volume scattering data of experiment and the
finite-volume spectrum of energy eigenstates in lattice QCD.
HEFT also provides phenomenological insight into the basis-state composition of the finite-volume
eigenstates via the state eigenvectors.
The Hamiltonian matrix is made finite through the introduction of finite-range regularisation.  The
extent to which the established features of this regularisation scheme survive in HEFT is examined.
In a single-channel $\pi N$ analysis, fits to experimental phase shifts withstand large variations
in the regularisation parameter, $\Lambda$, providing an opportunity to explore the sensitivity of
the finite-volume spectrum and state composition on the regulator.
While the L\"uscher formalism ensures
the eigenvalues are insensitive to $\Lambda$ variation
in the single-channel case,
the eigenstate composition varies with $\Lambda$;
the admission of short distance interactions diminishes single-particle contributions to the states.
In the two-channel $\pi N$, $\pi \Delta$ analysis, $\Lambda$ is restricted to a small range by the
experimental data. Here the inelasticity is particularly sensitive to variations in $\Lambda$ and
its associated parameter set.  This sensitivity is also manifest in the finite-volume spectrum for
states near the opening of the $\pi \Delta$ scattering channel. Future high-quality lattice QCD
results will be able to discriminate $\Lambda$, describe the inelasticity, and constrain a
description of the basis-state composition of the energy eigenstates.
Finally, HEFT has the unique ability to describe the quark-mass dependence of the finite-volume
eigenstates.  The robust nature of this capability is presented and used to confront current
state-of-the-art lattice QCD calculations.
\end{abstract}

\maketitle
\section{Introduction}
\label{sec:intro}
The calculation of scattering observables from first principles in \lqcd simulations
is facilitated by an understanding of the impact of the lattice finite volume on the spectrum of
observed states.  The most prominent method for converting quantities obtained through \lqcd into
physical scattering observables is that developed by L\"uscher~\rref{Luscher:1985dn,Luscher:1986pf,Luscher:1990ux}.
With only a single open scattering channel, L\"uscher's method can be used to calculate a scattering phase
shift from a single eigenstate generated on a finite volume through a relatively straightforward
process.

While this process has been generalised to cases such as multiple scattering
channels~\rref{He:2005ey,Lage:2009zv,Bernard:2010fp,Guo:2012hv,Hu:2016shf,Li:2012bi,Hansen:2012bj}, and three-body systems~\rref{Doring:2018xxx,Hansen:2019nir,Blanton:2019vdk}, it requires a parametrisation of the scattering observables and becomes significantly more complicated.
Through increases in computational power and algorithmic advances, \lqcd is now able to
consider physical quark masses, yielding, for example, results for the lowest-lying
resonance, the \( \Delta \)(1232)~\rref{PACS-CS:2008bkb,Andersen:2017una,Morningstar:2021ewk,Silvi:2021uya}.

Hamiltonian Effective Field Theory (HEFT) also allows for the conversion between lattice
QCD quantities and physical quantities, and may have advantages for systems involving multiple
scattering channels.
These advantages have been seen to manifest in studies of resonances such as the Roper \rref{Wu:2017qve}, and the \( \Lambda(1405) \) \rref{Hall:2014uca}.
Here the parametrisation is performed in constructing the Hamiltonian describing the scattering
process.  Single and non-interacting multi-particle basis states are then mixed in solving the
Hamiltonian system and insight into the composition of the states is contained in the energy
eigenvectors of the Hamiltonian.
On multiple occasions
\rref{Hall:2013qba,Wu:2014vma}, it has been shown that by isolating the pole term in the
eigenvalue equation for the Hamiltonian of HEFT, one can obtain an identical expression to the
L\"uscher quantisation up to exponentially suppressed terms in the lattice size \( L\,. \)
However insights into the composition of the states through analysis of the eigenvectors is
unique to HEFT.

In chiral perturbation theory ($\chi$PT), working within the power-counting regime (PCR), where higher order
terms of the chiral expansion make a negligible contribution, the independence of results on the
regularisation method is well understood.  Consider, the finite-range regularisation (FRR)
formalism where a momentum regulator, governed by $\Lambda$, is introduced into loop integrals to suppress the
effective-field contributions at large momenta.  Each $\Lambda$-dependent term in the PCR is
accompanied by a ($\Lambda$-dependent) counter term constrained by phenomenology.  In this way the
regulator cannot have any impact, provided one works within the PCR
\cite{Young:2002ib,Leinweber:2003dg,Leinweber:2005cm}.

In a nonperturbative extension of effective field theory (EFT), the process of renormalisation has
a significant impact.  The couplings between the basis states are significantly
renormalised, introducing a new degree of influence in the calculations.
As the regulator changes, the couplings are renormalised as they maintain fits to experimental data.
This is different from chiral perturbation theory where the couplings are defined and fixed in the chiral
limit.  The idea of a power-counting regime is lost as the Hamiltonian models the experimental
data.
We will illustrate how the role of a single particle basis state can be exchanged for
two-particle states within the Hamiltonian, provided the regulator allows strong short-distance
attractive interactions.

However, model independence is not completely lost.  The L\"uscher formalism embedded within HEFT
brings model independence by linking the scattering data to the finite-volume energy levels.  We
will explore how this relationship is independent of the manner in which the data is modelled via
the intermediate Hamiltonian.

Upon examining the quark-mass dependence of nonperturbative HEFT, one finds a useful degree of
model-independence for the finite-volume eigenvalues across the range of regulator parameters
considered.  Without contributions from a bare state however, the correct mass extrapolation cannot
be obtained purely through the interaction of two-particle states.

In this work we take the \( \Delta(1232) \) as a case study to explore the process of
regularisation of the nonperturbative extension of effective field theory, HEFT.  As the energy
eigenvectors describing the composition of the states are renormalisation dependent, our aim is to
understand the extent to which one can obtain insight into the structure of a resonance through the
application of HEFT.

We commence our report with a brief review of the FRR formalism in perturbation theory in
\sref{sec:FRRpt}.  Our aim here is to present the residual series expansion and connect its
role to HEFT.
In \sref{sec:HEFT} we present the details of the nonperturbative
HEFT approach. This is followed in \sref{sec:1c} by
an analysis of renormalisation in the single $\pi N$-channel calculation of the $\pi N$ phase shifts and
the corresponding finite-volume lattice energy levels.
It is here the
model independence provided by the L\"uscher formalism is manifest.  While the Hamiltonian and its
associated eigenvectors are $\Lambda$-dependent, the finite-volume energy eigenvalues are
constrained by experimental data.
In \sref{sec:2c} the analysis is extended to the two-channel case, where the $\pi \Delta$ coupled
channel is included, giving access to somewhat higher energies and allowing for a
  comparison with contemporary \lqcd results.
In this case, the inelasticity constrains the regulator parameter and a unique description of the
eigenstate composition emerges.
We finish with an outline of the conclusions in
\sref{sec:con}.
%
%
\section{Finite-Range Regularisation in Perturbation Theory}
\label{sec:FRRpt}

The $\Delta$ resonance has the formal quark-mass expansion ($m_q \propto m_\pi^2\, ,$
\cite{Gell-Mann:1968hlm})
\begin{align}
M_\Delta &= \left \{ a_0^\Lambda + a_2^\Lambda\, m_\pi^2 + a_4^\Lambda\, m_\pi^4 + \cdots \right
\} \nonumber \\
    &\quad +\Sigma_{\pi\Delta}(m_\pi^2,\,\Lambda) + \Sigma_{\pi N}(m_\pi^2,\,\Lambda) + \Sigma_{t \Delta}(m_\pi^2,\,\Lambda) \, .
\label{eq:formal}
\end{align}
The leading terms in $\{\cdots\}$ are referred to as the residual series expansion and it plays a
central role in the process of renormalisation.  This is indicated by the appearance of the
regulator parameter, $\Lambda$, as a superscript on the coefficients of the expansion.
The quantities $\Sigma(m_\pi^2,\,\Lambda)$ contain pion loop integrals for the $\Delta$ resonance
with intermediate states as described by the subscripts, where a subscript \( t \) denotes a
tadpole term.  The loop integrals are regulated by the parameter $\Lambda$, which can appear in a
dipole or exponential regulator or as a momentum cutoff in a theta function, etc.  These loop
integrals generate the leading and next-to-leading non-analytic quark-mass terms for the $\Delta$
self energy; they have model-independent coefficients with known values.

The full FRR expansion of Eq.~(\ref{eq:formal}) includes an ultra-violet (UV) completion of the
chiral expansion which ensures the loop integrals tend to zero for large pion masses. The UV
summation depends on both the form of the regularisation function and the regularisation parameter,
$\Lambda$.  As such, FRR provides a model for higher order terms of the chiral expansion, beyond
the leading nonanalytic terms.

The integrals can be evaluated analytically.  For example, explicit forms for a sharp cutoff
regulator are reported in Ref.~\cite{Leinweber:1999ig}.  To proceed with the process of
renormalisation, one then expands the integral results about the chiral limit~\cite{Hall:2010ai}.
One works within the PCR where the leading terms dominate and higher-order terms are suppressed by
powers of $m_\pi/\Lambda$.

Working in the heavy-baryon limit for simplicity of presentation, one observes a
polynomial analytic in $m_\pi^2$ and nonanalytic terms
\begin{align}
   \label{eqn:sigmaD}
   \Sigma_{\pi\Delta} &= b_0^\Delta\, \Lambda^3 + b_2^\Delta\, \Lambda\, m_\pi^2 + \chi_{\pi\Delta}\, m_\pi^3 + b_4^\Delta\,
                        \frac{m_\pi^4}{\Lambda} + \cdots \,,\\
   \label{eqn:sigmaN}
   \Sigma_{\pi N} &= b_0^N\, \Lambda^3 + b_2^N\, \Lambda\, m_\pi^2 + \chi_{\pi N}\, \frac{m_\pi^4}{\delta M}\,\log {m_\pi}
                    + b_4^N\, \frac{m_\pi^4}{\Lambda} \nonumber \\
                      &\quad + \cdots \,,\\
   \label{eqn:sigmaT}
   \Sigma_{t \Delta} &= b_2^{t}\, \Lambda^2\, m_\pi^2 + c_2\, \chi_{t \Delta}\, m_\pi^4\,\log
                       {m_\pi} +  {b_4^{t}}\, m_\pi^4 + \cdots \, .
\end{align}
Here $\delta M$ is the $\Delta$-$N$ mass splitting in the chiral limit, $\chi_{\pi \Delta}$,
$\chi_{\pi N}$, and $\chi_{t \Delta}$ denote the model independent chiral coefficients of the terms
that are nonanalytic in the quark mass, and $c_2$ is a renormalised low-energy coefficient,
discussed in the following.  The regulator dependence of the terms polynomial in
$m_\pi^2$ is explicit.

The process of renormalisation in FRR $\chi$EFT proceeds by combining the renormalisation-scheme
dependent coefficients to provide the physical low energy coefficients, denoted as $c_i$.
The $\Delta$ energy expansion has the form~\cite{Hall:2010ai}
\begin{eqnarray}
M_\Delta &=& c_0 + c_2\, m_\pi^2 + \chi_{\pi \Delta}\, m_\pi^3 +
c_4\, m_\pi^4 \nonumber \\
&&+ \!\left( \frac{\chi_{\pi N}}{\delta M} + c_2\, \chi_{t \Delta} \right) \, m_\pi^4\, \log {m_\pi} + \cdots \,.
\label{eqn:mDexpansion}
\end{eqnarray}
with the coefficients, $c_i$, given by
\begin{subequations}
\begin{eqnarray}
\label{eqn:c0norm}
c_0 &=& a_0^\Lambda + b_0^\Delta\, \Lambda^3 + b_0^N\, \Lambda^3 \,,\\
\label{eqn:c2norm}
c_2 &=& a_2^\Lambda + b_2^\Delta\, \Lambda + b_2^N\, \Lambda + b_2^{t} \, \Lambda^2\, ,\\
\label{eqn:c4norm}
c_4 &=& a_4^\Lambda + \frac{b_4^\Delta}{\Lambda} + \frac{b_4^N}{\Lambda} + b_4^{t}\,, \mbox{\,\,etc.}
\end{eqnarray}
\label{eqn:renorm}
\end{subequations}
Strategies for implementing these formal relations in practice are presented in
Ref.~\cite{Hall:2010ai}.

In this way, the FRR expansion
reproduces chiral perturbation theory
in the PCR.
Any dependence on the regulator is absorbed by the residual-series coefficients,
$a_i^\Lambda$.  In this way, the coefficients $c_i$ are scheme-independent quantities.

Of course the advantage of the FRR approach becomes apparent as one approaches the extent of the
PCR.  FRR provides a model for the small contributions from higher-order terms that are otherwise
absent in common massless renormalisation schemes.

The value of $c_0$ describes the $\Delta$ resonance in the chiral limit, and $c_2$ is related to
the sigma term of explicit chiral symmetry breaking.  The nonanalytic terms $m_\pi^3$ and
$m_\pi^4\,\log m_\pi$ have known, model-independent coefficients denoted by $\chi_{\pi N}$,
$\chi_{\pi \Delta}$ and $\chi_{t \Delta}$.
In practice, the coefficients, $a_i^\Lambda$, are determined by fitting to \lqcd results.

We note that the leading nonanalytic tadpole contribution, $c_2\, \chi_{t \Delta} \, m_\pi^4\, \log
{m_\pi}$, contains the renormalised coefficient $c_2$.  This reflects the origin of the tadpole
contribution in a term of the chiral Lagrangian proportional to the quark mass.  As $c_2\, m_\pi^2$
governs the leading quark-mass dependence of the chiral expansion, $c_2\, m_\pi^2$ appears as a
coefficient of the tadpole term, with the remaining factor of $m_\pi^2\, \log {m_\pi}$ arising in
the loop integral.

The value of $\Lambda$ determines the origin of the physics contributing to the renormalised
coefficients, $c_i$. For small $\Lambda \sim 1$ GeV, the regulated loop integrals do not contain
significant short-distance physics and can be associated with pion-cloud contributions.  By
preventing large momenta from flowing through the effective-field propagators, one avoids large
errors that need to be rectified in the residual series.

The residual series coefficients are short-distance-related quantities directly tied
to the cutoff as illustrated in Eqs.~(\ref{eqn:renorm}). The contributions are associated with a
bare-baryon core contribution.  A phenomenologically-motivated value for $\Lambda$ will leave only
small corrections to be contained within the residual series expansion.

Studies of renormalisation in FRR chiral effective field theory have shown the functional form of
the cutoff to be unimportant \cite{Young:2002ib,Leinweber:2003dg,Leinweber:2005cm}.  By observing
the flow of low-energy coefficients as a functions of the regulator parameter, $\Lambda$,
\cite{%
Hall:2010ai,
Hall:2011en,
Hall:2012pk,
Hall:2013oga
},
the scale of the dipole-regulator parameter was determined, $\Lambda \sim 1$ GeV.  The
phenomenologically motivated value of 0.8 GeV is associated with the induced pseudoscalar form factor of
the nucleon \cite{Guichon:1982zk}, the source of the pion cloud. Values varying by $\pm 0.2$ GeV are typically considered to explore
alternative resummations of the expansion and inform estimates of the systematic error.

The key advantage of FRR for the extrapolation of \lqcd results is that it provides a
mechanism to exactly preserve the leading nonanalytic terms of chiral perturbation theory,
including the values of the model-independent coefficients of the leading
nonanalytic terms, while addressing quark masses beyond the PCR.
This contrasts other popular approaches that draw on the nonanalytic terms of the expansion, but
relegate the model-independent coefficients of these terms to fit parameters.  In the absence of
FRR, the fit parameters of the nonanalytic terms may differ significantly from the known results of
chiral perturbation theory and the extrapolation does not correctly incorporate the known leading
nonanalytic behaviour.  In contrast, while FRR develops some degree of model dependence in
how the loop-integral contributions to the chiral expansion sum to zero as
  $m_\pi \rightarrow \infty$ it does preserve the correct leading and next-to-leading nonanalytic
behaviour of QCD.

In returning our attention to HEFT, it is important to consider where the physics lies in the
calculation.  If one makes a poor choice for the regulator, short distance physics will not be
correctly suppressed in the loop integrals, and the coefficients, $a_i^\Lambda$, will need to be
large in magnitude to correct and ensure the renormalised coefficients take
their scheme-independent physical values.  In
this case one would need to acknowledge that there is a significant role for the bare-baryon core
contribution. Moreover, one might be concerned that the Hamiltonian theory is missing important
physics that is put in by hand via the residual-series coefficients.  In this case, intuition
obtained from the Hamiltonian theory may have relatively poor value.

On the other hand, a good choice for the regulator can allow the residual series coefficients to
become small at higher orders, presenting the possibility that the first two or three terms of the
residual series are sufficient to describe the results from \lqcd.  Moreover, the Hamiltonian
itself contains the correct physics such that insight into the structure of the states obtained
from the energy eigenvectors of the theory is more robust.

As one moves from a perturbative EFT to a nonperturbative extension of EFT incorporating the
L\"uscher formalism, the Hamiltonian will take on a model dependence as it is constrained to fit
experimental data.  As the regularisation is changed, the coupling parameters are renormalised and
optimised to describe the scattering data.  Model independence will not be through the
consideration of a PCR, but rather through the L\"uscher formalism linking
scattering data to finite-volume energy levels.  Here the Hamiltonian serves to mediate between the
infinite and finite-volume worlds.  The eigenvectors of the Hamiltonian are model dependent and
will evolve with the regularisation parameter.  It will be interesting to learn the way in which
the composition of the finite-volume energy eigenstates evolves.  Finally, the L\"uscher formalism
only provides model independence at the physical point.  As HEFT provides a formalism to link
different quark masses, it will be paramount to learn the extent of model dependence in the
quark-mass evolution of the finite-volume energy levels.

Finally, we note that when working at a fixed pion mass such as the physical pion mass, the
residual series of Eq.~(\ref{eq:formal}) sums to a single coefficient. We will refer to this as
the bare-baryon mass, $m_\Delta^{(0)}$, and associate it with the bare-baryon basis state.

\section{Hamiltonian Framework}\label{sec:HEFT}
\subsection{Hamiltonian Model} \label{sec:HamMod}
\begin{center}
  \begin{figure}
    \includegraphics[width=0.45\textwidth]{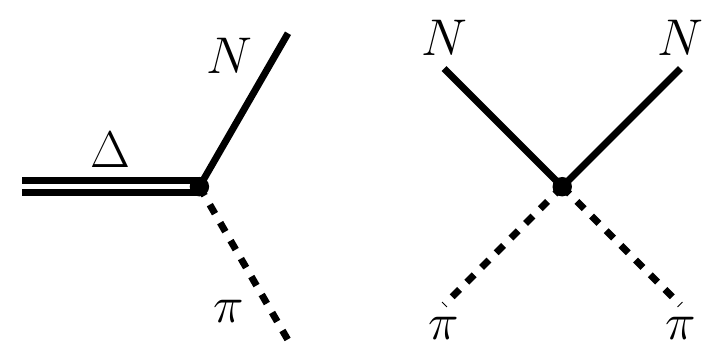}
    \caption{Self-energy (left) and two-particle interaction (right) contributions to the \(\Delta\) mass.}
    \label{fig:Sigma_diagrams}
  \end{figure}
\end{center}

In the rest frame, the Hamiltonian for an interacting system can be represented by the form
\begin{equation}
  H = H_0 + H_I \, ,
\end{equation}
where \( H_0 \) is the free, non-interacting Hamiltonian, and \( H_I
\) is the interaction Hamiltonian.  In the HEFT formalism we allow for
a single-particle bare-baryon basis state \( \ket{B_0} \), which may
be thought of as a quark model state (a state in the $P$-space in the
notation of Ref.~\cite{Thomas:1982kv}).  With coupled two-particle
channels \( \ket{\alpha}\,, \) such as $\pi N$ and $\pi \Delta$, \(
H_0 \) can be expressed as
\begin{align}
  H_0 &= \ket{B_0} m_{B_0} \bra{B_0} + \sum_{\alpha} \int\dkt \nonumber\\
      &\ket{\alpha(\vec k)} \left[ \sqrt{m_{\alpha_B}^2 + k^2} + \sqrt{m_{\alpha_M}^2 + k^2} \right] \bra{\alpha(\vec k)} \, ,
\label{eq:H0}
\end{align}
where \( m_{\alpha_B} \) and \( m_{\alpha_M} \) are the baryon and meson masses respectively in channel \( \alpha\,, \) and \( m_{B_0} \) is the mass of the bare basis state.
In general, \( H_I \) is governed by two types of interactions, examples of which are gives in \fref{fig:Sigma_diagrams}.  The
first, which is denoted by \( g\,, \) represents the vertex
interaction between the bare state \( B_0 \), and the
two-particle basis states \( \alpha\,,
\)
\begin{align}
  g = \sum_{\alpha}\, \int\dkt &\Bigl\{ \ket{B_0} G_{\alpha}^{B_0}(\vec k) \bra{\alpha(\vec k)}  \Bigr. \nonumber\\
                               &\qquad + \Bigl. \ket{\alpha(\vec k)} \left.G_{\alpha}^{B_0}\right.^{\dagger}(\vec k) \bra{B_0} \Bigr\}
\,,
\label{eq:Hg}
\end{align}
where \( G_{\alpha}^{B_0} \) is the momentum-dependent strength of the interaction between the bare
state and each two-particle state.  The momentum-dependence of these couplings is selected to
reproduce the established results of chiral perturbation theory (\( \chi \)PT).
The second type of interaction represents the coupling between two different two-particle basis
states \( \alpha \) and \( \beta \) with momentum-dependent interaction strength
\( V_{\alpha\beta}\,, \) and is given by
\begin{equation}
  v = \sum_{\alpha\,\beta}\, \int\dkt\int\dkt^{'} \ket{\alpha(\vec k)} V_{\alpha\beta}(\vec k,\vec k') \bra{\beta(\vec k^{'})} \, .
\label{eq:Hv}
\end{equation}
The interaction Hamiltonian is therefore given by
\begin{equation}
  H_I = g + v\,. \label{eq:HI}
\end{equation}

\subsection{Finite-Range Regularisation}
In order to work within a finite Hilbert space, we require a renormalisation scheme.  One such
renormalisation scheme is FRR, which has been shown to reproduce
other schemes, such as dimensional regularisation, while in the PCR of
$\chi$PT (\( m_{\pi} \sim m_{\text{phys}} \)) \rref{Young:2002ib}.

Finite-range regularisation introduces a regulator, \( u(k,\Lambda)\,,
\) a function which cuts off the UV
contributions at a rate governed by the
regulator parameter \( \Lambda\,. \) While in
principle, regulators such as a sharp cutoff can be used, it is
desirable to have a smooth regulator which phenomenologically
respects the shape of the source.  For this study, both a dipole
regulator of the form
\begin{equation}
  u(k,\Lambda) = \left( 1 + \frac{k^2}{\Lambda^2} \right)^{-2} \, ,
  \label{eq:dip_reg}
\end{equation}
and a Gaussian regulator of the form
\begin{equation}
  u(k,\Lambda) = \exp\left( -\frac{k^2}{\Lambda^2} \right)\,,
  \label{eq:Gaussian_reg}
\end{equation}
are considered.  As illustrated in Sec.~\ref{sec:FRRpt}, the choice of functional
form between these or a sharp cutoff is irrelevant in the power counting regime of $\chi$PT.

The FRR expansion contains a resummation of higher-order terms that come into play as one works
beyond the PCR, extending the range of
utility~\cite{Young:2002ib,Leinweber:2003dg,Leinweber:2005cm}.  The resummation ensures the FRR
loop-integral contributions are smooth and approach zero for large pion masses, providing a natural
explanation for the slow variation with increasing quark mass observed in \lqcd results.  FRR
provides a mechanism to exactly preserve the leading nonanalytic terms of chiral perturbation
theory, including the values of the model-independent coefficients of the nonanalytic terms, even
when working beyond the PCR.  As one addresses larger quark masses, $\Lambda$ can take on a
physical role modelling the physical size of the particles \cite{Leinweber:2003dg}.

\subsection{Renormalisation of the Coupling} \label{sec:renormalisation}
Following the approach outlined in \refref{Miller:1980hp}, for a system with a single bare state such as that described in this paper, the full propagator \( A(E) \) takes the form
\begin{equation}
  A(E) = \frac{1}{E - m_{B_0} - \Sigma(E)} \, ,
\end{equation}
where \( \Sigma(E) \) is the compilation of all self-energy diagrams.
In particular \( \Sigma(E) \) is taken such that \( A(E) \) should contain a pole at the physical mass of the desired resonance.
As we are only interested in diagrams which yield dominant contributions near the pole position, we need only consider the region about the resonance, where \( m = m_{B_0} + \Sigma(m) \).
Therefore \( A(E) \) can be rewritten as
\begin{align}
  A(E)^{-1} &= E - m - \left( \Sigma(E) - \Sigma(m) \right)\,, \nonumber\\
            &= \left( E - m \right) \left( 1 - \frac{\Sigma(E)-\Sigma(m)}{E-m} \right)\,.
\end{align}
Expanding about the resonance position gives an expression of the form
\begin{align}
  A(E)^{-1} &= \left( E - m \right)\left( 1 - \Sigma'(m) - \frac{\Sigma^{R}(E)}{E-m} \right)\,,\nonumber\\
  &= \left( E - m \right)\left( 1 - \Sigma'(m) \right) - \Sigma^{R}(E)\,,
\end{align}
where $\Sigma'(m)$ is the first derivative of $\Sigma(E)$ evaluated at the physical-mass expansion
point $m$, and \( \Sigma^R(E) \) is defined to contain all higher-order terms in the self-energy.
Finally, defining the renormalised self-energy \( \tilde{\Sigma}(E) = \left\{ 1 - \Sigma'(m) \right\}^{-1}\,\Sigma^{R}(E)\,, \) the propagator may be expressed as
\begin{equation}
  A(E) = \frac{\left\{ 1 - \Sigma'(m) \right\}^{-1}}{E - m - \tilde{\Sigma}(E)}\,.
\end{equation}
This form naturally reveals that the overall propagator has been renormalised by a factor of \( \left\{ 1 - \Sigma'(m) \right\}^{-1}, \) and within the new self-energy the couplings will also be renormalised by the same factor.
In a nonperturbative extension of EFT, this renormalisation of the coupling can become significant.

\subsection{Infinite-Volume Scattering} \label{sec:infVolScat}
In order to constrain bare state masses and potential coupling strengths, we can fit the scattering phase shifts and inelasticities calculated via the \( T \)-matrix.
This can be obtained by solving the coupled-channel integral equations,
\begin{align}
  T_{\alpha\beta}(k,k';E) &= \tilde{V}_{\alpha\beta}(k,k',E) \nonumber\\
                          &+ \sum_{\gamma}\int dq\,q^2\, \frac{\tilde V_{\alpha\gamma}(k,q,E)\, T_{\gamma\beta}(q,k';E)}{E - \omega_\gamma(q)  + i\epsilon}\,, \label{eq:BS}
\end{align}
where \( \omega_{\gamma}(q) = \sqrt{q^2 + m_{\gamma_M}^2} + \sqrt{q^2 + m_{\gamma_B}^2}\,. \)
We have also defined the coupled-channel potential \( \tilde V_{\alpha\beta} \) for some bare state \( B_0 \) as
\begin{equation}
  \tilde{V}_{\alpha\beta}(k,k',E) = \frac{G_{\alpha}^{B_0\,\dagger}(k)\,G_{\beta}^{B_0}(k')}{E - m_{B_0}} + V_{\alpha\beta}(k,k') \, .
\label{eq:ccV}
\end{equation}
The phase shifts and inelasticity however are extracted from the unitary \( S \)-matrix, which is related to the \( T \)-matrix by
\begin{equation}
  S_{\alpha\beta}(E) = \delta_{\alpha\beta} - 2i\pi \sqrt{\rho_{\alpha}\,\rho_{\beta}}\, T_{\alpha\beta}(k_{\text{on},\alpha}, k_{\text{on},\beta}; E) \, ,
\label{eq:Smat}
\end{equation}
where \( k_{\text{on},\alpha} \) is the on-shell momentum in channel \( \alpha\,, \)
and \( \rho_\alpha \) is defined as
\begin{equation}
  \rho_{\alpha} = \frac{\sqrt{k_{\text{on},\alpha}^2 + m_{\alpha_M}^2}\, \sqrt{k_{\text{on},\alpha}^2 + m_{\alpha_B}^2}}{E}\, k_{\text{on},\alpha} \, .
\label{eq:rho}
\end{equation}
The inelasticity, \( \eta_\alpha \), and phase shift, \( \delta_\alpha \), are then calculated from
\begin{equation}
  S_{\alpha\alpha}(E) = \eta_\alpha \exp(2i \delta_\alpha)\,.
\end{equation}
Using this formalism, the position of any poles in the \( S \)-matrix can be found by solving for the complex energy \( E \) which satisfies \( T(k, k'; E)^{-1} = 0\,. \)

\subsection{Finite-Volume Matrix Method} \label{sec:finVol_method}
On a three-dimensional, cubic lattice of volume \( L^3\,, \) the allowed momentum is discretised to
\begin{equation}
  \vec{k_{\vec n}} = \frac{2\pi}{L}\, \vec{n}\,, \quad \vec{n}=(n_x\,,n_y\,,n_z) \,,
\label{eq:mom_disc}
\end{equation}
where \( n_x\,, n_y\,, \) and \( n_z \) can take any integer values.
As a result of this, the integrals over momentum in \eref{eq:H0} to \eref{eq:Hv} undergo discretisation of the form
\begin{equation}
  \int \dkt \rightarrow \sum_{\vec n\in \mathbb{Z}^3}\, \left(\frac{2\pi}{L}\right)^3 \,. \label{eq:int_disc}
\end{equation}
For \( P \)-wave scattering however, at a sufficiently large \( L \) we can approximate spherical symmetry and consider only the degenerate momentum states.
For a discussion on the effects of this approximation and partial wave mixing, see \refref{Li:2019qvh}.
These degenerate momentum states are labelled \( k_n\,, \) where we have defined the integer \( n = n_x^2 + n_y^2 + n_z^2\,. \)
We can represent the degeneracy of each \( k_n \) by defining a function \( C_3(n)\,, \) which counts the number of ways the squared integers \( n_x^2\,,n_y^2\,, \) and \( n_z^2 \) can sum to each \( n\,. \)
Some example values of this function are \( C_3(2) = 12\,, \) and \( C_3(7) = 0\,, \) as there are no combinations of square integers that sum to 7.
Using this definition, the three-dimensional finite sums undergo the transformation
\begin{equation}
  \sum_{\vec n \in \mathbb{Z}^3} \rightarrow \sum_{n \in \mathbb{Z}} C_3(n) \, . \label{eq:C3n_trans}
\end{equation}

As our regulator parameter \( \Lambda \) provides a momentum cutoff, the Hamiltonian
matrix can have a finite size.
We define $k_{\text{max}}$ as the maximum momentum to be considered in the calculation.  We seek a value
sufficiently high compared to the regulator mass such that variation of $k_{\text{max}}$ does not
change the Hamiltonian solution.
In doing this, we refer to the magnitude of the regulator at $k_{\text{max}}$ as $u_{\min}$.
The value of $u_{\min}$ is chosen to minimise the size of the matrix to reduce computational
requirements while ensuring convergence in the evaluation of the contributions from all significant
basis states.

\begin{figure}[t]
  \centering
  \includegraphics[width=0.48\textwidth]{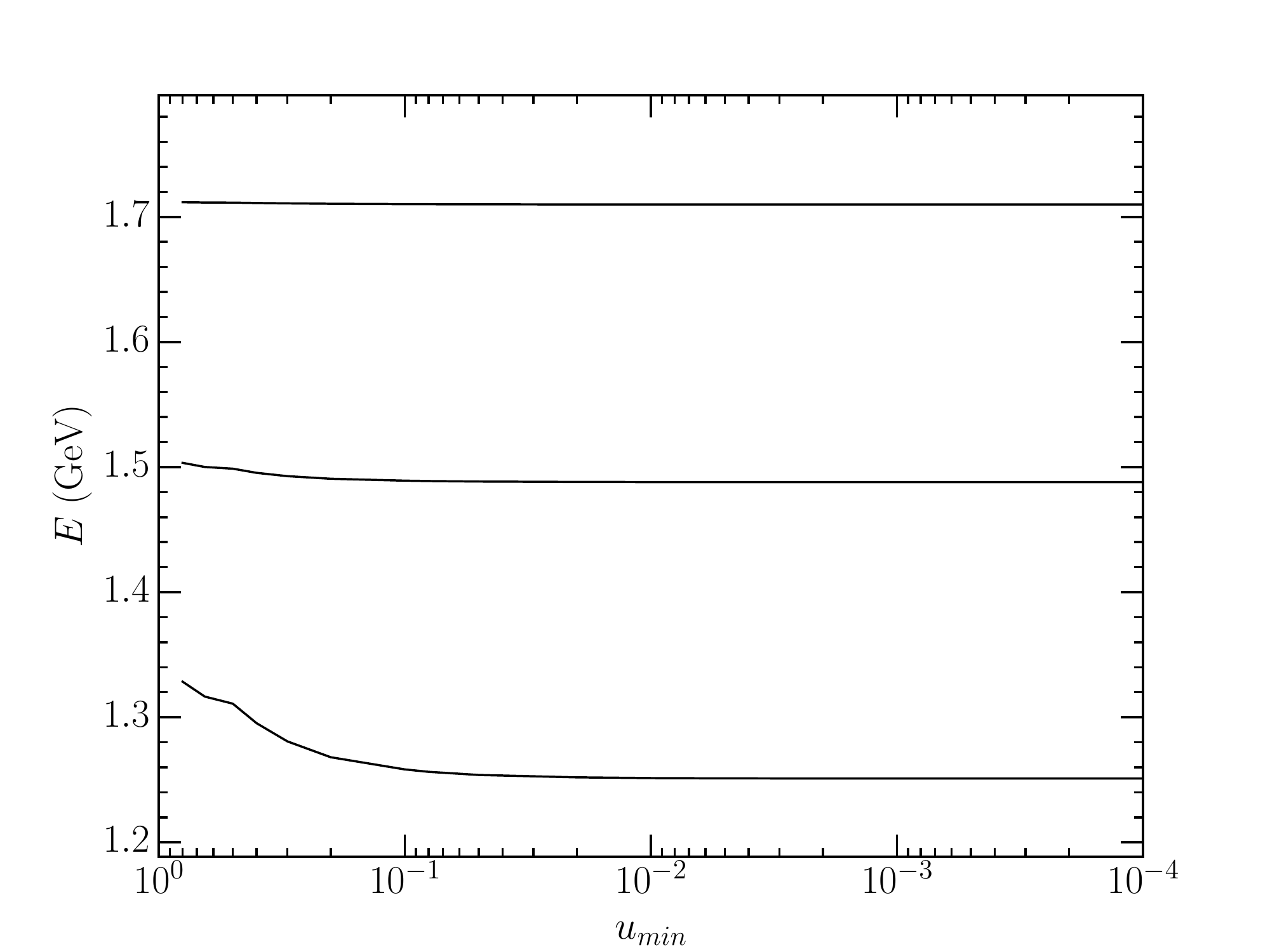}
  \caption{Dependence of Hamiltonian energy eigenvalues for the one-channel analysis of
    \sref{sec:1c} on \( u_{\min} \), governing the maximum momentum to be considered in
    constructing the finite-volume Hamiltonian.  Our selection of \( u_{\min} = 10^{-2} \) ensures
    a robust consideration of high-momentum basis states regulated by $u(k^2)$. }
  \label{fig:umin}
\end{figure}

A value of \( u_{\min} = 10^{-2} \) is selected to balance these two requirements.
Reducing the minimum value of the regulator any further significantly increases the size of the
Hamiltonian and therefore the computational requirements without providing a notable change to the
finite-volume eigenvalues.  The effect of varying \( u_{\min} \) on low-lying energy eigenvalues
for a one-channel analysis (described in \sref{sec:1c}) is shown in \fref{fig:umin}.

Inserting \( u_{\min} \) into equation \eref{eq:dip_reg} and solving for the resulting
\( k_{\max} \) gives a maximum momentum of
\begin{equation}
  k_{\max} = \Lambda \sqrt{u_{\min}^{-\frac 12} - 1} \, .
\end{equation}
This can then be used to solve for the size of the Hamiltonian matrix from \eref{eq:mom_disc},
\begin{equation}
  n_{\max} = \left( \frac{k_{\max}\,L}{2\pi} \right)^2\,.
\end{equation}
The same process can be repeated to calculate the size of the Hamiltonian using
the Gaussian regulator defined in \eref{eq:Gaussian_reg}.

Finally,
due to the discretisation process, the potentials in \eref{eq:Hg} and \eref{eq:Hv} undergo a scaling due to finite-volume factors.
These finite-volume potentials are labelled as \( \bar{G}_{\alpha}^{B_0}(k) \) and \(
\bar{V}_{\alpha\beta}(k,k')\,, \) and the relationship between the finite and infinite volume
potentials will be outlined in the next section.

\subsection{Finite-Volume Factors} \label{sec:FinVol_facts}
In order to calculate the scaling factors due to the finite volume, we consider the relationship between poles in the \( S \)-matrix, and solutions to the eigenvalue equation of the Hamiltonian.
Considering a simple toy system, with a single bare state, and a single two-particle scattering state which only couples to the bare state with strength \( G(k)\,. \)
This scenario is one typically considered in leading one-loop \( \chi \)PT calculations.
Using the notation defined in \sref{sec:finVol_method}, in finite-volume this interaction strength can be written as \( \bar{G}(k)\,. \)
The Hamiltonian for such a system therefore takes the form
\begin{equation}
  H = \begin{pmatrix}
    m_{B_0} & \bar{G}(k_1) & \bar{G}(k_2) & \hdots & \\
    \bar{G}(k_1) & \omega_{\alpha}(k_1) & 0 & \hdots & \\
    \bar{G}(k_2) & 0 & \omega_{\alpha}(k_2) & \ddots & \\
    \vdots & \vdots & \ddots & \ddots & \\
  \end{pmatrix}\,.
\end{equation}
Due to the sparse nature of this matrix, an exact expression can be written for solutions of the eigenvalue equation \( \left| H - E\,\mathbb{I}\right|=0\,, \) giving
\begin{equation}
  E = m_{B_0} - \sum_{n}\frac{\bar{G}^2(k_n)}{\omega(k_n) - E}\,. \label{eq:fin_eigs}
\end{equation}
It is worth noting that upon replacing \( E \) on the RHS with the renormalised mass \( m_{B}\,, \) one can make contact with finite-volume \( \chi \)PT, thus defining the relationship between \( G(k) \) and \( \bar{G}(k)\,. \)

To define this relationship, we return to infinite-volume.
For the simple system defined in this section, the absence of any interactions between different two-particle states means that \eref{eq:ccV} reduces to
\begin{equation}
  \tilde{V}(k,k';E) = \frac{G(k)\, G(k')}{E - m_{B_0}}\,.
\end{equation}
For a separable potential such as this, the associated \( T \)-matrix is also separable, and is able to be written as
\begin{equation}
  T(k,k';E) = G(k)\,t(E)\,G(k')\,,
\end{equation}
Substituting this into \eref{eq:BS} therefore gives an expression for \( t(E) \) in the form
\begin{equation}
  t(E) = \left[m_{B_0} - E - \int_0^{\infty} \dq q^2\, \frac{G^2(q)}{E - \omega(q) + i\epsilon}\right]^{-1} \,.
\end{equation}
As the \( S \)-matrix is proportional to \( t(E)\,, \) poles in the \( S \)-matrix can be found by solving for \( t^{-1}(E) = 0\,, \) giving
\begin{equation}
  E = m_{B_0} - \int_0^{\infty} \dq q^2\, \frac{G^2(q)}{E - \omega(q) + i\epsilon}\,. \label{eq:inf_E}
\end{equation}
Using the fact that for a spherically symmetric momentum space,
\begin{equation}
  \int\dq q^2 = \int\dq q^2 \int\frac{d\Omega}{4\pi} = \int\frac{\dqt}{4\pi}\,,
\end{equation}
we can make use of \eref{eq:int_disc} and \eref{eq:C3n_trans} to obtain an expression for the finite-volume energies in terms of the infinite-volume potentials,
\begin{equation}
  E = m_{B_0} - \sum_{n} \frac{C_3(n)}{4\pi}\, \left(\frac{2\pi}{L}\right)^3\, \frac{G^2(k_n)}{\omega(k_n) - E}\,.
\end{equation}
By comparing this expression to \eref{eq:fin_eigs}, it can be seen that the finite and infinite-volume potentials are related according to
\begin{equation}
  \bar{G}^2(k_n) = \frac{C_3(n)}{4\pi}\, \left(\frac{2\pi}{L}\right)^3\, G^2(k_n)\,.
\end{equation}
Having found this relation, we can therefore return to the general notation for interaction strengths defined in \sref{sec:HamMod}, giving the finite-volume potentials
\begin{align}
  \bar G_{\alpha}^{B_0}(k_n) &= \sqrt{\frac{C_3(n)}{4\pi}}\left(\frac{2\pi}{L}\right)^{\frac 32}\, G_{\alpha}^{B_0}(k_n)\,, \\
  \bar V_{\alpha\beta}(k_n,k_m) &= \sqrt{\frac{C_3(n)}{4\pi}}\sqrt{\frac{C_3(m)}{4\pi}}\left(\frac{2\pi}{L}\right)^3\, V_\beta^\alpha(k_n,k_m)\,.
\end{align}

\section{Single Channel Analysis}
\label{sec:1c}

\subsection{Fitting Experimental Data}
In order to generate a finite-volume energy spectrum, we can obtain values for the bare mass and potential coupling strengths by fitting experimental phase shifts.
In the simplest case, we can consider the \( \pi N \) system to be described by a single bare state,
and with only the \( \pi N \) scattering channel contributing.
In this case, we fit to experimental data below the \( \pi\Delta \) threshold
at approximately 1350 MeV.
For the interaction between the bare \( \Delta \) and the \( \pi N \) scattering state,
the coupling from \eref{eq:Hg} is taken from \refref{Hall:2013qba},
and has the form
\begin{equation}
  G_{\pi N}^{\Delta}(k) = \frac{g_{\pi N}^{\Delta}}{m_\pi^{\rm phys}}\, \frac{k}{\sqrt{\omega_\pi(k)}}\, u(k,\Lambda) \, , \label{eq:g}
\end{equation}
where \( \, \omega_{\pi}(k) = \sqrt{k^2 + m_\pi^2}\, \), \( u(k,\Lambda) \)
is the regulator defined in \eref{eq:dip_reg}, and the inclusion of $m_\pi^{\rm
    phys}$ allows the coupling $g_{\pi N}^{\Delta}$ to be dimensionless.

For the \( \pi N \)-\( \pi N \) interaction of \eref{eq:Hv}, the separable
potential from \refref{Liu:2015ktc} is used, which takes the form
\begin{equation}
  V_{\pi N, \pi N}(k,k') = \frac{v_{\pi N,\pi N}}{( {m_\pi^{\rm phys}} )^2}\, \frac{k}{\omega_\pi(k)}\, \frac{k'}{\omega_\pi(k')}\,
u(k,\Lambda)\, u(k',\Lambda) \, .
\label{eq:V}
\end{equation}
While in principle the regulator parameter in \( G_{\pi N}^{\Delta}(k) \) and in \( V_{\pi N,\pi
  N}(k,k') \) can take different values, in this study they will be fixed to the same value to
simplify the analysis.

Inserting these into the relativised Lippmann-Schwinger equation from \eref{eq:BS},
we can extract the \( \pi N \) phase shift \( \delta_{\pi N}\,. \)
Using these phase shifts, and choosing \( \Lambda = 0.8 \) GeV for now,
we can fit \( \pi N \) scattering data, such as that from \refref{Arndt:1985vj,Workman:2012hx}.
\begin{table*}
  \centering
  \caption{Single-channel fit parameters constrained to the WI08 solution of the $P_{33}$ \( \pi N \) scattering data
    \cite{site:SAID,Workman:2012hx}. Fits I-III contain a single-particle basis state $\ket{\Delta_0}$, while Fit IV does not.}
  \begin{center}
    \begin{ruledtabular}
      \begin{tabular}{ccccc}
        & \multicolumn{3}{c}{With \( \ket{\Delta_0} \)} & No \( \ket{\Delta_0} \) \\
        \noalign{\smallskip}
        \cline{2-4} \cline{5-5}
        \noalign{\smallskip}
        Parameter & Fit I & Fit II & Fit III & Fit IV \\
        \noalign{\smallskip}
        \hline
        \noalign{\smallskip}
        \( m_{\Delta}^{(0)} / \text{ GeV} \) & 1.3589 & 1.4965 & 1.4700 & - \\
        \( g_{\pi N}^{\Delta} \) & 0.1762 & 0.0818 & 0.0101 & - \\
        \( v_{\pi N,\pi N} \) & -0.0286 & -0.0238 & -0.0090 & -0.0029 \\
        \( \Lambda / \text{ GeV} \) & 0.8000 & 1.6000 & 4.0000 & 8.0000 \\
        \noalign{\smallskip}
        \hline
        \noalign{\smallskip}
        DOF & 13 & 13 & 13 & 15 \\
        \( \chi^2 \) & 236.81 & 230.85 & 194.68 & 24373.42 \\
        \( \chi^2 / \text{DOF} \) & 18.22 & 17.76 & 14.98 & 1624.90 \\
        \noalign{\smallskip}
        \hline
        \noalign{\smallskip}
        \( \alpha_{2} / \text{ GeV}^{-1} \) & 1.092 & 0.655 & 0.370 & - \\
        \( \alpha_{4} / \text{ GeV}^{-3} \) & -0.832 & -0.231 & 0.375 & - \\
        \noalign{\smallskip}
        \hline
        \noalign{\smallskip}
        Pole / GeV & \( 1.211 - 0.049i \) & \( 1.210 - 0.049i \) & \( 1.209 - 0.049i \) & \( 1.205 - 0.045i \) \\
      \end{tabular}
    \end{ruledtabular}
  \end{center}
  \label{tab:1c}
\end{table*}

\begin{figure}
  \centering
  \includegraphics[width=0.46\textwidth]{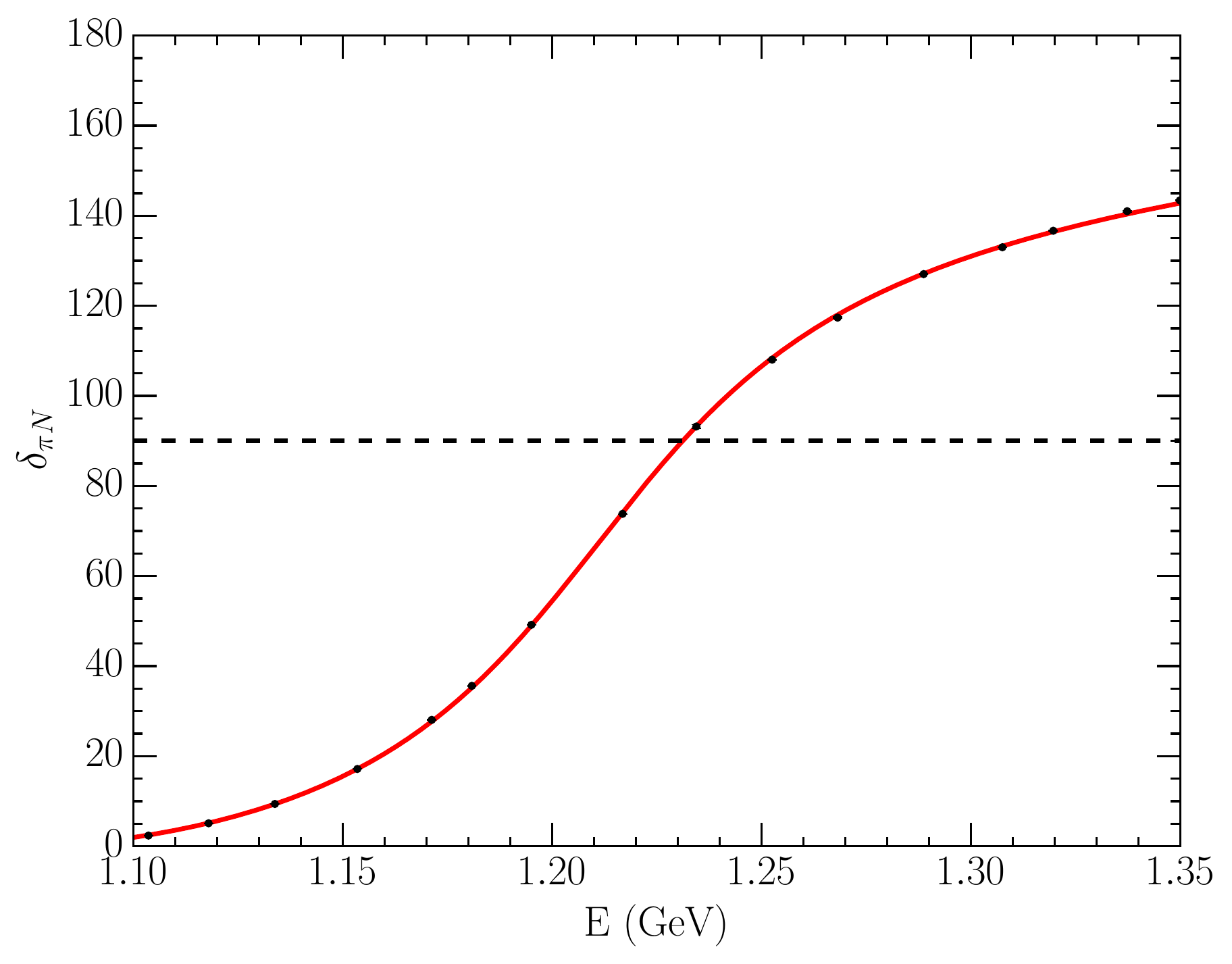}
  \caption{\(P\)-wave \( \pi N \) phase shifts
    for a system with a bare state, where the solid points are experimental data obtained from
    \refref{site:SAID,Workman:2012hx}, the solid line is the fit using HEFT to the data, and the dashed line
    represents a phase shift of 90 degrees. The parameter set producing this curve is given by Fit
    I of \tref{tab:1c}, and gives a \( \chi^2 \)/DOF of 18.22.}
  \label{fig:1b1c_phase_800MeV}
\end{figure}

The parameter set for this fit can be seen in \tref{tab:1c} where a bare state is included, and the
phase shifts corresponding to this fit are illustrated in \fref{fig:1b1c_phase_800MeV}.

Visually, this produces a good fit, and as such it may be surprising that the \(  \chi^2 \) per degree of freedom (DOF) is 18.22.
The origin of the large $\chi^2$/DOF value is in the extraordinary statistical precision of the $\pi N$
scattering data, obtained in a fixed-energy analysis. However, there is additional
systematic uncertainty that is not reflected in the statistical error bars.  The fixed-energy analysis encounters
systematics which give rise to significant fluctuations in the data as a function of energy on the scale
of the statistical errors themselves such that the data are incompatible with a smooth curve.

While many authors do not report a $\chi^2$, we note Meissner {\it et al.} \cite{Fettes:1998ud} assigned a
relative error of 3\% to the scattering data and quote $\chi^2$/DOF values exceeding 0.77 for fits
to 1.2 GeV.  Similarly in Ref.~\cite{Meissner:1999vr}, a 5\% error is assigned and $\chi^2$/DOF values exceeding
0.78 for fits to 1.3 GeV are reported.  If we take a similar approach, the introduction of 3\%
uncertainties provides a $\chi^2$/DOF of 0.07 for fits to 1.35 GeV with 23 DOF. Similarly 5\%
uncertainties provide a $\chi^2$/DOF of 0.02.  In this light, our fits are excellent.  Indeed the
introduction of 1\% uncertainties is sufficient to reduce our $\chi^2$/DOF $\lesssim 1$.

Empirically, our fits also compare well with the \( P_{33} \) results of both
\refref{Juli_D_az_2007} and \refref{Sato:2009de}.

In addition, other quantities such as the pole position can be considered for comparison. As can be
seen in \tref{tab:1c}, using the 0.8 GeV fit (Fit I) we calculate a pole position of \( 1.211 -
0.049i\, \) GeV.  This is in excellent agreement with the pole position of the \( \Delta \) as
quoted by the Particle Data Group (PDG)\cite{10.1093/ptep/ptaa104}, which takes the value of
approximately \( 1.210 - 0.050i\, \) GeV.

\subsection{Finite-Volume Dependence}
Using the parameters found by fitting the scattering data, the matrix Hamiltonian can be constructed.
For the single-channel system, the free Hamiltonian from \eref{eq:H0} can be written as
\begin{equation}
  H_0 = \text{diag}\left( m_{\Delta}^{(0)}, \,\omega_{\pi N}(k_1), \,\omega_{\pi N}(k_2), \,\ldots \right) \, ,
\end{equation}
where \( \, \omega_{\pi N}(k_i) = \sqrt{k^2 + m_\pi^2} + \sqrt{k^2 + m_N^2} \, . \)
The interaction Hamiltonian from \eref{eq:HI} can be written in matrix form as
\begin{equation}
  H_I =
  \begin{pmatrix}
   0 & \bar{G}_{\pi N}^{\Delta}(k_1) & \bar{G}_{\pi N}^{\Delta}(k_2) & \hdots \\
   \bar{G}_{\pi N}^{\Delta}(k_1) & \bar{V}_{\pi N,\pi N}(k_1,k_1) & \bar{V}_{\pi N,\pi N}(k_1,k_2) & \hdots \\
   \bar{G}_{\pi N}^{\Delta}(k_2) & \bar{V}_{\pi N,\pi N}(k_2,k_1) & \bar{V}_{\pi N,\pi N}(k_2,k_2) & \hdots \\
   \vdots & \vdots & \ddots & \vdots
  \end{pmatrix} \, ,
\end{equation}
and so the full Hamiltonian can be constructed as \( \, H = H_0 + H_I\,. \)
In the simplest possible case where \( \, v_{\pi N,\pi N} = 0\,, \) and therefore the only interaction present is
the \( \,\Delta_0 \rightarrow \pi N\, \) vertex, the eigenvalues of the matrix Hamiltonian can be solved exactly\rref{Hall:2013qba}.
Solving the eigenvalue equation \( \, \left|H - E\,\mathbb{I}\right| = 0\,,  \) the eigenvalues are found as solutions of
\begin{equation}
  E = m_{\Delta}^{(0)} - \sum_{n=1}^{n_{\text{max}}} \frac{\bar{G}_{\pi N}^{\Delta}(k_n)^2}{\omega_{\pi N}(k_n) - E} \, .
\end{equation}

We note that by taking the limit for this equation where \( L, n_{\text{max}} \rightarrow \infty\,, \) and
associating the energy \( E \) on the right-hand side with the renormalised \( \Delta \) mass \(
m_{\Delta}\,, \) this expression for the eigenvalues is restored to the one-loop correction to the
\( \Delta \) mass,
\begin{equation}
  m_{\Delta} = m_{\Delta}^{(0)} - \left( \frac{g_{\pi N}^{\Delta}}{m_\pi^{\rm phys}} \right)^2\, \int_0^\infty \frac{k'^4\,u(k',\Lambda)^2\, dk'}{\omega_\pi(k')\left[m_{\Delta} - \omega_{\pi N}(k') + i\epsilon\right]}\,.
\end{equation}

\begin{figure}
  \centering
  \includegraphics[width=0.45\textwidth]{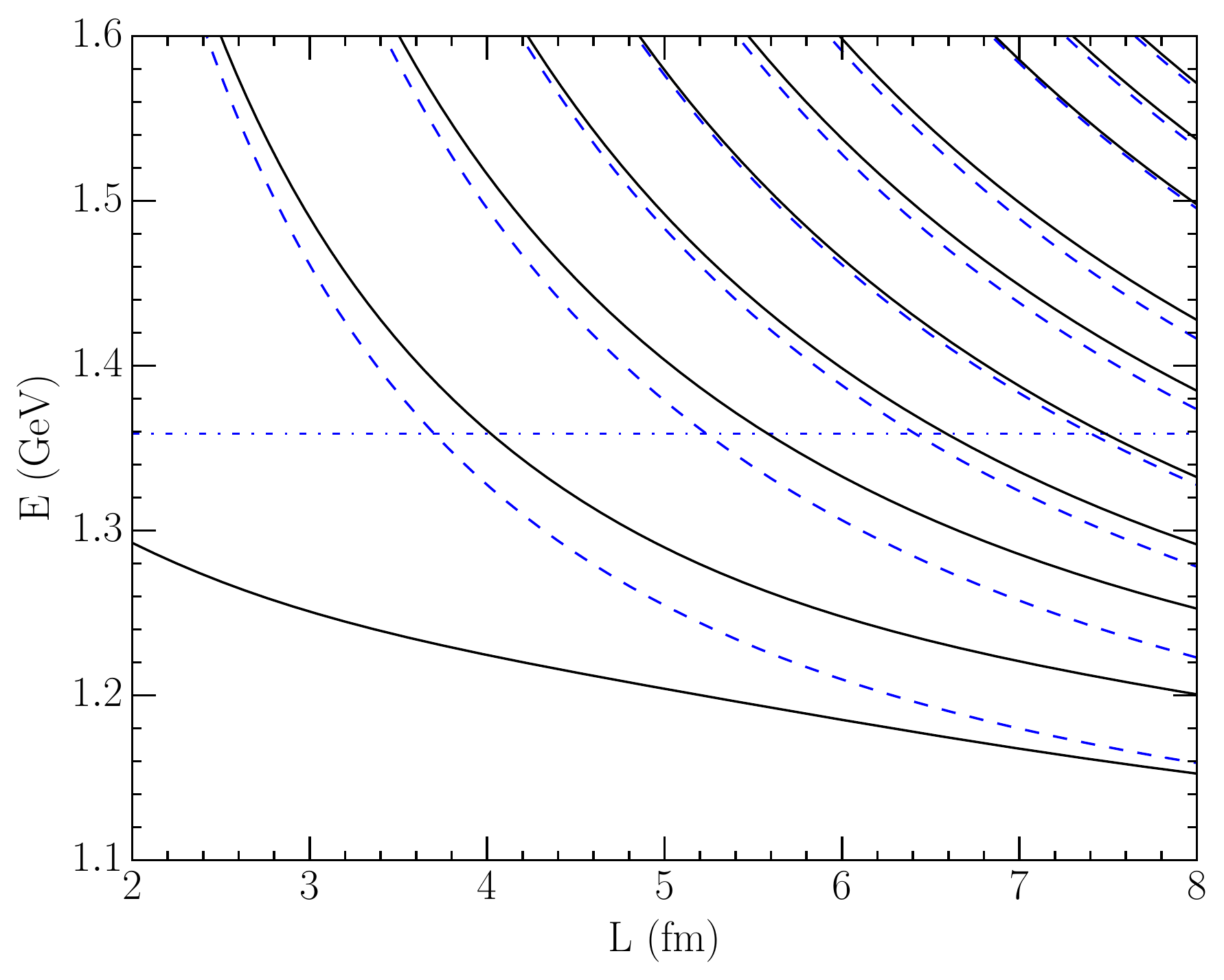}
  \caption{Lattice volume dependence of the energy eigenvalues of the Hamiltonian. The solid lines
    represent the energy eigenvalues following from Fit I of \tref{tab:1c}. The horizontal dot-dashed line is the bare mass and the curved dashed lines are the non-interacting \( \pi N \) basis states at \( k = 2\pi/L, 2\sqrt{2}\pi/L, \ldots \)}
  \label{fig:1b1c_EvL_800MeV}
\end{figure}

\begin{figure}
  \centering
  \includegraphics[width=0.45\textwidth]{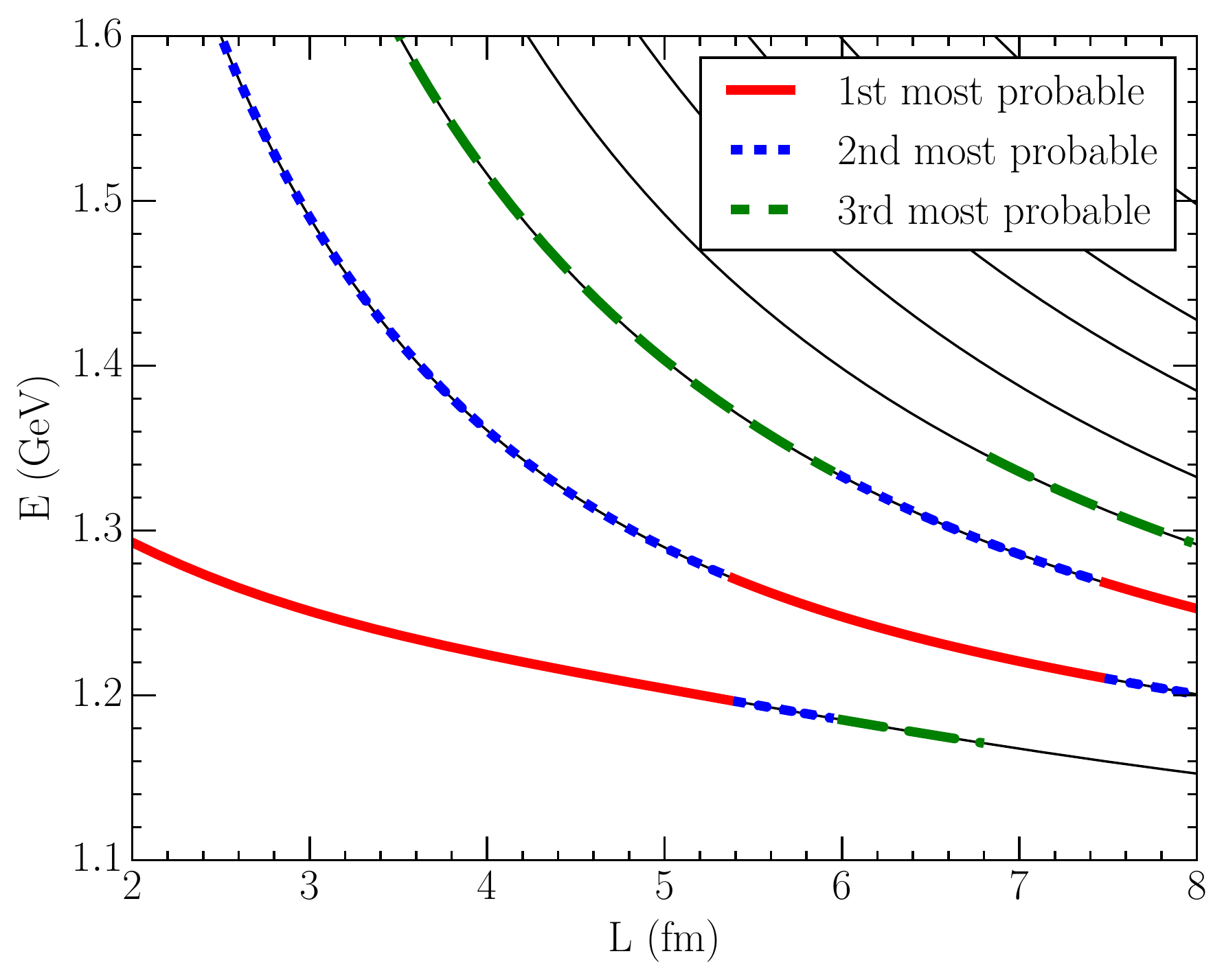}
  \caption{Lattice volume dependence of the energy eigenvalues of the Hamiltonian from Fit I of
    \tref{tab:1c}. The solid (red), short-dashed (blue) and long-dashed (green) highlights on the
    eigenvalues correspond to the states with the largest, second-largest and third-largest
    contribution from the bare basis state $\ket{\Delta_0}$ respectively.}
  \label{fig:1b1c_EvL_Lam800MeV_bare}
\end{figure}

We use a numerical routine to solve for the eigenmodes of $H$.
Varying the lattice volume \( L\,, \) we can generate the finite-volume energy spectrum for this system, as seen in \fref{fig:1b1c_EvL_800MeV}.
In order to observe the contributions from the single-particle basis state, $\ket{\Delta_0}$, to the energy eigenvalues, it is convenient to highlight the states which have the largest contribution from $\ket{\Delta_0}$.
This can be seen in \fref{fig:1b1c_EvL_Lam800MeV_bare}, where the three different highlighting
methods show the states with the first, second and third highest probabilities for the
single-particle $\ket{\Delta_0}$ basis-state contribution.

In \sref{sec:1c_LQCD}, these states will be identified as eigenstates having the largest overlap
with \lqcd eigenstates excited by three-quark interpolating fields, and therefore can be
considered the states which are first, second and third most likely to be observed in a \lqcd
calculation with three-quark operators.

\subsection{Dipole Regulator Dependence}
\label{sec:1c_reg_dep}

\begin{figure}
  \centering
  \includegraphics[width=0.46\textwidth]{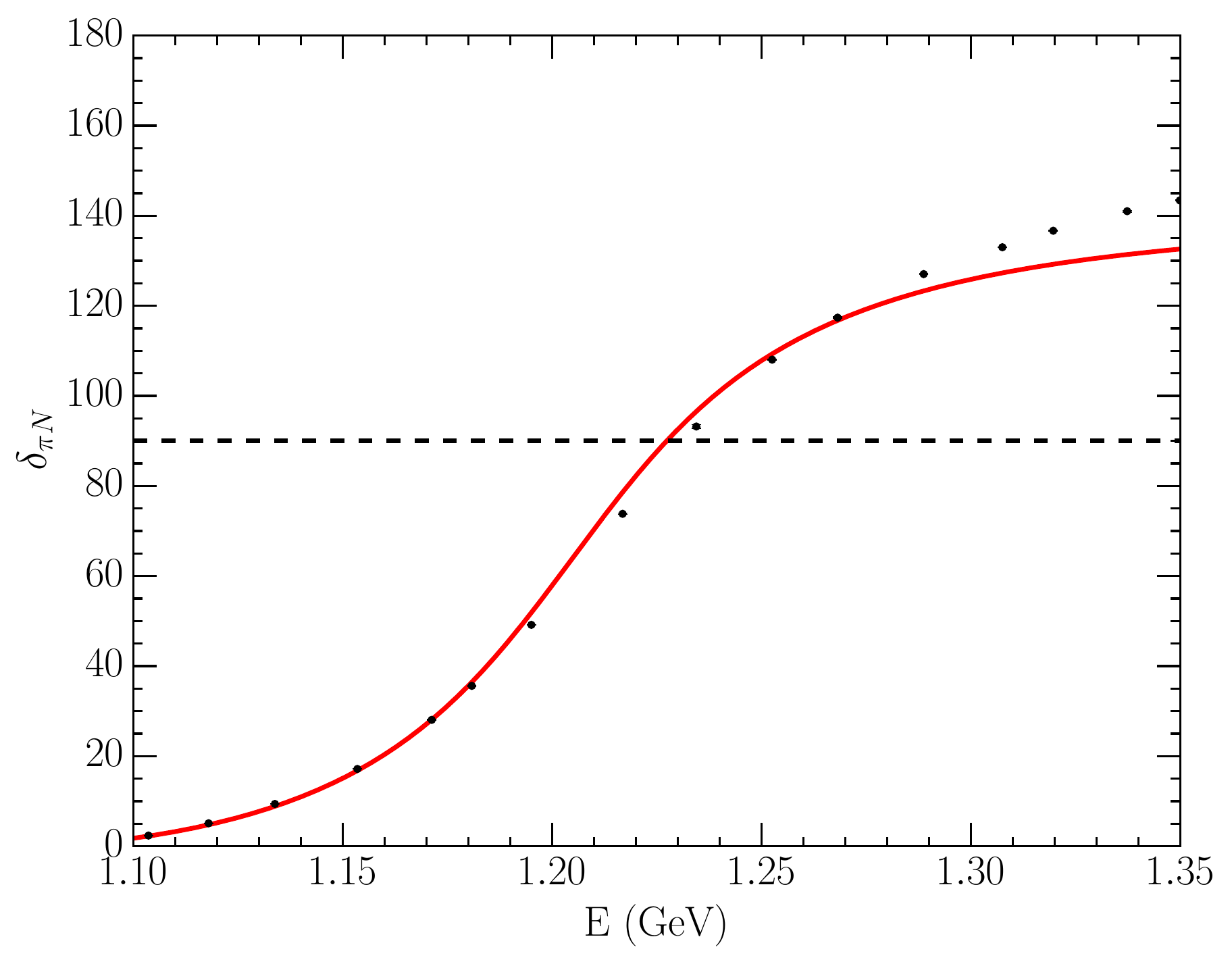}
  \caption{\(P\)-wave \( \pi N \) phase shifts for a system with no single-particle state, where the solid points are experimental data obtained from \refref{site:SAID,Workman:2012hx}, the solid line is the fit using HEFT to the data, and the dashed line represents a phase shift of 90 degrees. The parameter set producing this curve is given by Fit IV of \tref{tab:1c}.}
  \label{fig:0b1c_phase_8000MeV}
\end{figure}

The incorporation of the L\"uscher formalism within HEFT ensures the eigenvalues of the Hamiltonian
in HEFT will be $\Lambda$ independent provided the experimental data is described accurately by the
Hamiltonian model.
To test this we fit the scattering data for values of
\( \Lambda \) varying from 0.6 GeV to 8.0 GeV for a dipole form factor.  The
upper limit selected here is interesting as one can describe the experimental data reasonably well
in the vicinity of the resonance region without a single-particle basis state as
illustrated in Fig.~\ref{fig:0b1c_phase_8000MeV}.

Using each fit to the experimental data for $E \alt 1350$ MeV, we can then solve for the eigenvalues of the finite-volume
Hamiltonian and plot the lowest lying states to check for any \( \Lambda \)-dependence of these
states.

\begin{figure}
  \centering
  \includegraphics[width=0.46\textwidth]{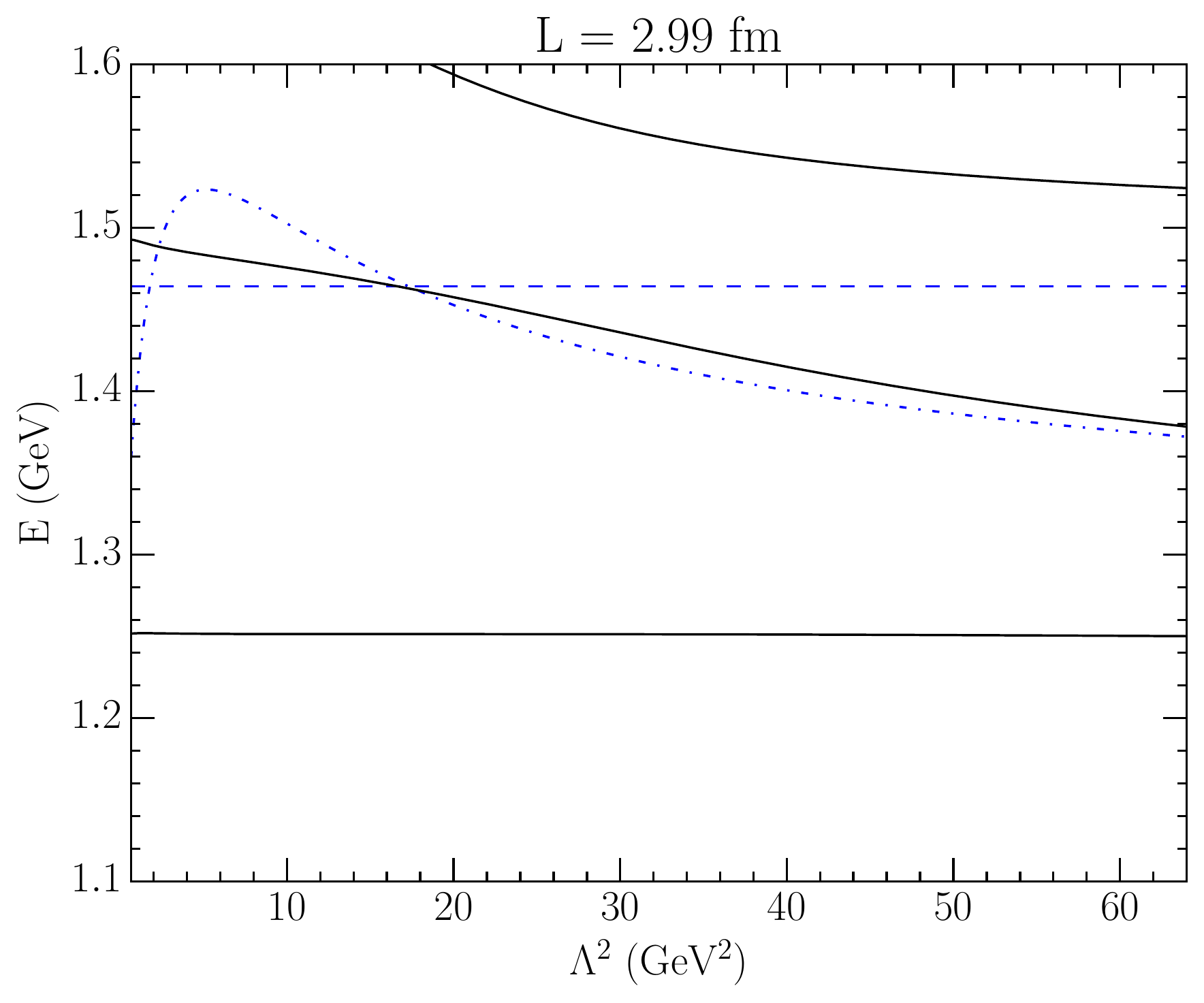}~\\
  \includegraphics[width=0.46\textwidth]{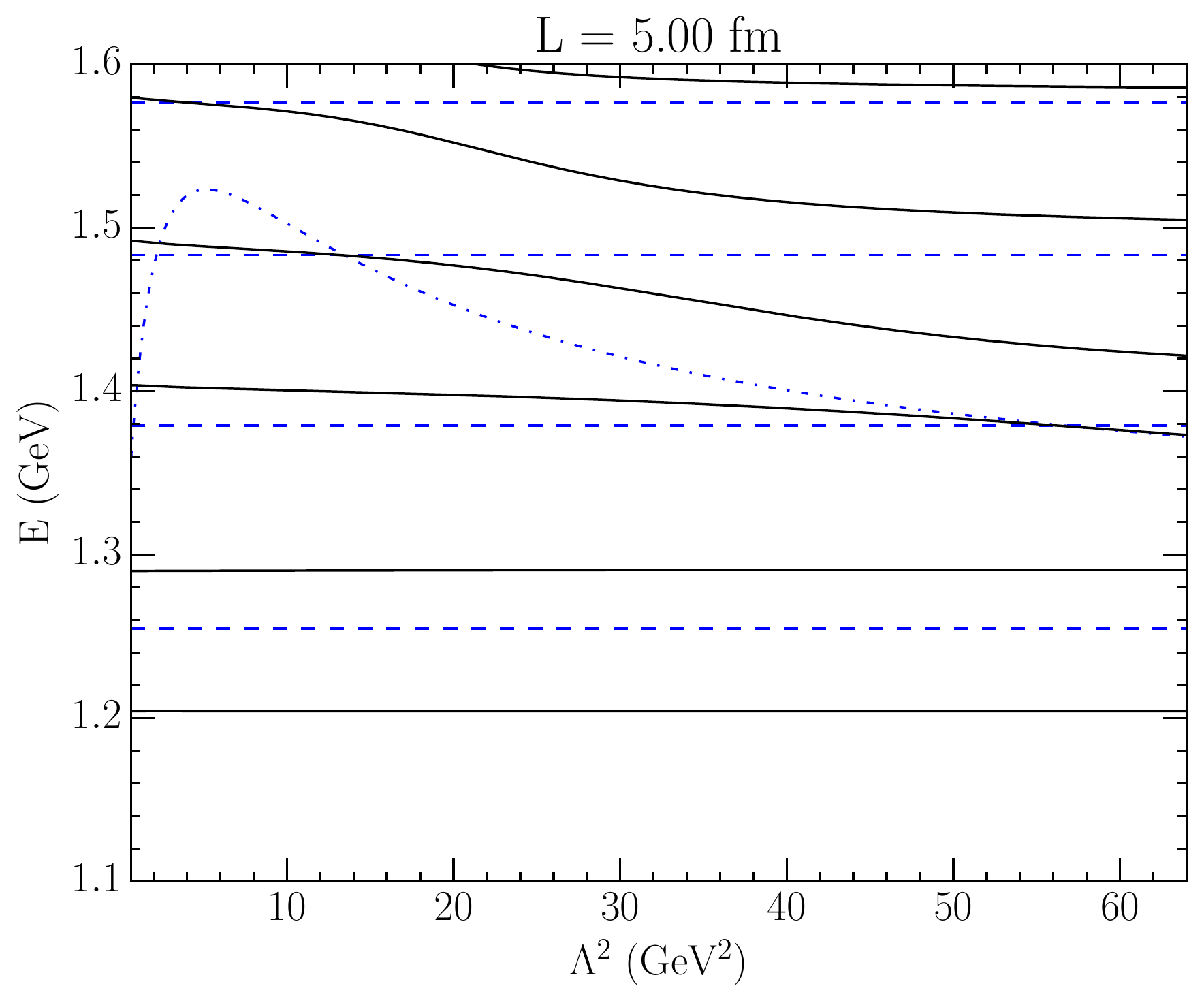}
  \caption{Dependence of the lowest lying eigenvalues of the finite-volume Hamiltonian on the
    regulator parameter \( \Lambda \) for two different lattice sizes, where \( \Lambda \) is
    varying from 0.6 to 8.0 GeV. The solid (black) lines are the eigenvalues, the horizontal dashed
    (blue) lines are \( \pi N \) basis states, and the curved dot-dashed (blue) line is the mass of
    the bare \( \Delta\,. \) The Hamiltonian was constrained to experimental data with $E \alt
    1350$ MeV.  }
  \label{fig:1b1c_EvLambda}
\end{figure}

Indeed, as we see in \fref{fig:1b1c_EvLambda}, the lowest lying states on both lattice volumes are
\( \Lambda \)-independent provided that these eigenstates lie
within $E \alt 1350$ MeV, where the theory is constrained to fit the phase shift data.

While there is no observable \( \Lambda \)-dependence for the eigenvalues in the energy region
where the HEFT has been constrained by data, we are also interested in how the regulator parameter
could affect the physical interpretation of the \lqcd results.

In particular, it is of interest to see how the location of the state dominated by the
single-particle basis state, $\ket{\Delta_0}$, is
affected by \( \Lambda\,. \)
This can be investigated by illustrating the values of the eigenvectors from the Hamiltonian matrix.
As the eigenvectors represent the contribution of each basis state to the final eigenstate, we can plot these as
a function of \( \Lambda \) to observe how the $\ket{\Delta_0}$ contribution to each eigenstate depends on
the regulator parameter.

\begin{figure}
  \centering
  \includegraphics[width=0.46\textwidth]{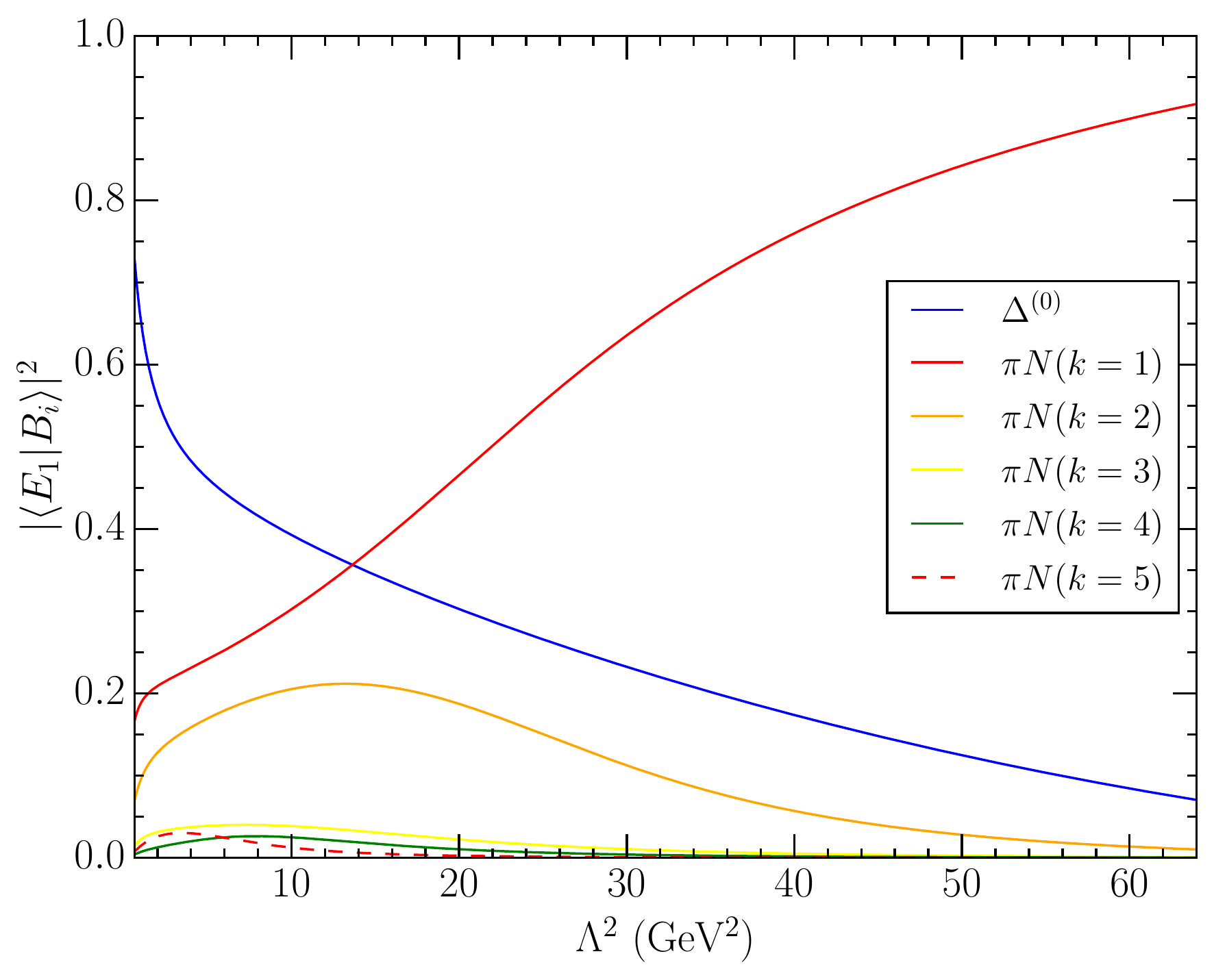}
  \caption{Dependence of the energy-eigenstate basis-state structure on the regulator parameter \( \Lambda\,, \) where \( \Lambda \) is varying from 0.6 to 8.0 GeV, for a lattice size of \( L = 2.99 \) fm.
    Only the ground state is shown, as all higher eigenstates lie above the fitting threshold of \( E = 1350 \) MeV and thus are not physically constrained.}
  \label{fig:1b1c_EvectvLambda_3fm}
\end{figure}
\begin{figure}
  \centering
  \includegraphics[width=0.46\textwidth]{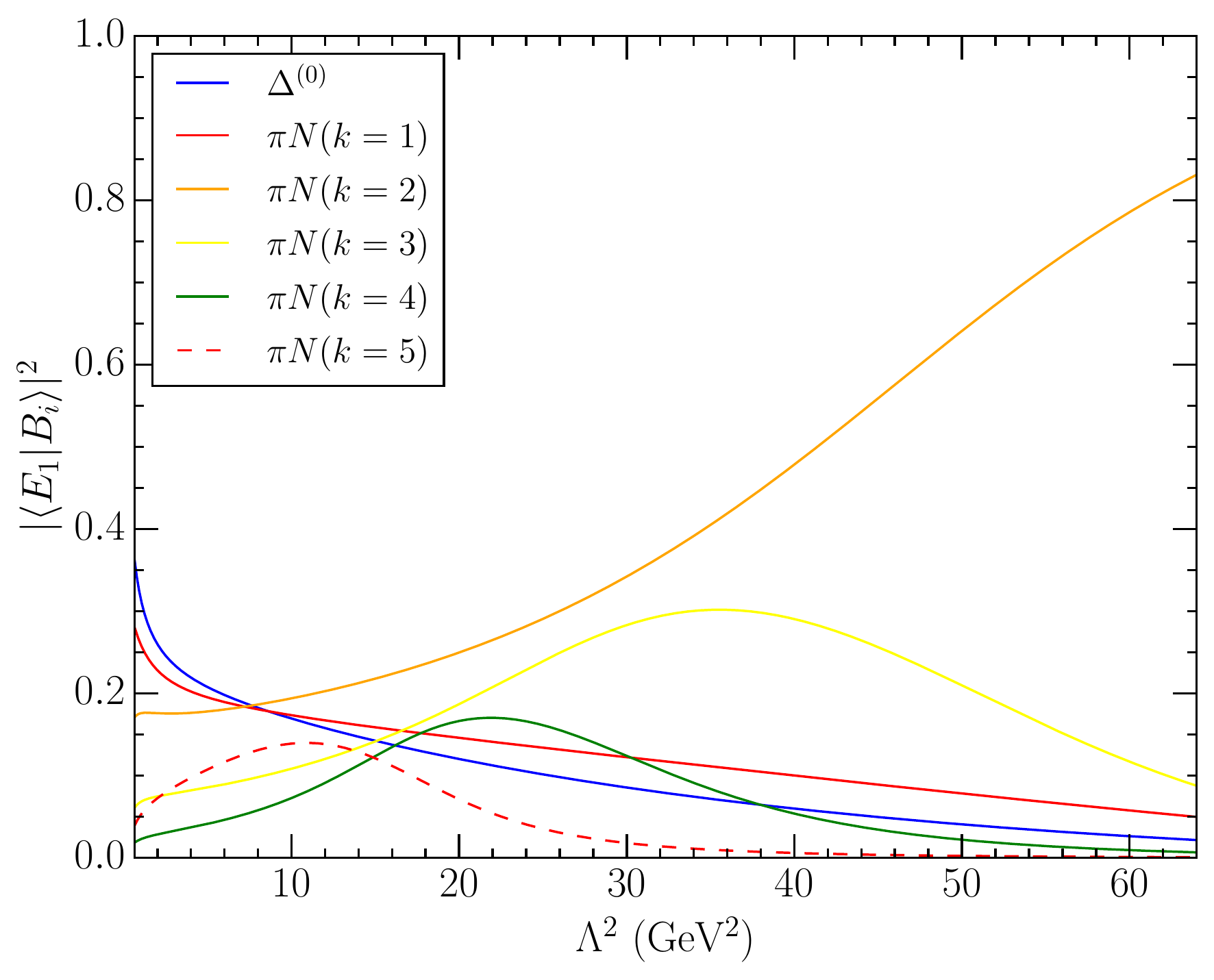}~\\
  \includegraphics[width=0.46\textwidth]{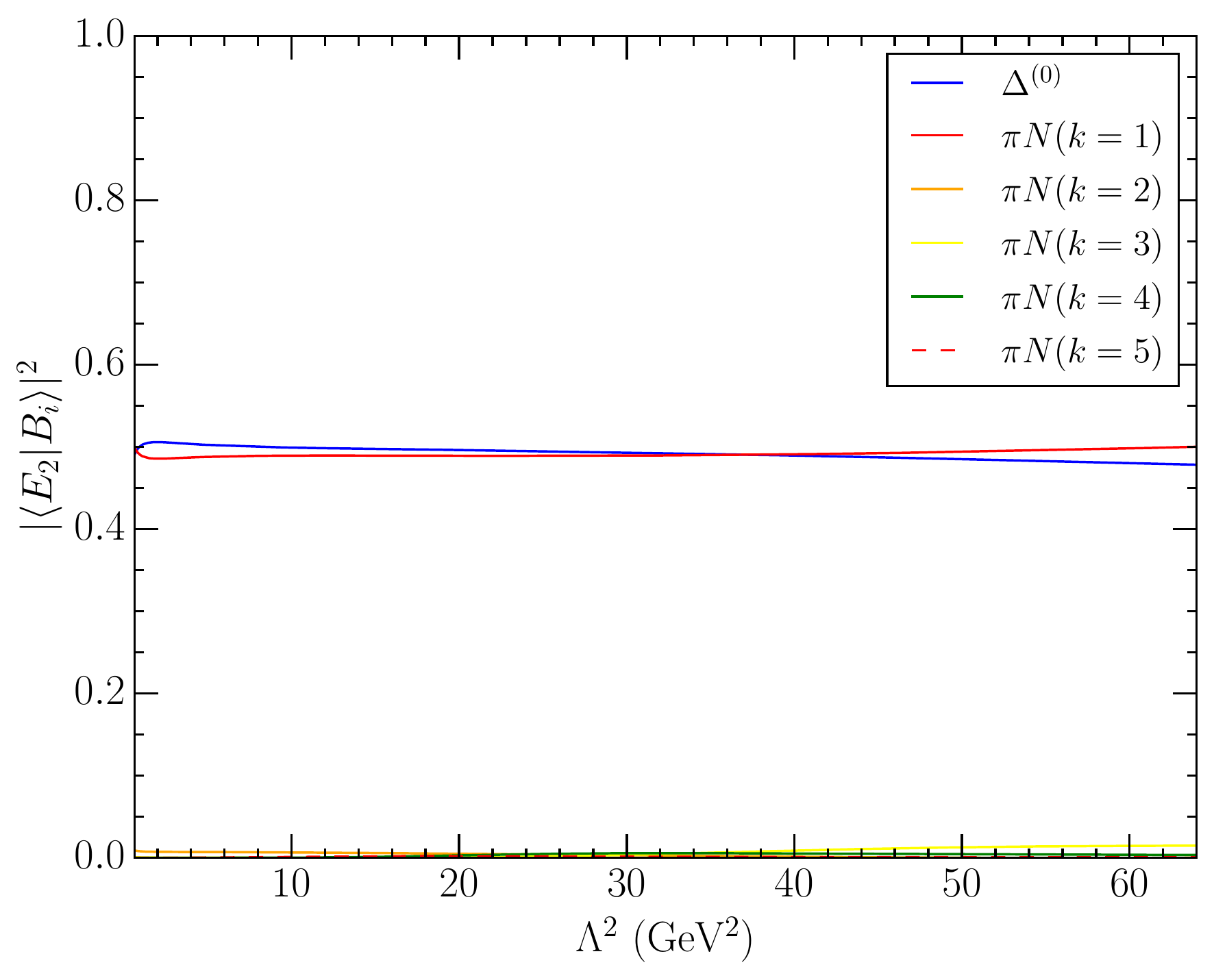}
  \caption{Dependence of the energy-eigenstate basis-state structure on the regulator parameter \(
    \Lambda\,, \) where \( \Lambda \) is varying from 0.6 to 8.0 GeV, for a lattice size of \( L=5
    \) fm. The two lowest-lying energy eigenstates of the finite-volume Hamiltonian are
    investigated.  The upper plot shows the eigenvectors for the ground state, while the lower plot
    shows the first excited state.}
  \label{fig:1b1c_EvectvLambda_5fm}
\end{figure}

In \fref{fig:1b1c_EvectvLambda_3fm} and \fref{fig:1b1c_EvectvLambda_5fm}, we show the \( \Lambda \)-dependence of the two lowest lying
eigenstates for two different lattice volumes.  It is clear that unlike the eigenvalues, the
eigenvectors have a strong \( \Lambda \)-dependence and, in fact, the position of the state which
is dominated by the bare \( \Delta \) is not always \( \Lambda \)-independent.  This is
particularly clear at smaller lattice volumes, where the bare contribution to the ground state
decreases as \( \Lambda \) increases.
For the larger volume at L = 5.0 fm, the first excited state varies between being associated
with the the single-particle contribution and the bare state contribution, and is more stable to \( \Lambda \) variation.

Probing the \( \Lambda \) dependence of these eigenvectors also shows how, as \( \Lambda \)
increases, the contribution from the bare \( \Delta \) becomes distributed throughout the higher
eigenstates, rather than being concentrated in the ground state.  As can be seen in \tref{tab:1c},
the strength of the coupling \( g_{\pi N}^{\Delta} \) required to describe the scattering data at
large values of \( \Lambda \) significantly decreases, and indeed it seems as though a bare \(
\Delta \) may not be required at all.  In fact, with a very large regulator parameter, the
scattering data can be fit just as well with and without a bare
state~\cite{Miller:1979kg,Theberge:1980ye}.  It was found that \( \Lambda = 8.0 \) GeV is the
smallest value of \( \Lambda \) which gives a good description of the data.
This fit is labelled Fit IV in \tref{tab:1c}, and is illustrated in \fref{fig:0b1c_phase_8000MeV}.
Here we only have \( v_{\pi N \pi N} \) as a free parameter, the value of which can be found in
\tref{tab:1c}.

Comparing \fref{fig:1b1c_phase_800MeV} and \fref{fig:0b1c_phase_8000MeV}, both fits give a good
description of the data near the resonance position, though the inclusion of a single-particle
basis state improves the description of the data at higher energies.

While these two scenarios can reproduce scattering data at the physical pion mass, an important
strength of HEFT lies in its capacity to address and interpret \lqcd results in the
region beyond the physical pion mass.  By observing the pion mass dependence of these states in
\lqcd versus the predictions of HEFT for different choices of interactions, one can obtain
some insight into which system of interactions, and hence which physical picture, best describes
the \( \Delta\,. \)

\subsection{Comparison with \lqcd}
\label{sec:1c_LQCD}
To generalise these finite-volume energies to larger-than-physical pion masses, we take the pion
mass dependence of the bare state to vary in the standard manner including
  terms to order \( m_\pi^4\,, \) taking the form
\begin{align}
m_{\Delta}^{(0)} = \left.m_{\Delta}^{(0)}\right|_{\text{phys}} &+ \alpha_{2}\,
\left ( m_\pi^2 - \left.m_\pi^2\right|_{\text{phys}} \right ) \nonumber\\
                 & + \alpha_{4}\, \left ( m_\pi^4 - \left.m_\pi^4\right|_{\text{phys}} \right )\,.
\label{eq:alpha0def}
\end{align}
The values for the \( \alpha_{2} \) and \( \alpha_4 \) are found by performing a two-parameter fit to \lqcd
results for the ground state \( \Delta \) at a lattice size of $L$ = 2.99 fm, as given by the PACS-CS
Collaboration~\rref{PACS-CS:2008bkb}.  At this volume the ground state is dominated by the three-quark-like bare state, as can be seen in \fref{fig:1b1c_EvL_Lam800MeV_bare}.

In considering pion masses away from the chiral limit, we adopt the approach of $\chi$PT where
couplings are fixed and variation with pion mass is contained within the higher-order terms of the
expansion \cite{McGovern:1998tm}.  Chiral limit couplings are approximated by our analysis of
scattering data necessarily at the physical point.

\begin{figure*}
  \centering
  \includegraphics[width=0.46\textwidth]{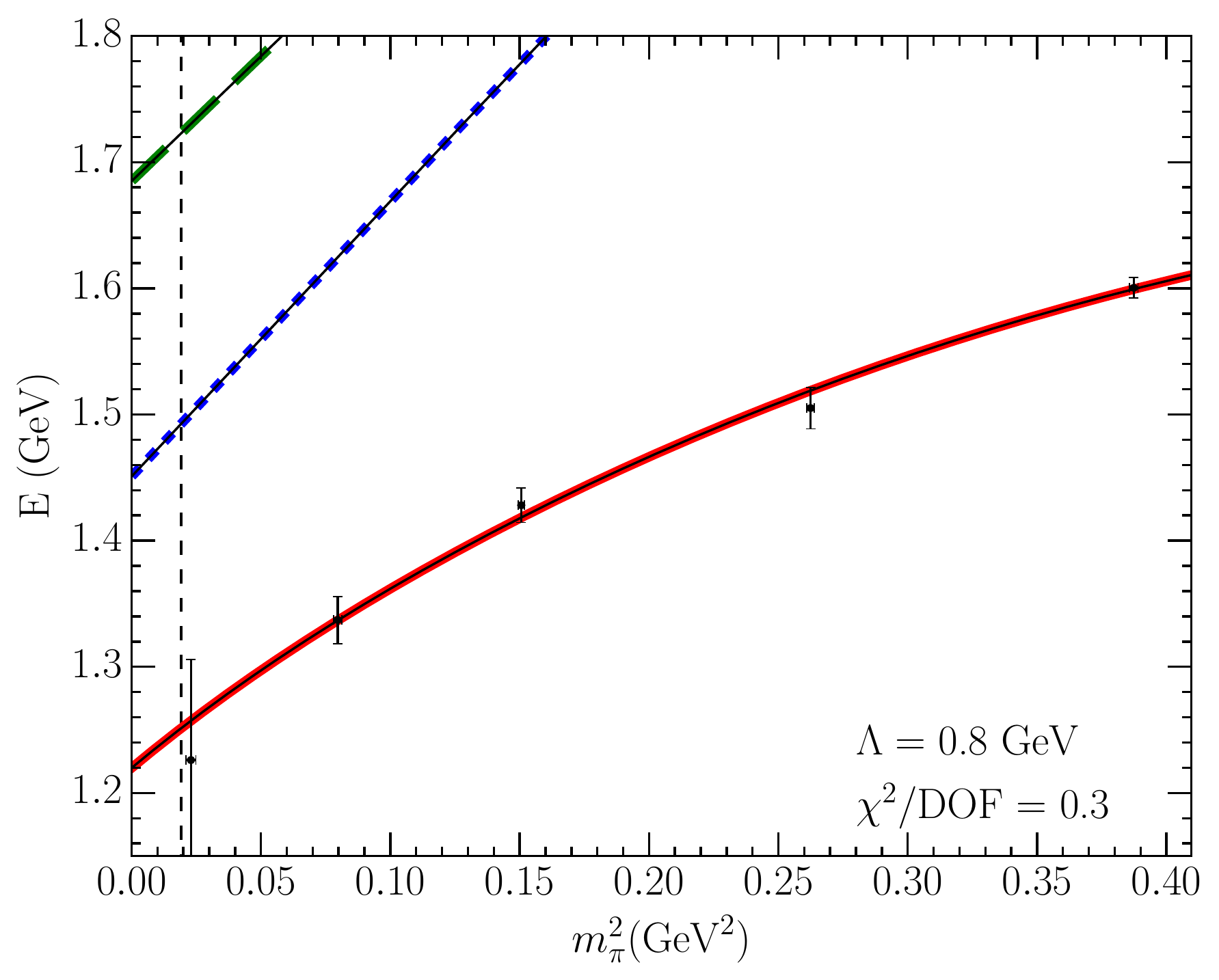}
  \includegraphics[width=0.46\textwidth]{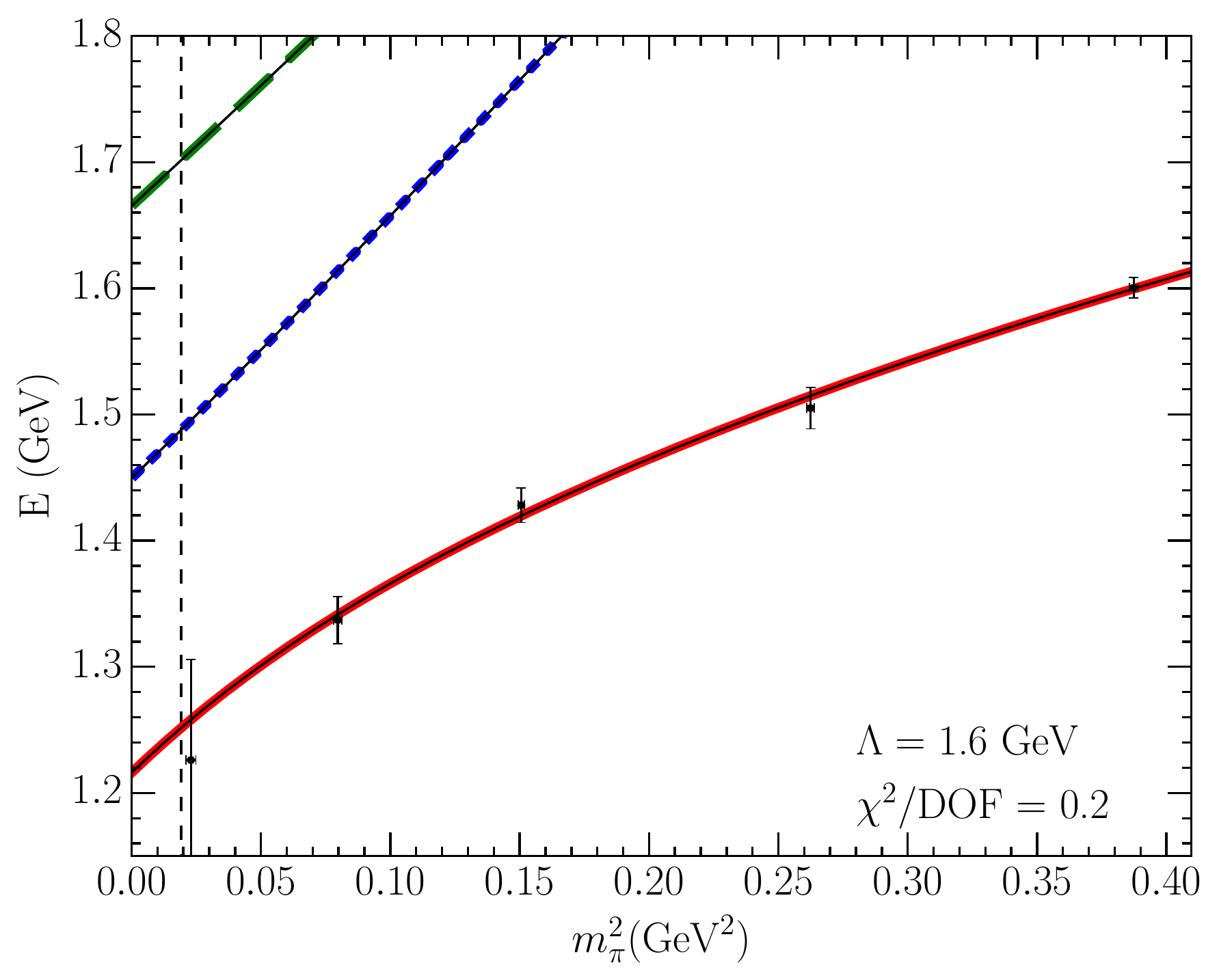}~\\
  \includegraphics[width=0.46\textwidth]{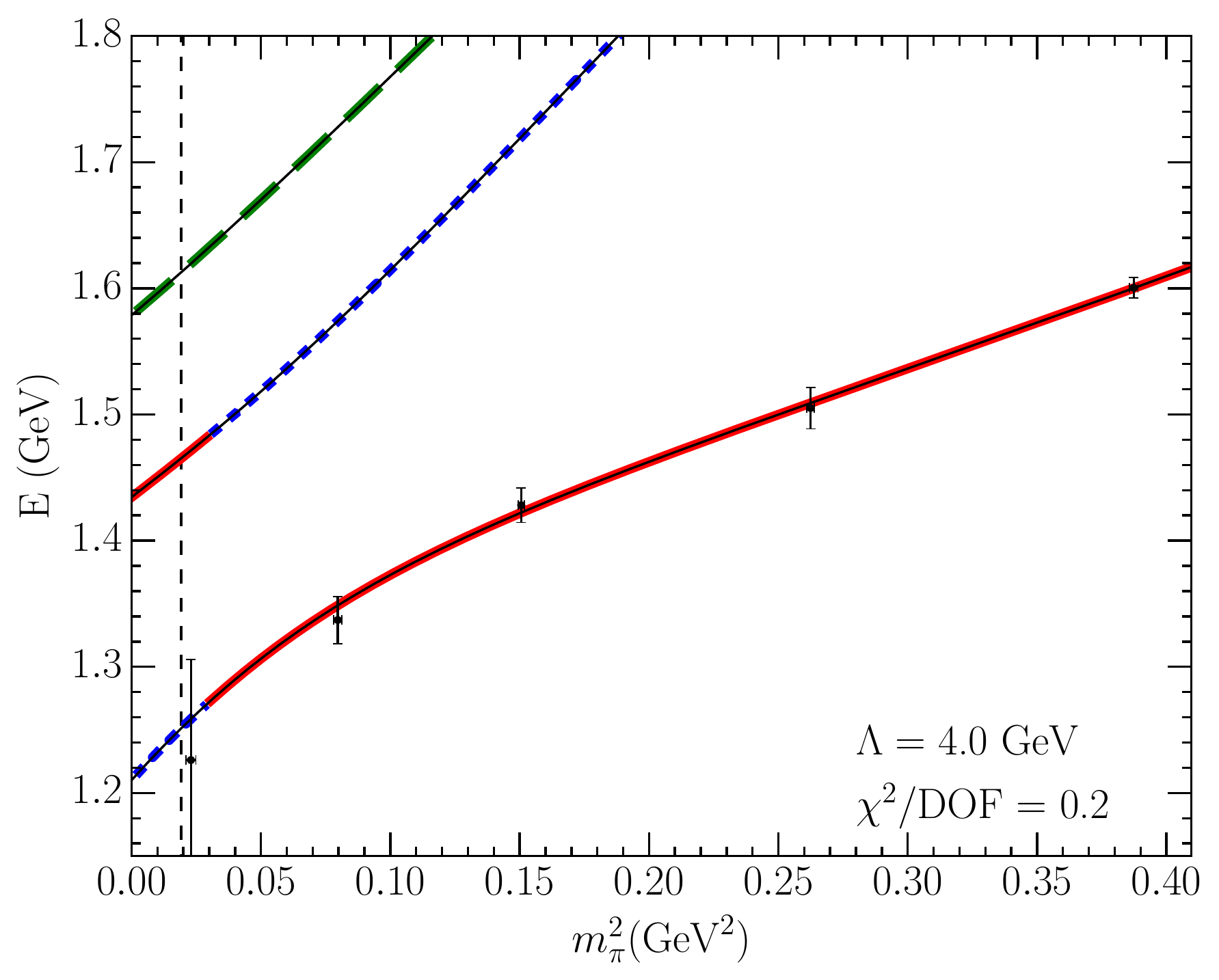}
  \includegraphics[width=0.46\textwidth]{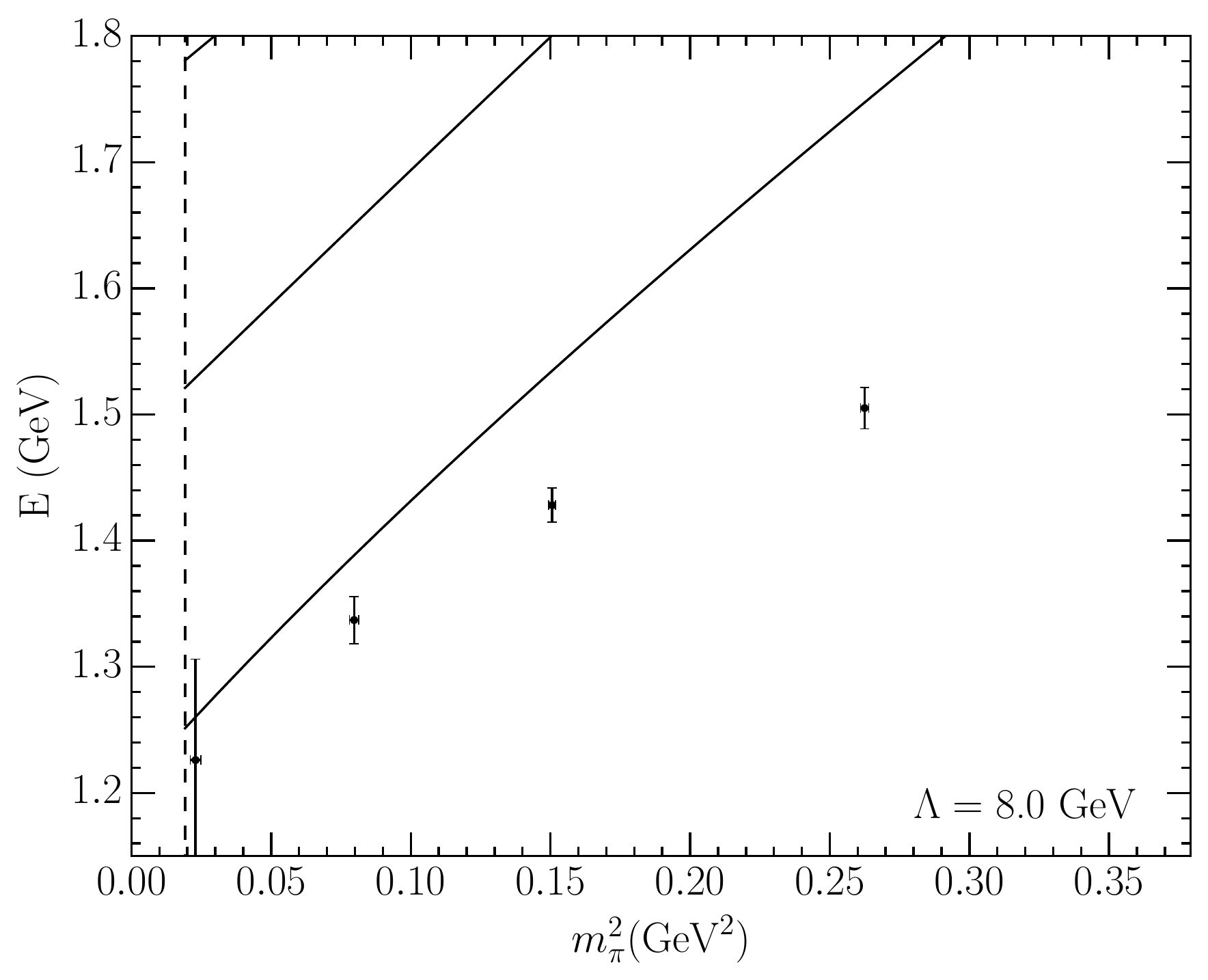}
  \caption{Pion mass dependence of the finite-volume HEFT eigenvalues at $L$ = 2.99 fm for
    increasing values of the regulator parameter \( \Lambda\,. \)  No bare basis state is
    present for \( \Lambda = 8.0 \) GeV.  The parameters for these fits are given
      by their corresponding entries in \tref{tab:1c}.  The solid black curves illustrate the
    finite-volume energy levels predicted by HEFT from fits to experimental phase shifts.  These
    lines are dressed by solid (red), short-dashed (blue) and long-dashed (green) highlights indicating
    states with the largest, second-largest and third-largest contribution from the bare basis
    state $\ket{\Delta_0}$ respectively.  Lattice QCD results for lowest-lying \( \Delta \) masses,
    denoted by the (black) points, are from the PACS-CS collaboration~\rref{PACS-CS:2008bkb}.
    The quoted $\chi^2$/DOF for each plot are for the lowest-lying energy
      eigenvalue with respect to the PACS-CS data points.  As these lattice results follow from
    local three-quark operators, they are expected to lie on a solid (red) energy eigenstate
    in the first case and perhaps on a short-dashed (blue) energy eigenstate when
      it is the lowest lying state of the spectrum.  The
    vertical dashed (black) line illustrates the physical pion mass.  }
  \label{fig:1b1c_Evmpi_3fm}
\end{figure*}

We begin by considering $\Lambda = 0.8$ GeV and constraining $\alpha_2$ and $\alpha_4$
by a fit to the \lqcd results.  The fit value is reported in \tref{tab:1c} and the fit is
illustrated in the upper-left panel of \fref{fig:1b1c_Evmpi_3fm}.  This process is repeated for \(
\Lambda = 1.6 \) and \( \Lambda = 4.0 \).  With the simple residual series of
\eref{eq:alpha0def}, the \lqcd data is described almost equally well with any of the three
regulator parameters selected.

In the bottom-right panel of \fref{fig:1b1c_Evmpi_3fm}, the \( \Lambda = 8.0 \) GeV case is
demonstrated.  As shown in \fref{fig:0b1c_phase_8000MeV}, at 8 GeV a bare state is no longer
required to describe the experimental scattering data.  As a result, there is no opportunity for
residual-series contributions, and the pion-mass extrapolation is performed without a bare mass.
As demonstrated here, without a bare state the correct energy eigenvalue is only obtained at the
physical point, and a system without a bare state is completely unsuitable for any extrapolations
from the physical point.  This is in agreement with conclusions from other analyses such as
\refref{Ren_2020}, where as shown in Fig. 2, the \( P_{33} \) scattering data is unable to be
reproduced without the introduction of an explicit degree of freedom for the \( \Delta
\)(1232).

\begin{figure}
  \centering
  \includegraphics[width=0.46\textwidth]{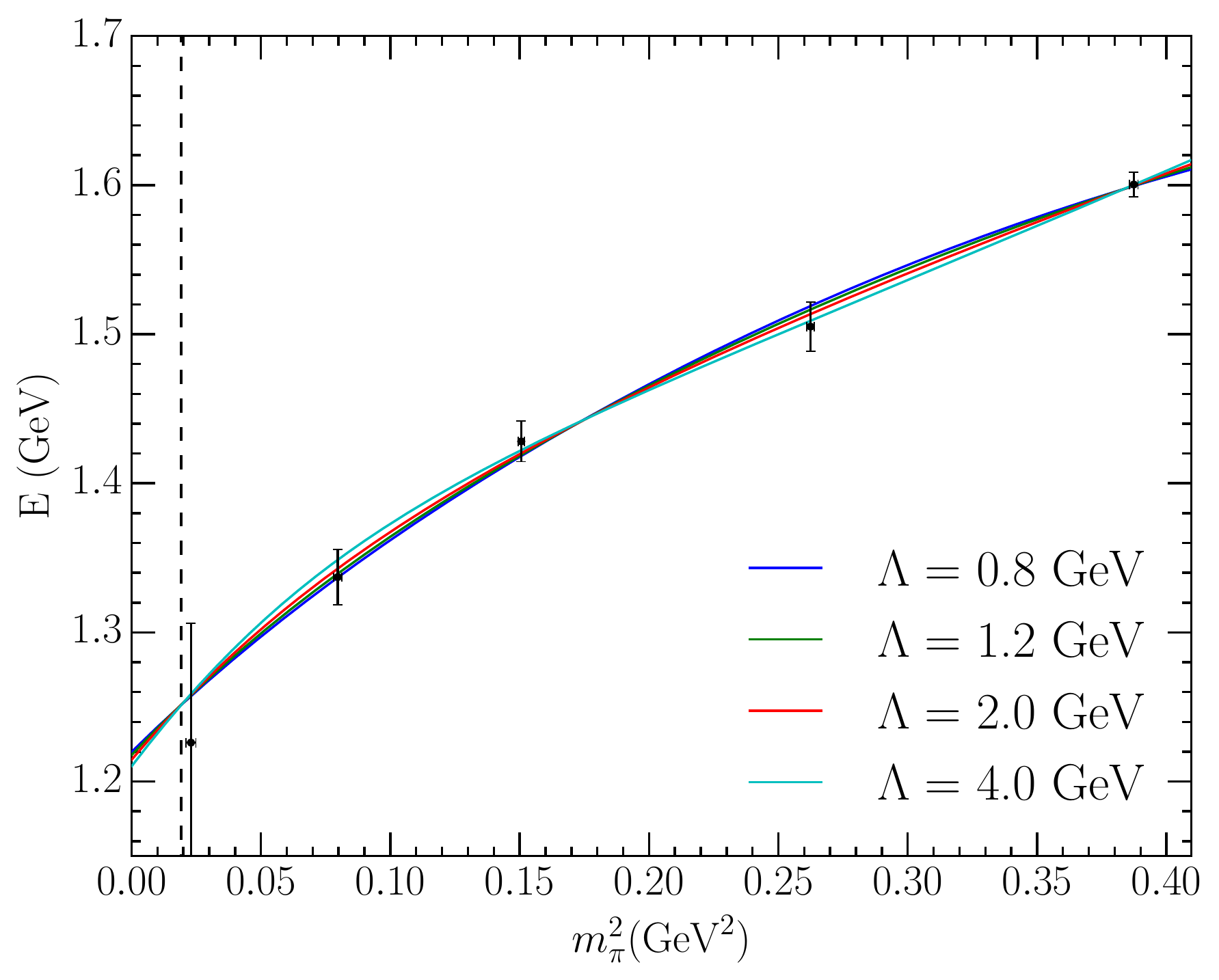}
  \caption{Pion mass dependence of the lowest-lying finite-volume HEFT eigenvalue at \( L = 2.99 \)
    fm using a dipole regulator, where the data points are the PACS-CS data.  Several parameter
    sets corresponding with each value of \( \Lambda \) are overlapped, each with a corresponding
    bare mass expansion fit to the PACS-CS data.}
  \label{fig:1b1c_reg_comparison}
\end{figure}

To further illustrate the robust nature of the pion-mass extrapolations away from the physical
point, we superpose results for four different values of the regulator parameter in
\fref{fig:1b1c_reg_comparison}.  While $\Lambda = 4.0$ GeV is not physically motivated, the
variation in the curves remains small.

In summary, a wide range of values for the regulator parameter \( \Lambda \) are able to give the
correct pion-mass extrapolation in accordance with the \lqcd data from PACS-CS.  As there is no
preference between these different values to be found in comparison with \lqcd, there is a freedom
to choose a value for \( \Lambda \) which suits other requirements.  A value of \( \Lambda = 0.8 \) GeV is
both in accord with findings from other models, such as the cloudy bag
model~\cite{Theberge:1980ye,Thomas:1982kv}, and is small enough such that computational
requirements are minimised.

\subsection{Gaussian Regulator Dependence}

\begin{figure}
  \centering
  \includegraphics[width=0.46\textwidth]{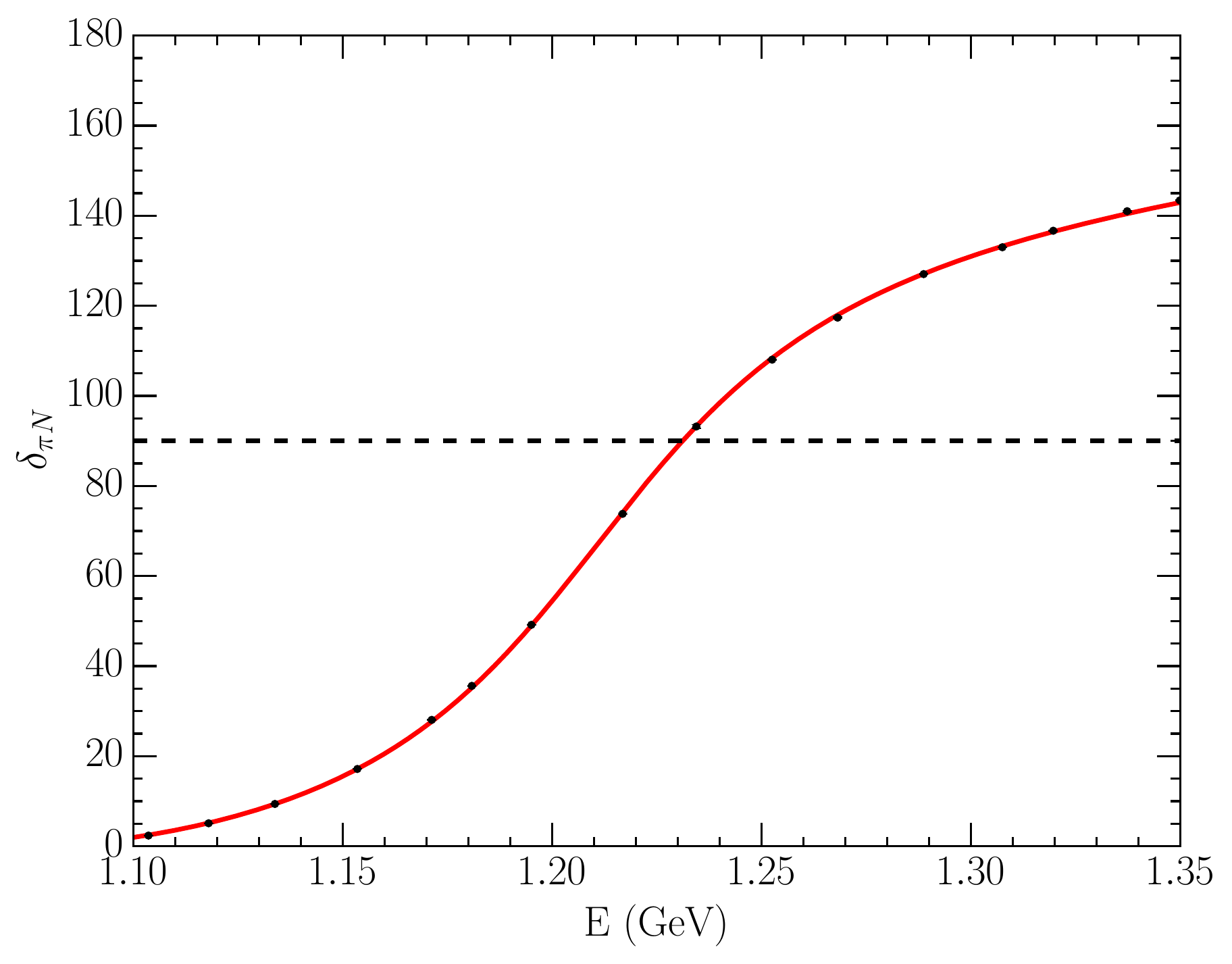}
  \caption{\(P\)-wave \( \pi N \) phase shifts
    for a Gaussian regulator, where the solid points are experimental data obtained from
    \refref{site:SAID,Workman:2012hx}, the solid line is the fit using HEFT to the data, and the dashed line
    represents a phase shift of 90 degrees. A Gaussian regulator with \( \Lambda = 0.8 \) GeV gives a \( \chi^2 \)/DOF of 18.8.}
  \label{fig:1b1c_gauss_inf}
\end{figure}

As a simple test of the model-dependence of HEFT, the analysis using a dipole regulator can be
repeated in part using a Gaussian regulator, as defined in \eref{eq:Gaussian_reg}.  Beginning with
\( \Lambda = 0.8 \) GeV, a similar quality description of the experimental scattering data can be
obtained, as seen in \fref{fig:1b1c_gauss_inf}, with a \( \chi^2 \)/DOF of 18.8.

Again, we note that if we follow Refs.~\cite{Fettes:1998ud} and \cite{Meissner:1999vr} and assign a
3\% or 5\% uncertainty to the experimental scattering data, our fit is shown to provide a superior
description of the experimental scattering data.  Whereas Refs.~\cite{Fettes:1998ud} and
\cite{Meissner:1999vr} report $\chi^2$/DOF values exceeding 0.77, our fit provides smaller values.
The introduction of 3\% uncertainties provides a $\chi^2$/DOF of 0.08 for fits to 1.35 GeV with 23
DOF.  Similarly 5\% uncertainties provide a $\chi^2$/DOF of 0.03.  Again, the introduction of 1\%
uncertainties is sufficient to reduce our $\chi^2$/DOF $\lesssim 1$.

\begin{figure}
  \centering
  \includegraphics[width=0.46\textwidth]{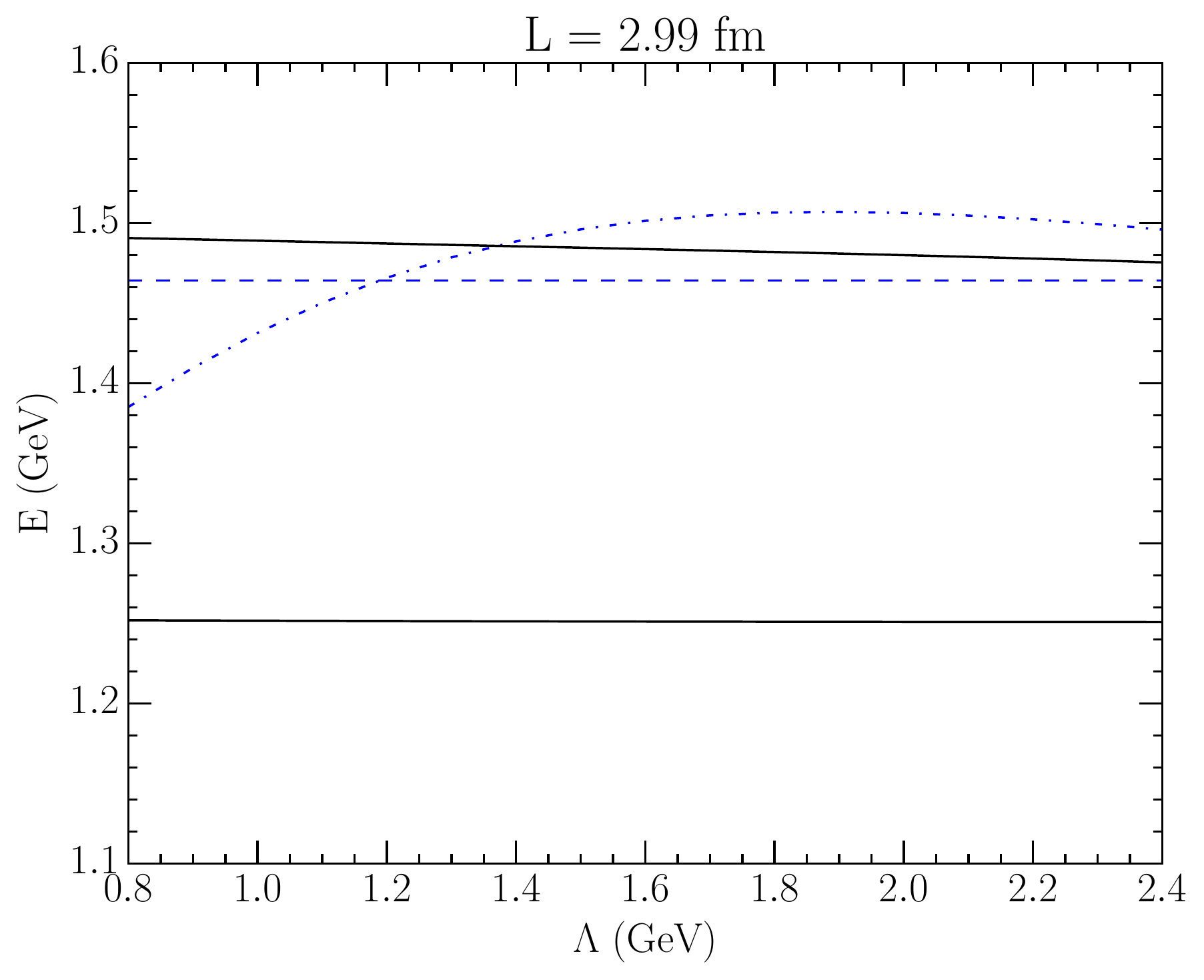}
  \caption{Dependence of the lowest lying eigenvalues of the finite-volume Hamiltonian on \(
    \Lambda \) for a Gaussian regulator at \( L = 2.99 \) fm. The solid (black) lines are the
    eigenvalues, the horizontal dashed (blue) line is the \( \pi N(k=1) \) basis state, and the
    curved dot-dashed (blue) line is the mass of the bare \( \Delta\,. \) The Hamiltonian was
    constrained to experimental data with $E \alt 1350$ MeV.}
  \label{fig:1b1c_gauss_EvLambda}
\end{figure}

As shown in \fref{fig:1b1c_gauss_EvLambda}, varying \( \Lambda \) over a modest range of 0.8 GeV to
2.4 GeV, the ground-state energy is similarly invariant
as demanded by the L\"uscher formalism contained within HEFT.
By comparing with \fref{fig:1b1c_EvLambda} it is clear that within the fitting region of \( E \leq
1350 \) MeV, both Gaussian and dipole functional forms are equivalent.

\begin{figure}
  \centering
  \includegraphics[width=0.46\textwidth]{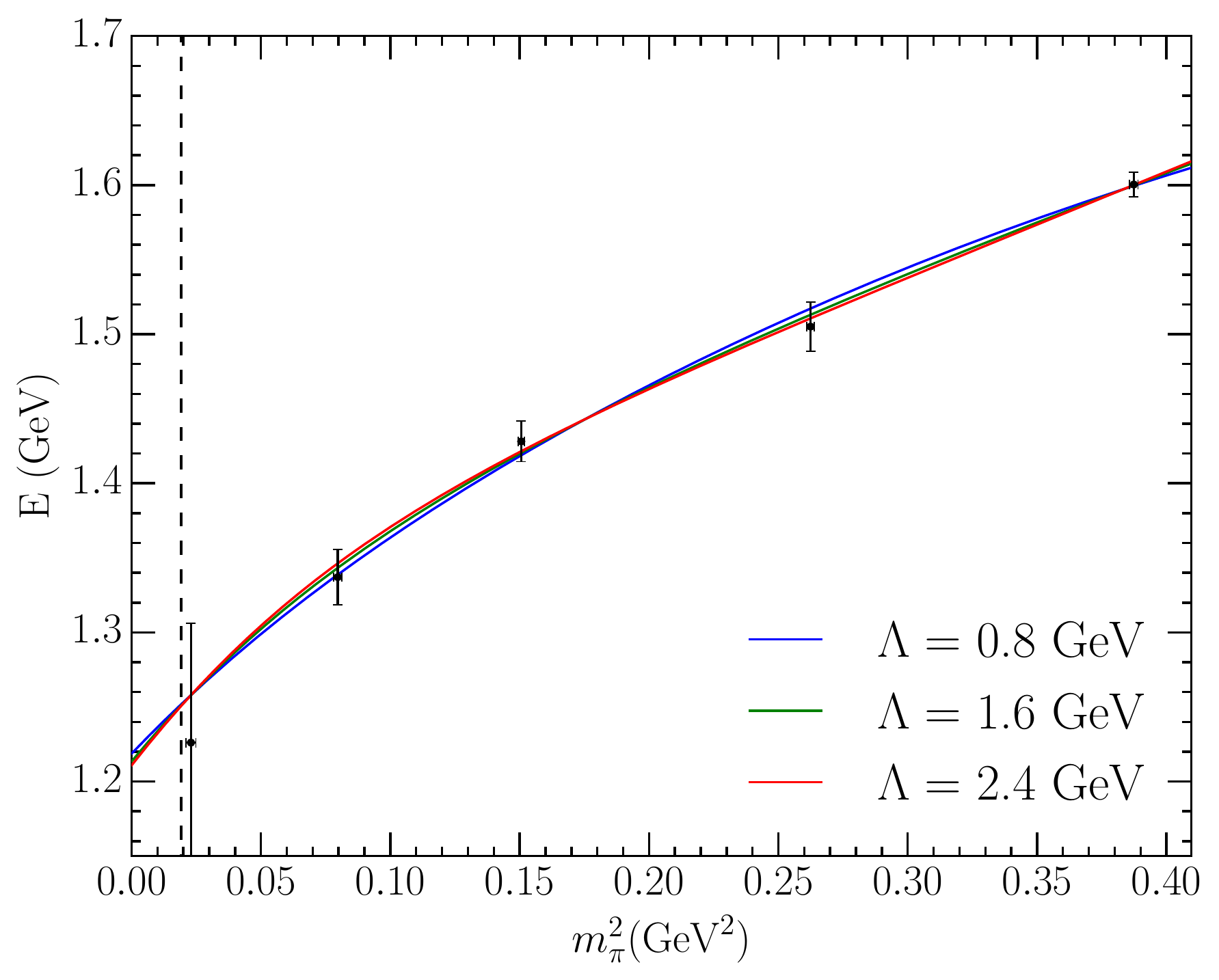}
  \caption{Pion mass dependence of the lowest-lying finite-volume HEFT eigenvalue at \( L = 2.99 \)
    fm using a Gaussian regulator, where the data points are the PACS-CS data.  Four parameter
    sets corresponding to each value of \( \Lambda \) are overlapped, each with a corresponding
    bare mass expansion fit to the PACS-CS data.}
  \label{fig:1b1c_Evmpi_Gaussian}
\end{figure}

Finally, by extending to unphysical pion masses and comparing with the PACS-CS data at \( L = 2.99 \) fm, it can be seen in \fref{fig:1b1c_Evmpi_Gaussian} that a Gaussian form factor is able to obtain a similarly good description of the \lqcd data, with \( \chi^2 \)/DOFs ranging from 0.2 - 0.4.
In light of the model-independence demonstrated, this study will only utilise a dipole regulator
for the forthcoming analysis.

\section{Two-Channel Analysis}
\label{sec:2c}
\subsection{Fitting Experimental Data}

The power of the constraints provided by experimental scattering data on the predictions of HEFT
via the L\"uscher formalism is manifest in Fig.~\ref{fig:1b1c_EvLambda}.  And while this model
independence applies only at the physical point, Figs.~\ref{fig:1b1c_reg_comparison} and
\ref{fig:1b1c_Evmpi_Gaussian} illustrate only a subtle model dependence in describing the
quark-mass dependence of the finite-volume energies.
Thus, it is desirable to extend the energy range considered beyond
the \( \pi\Delta \) threshold in an effort to describe excitations of the $\Delta$, in addition to
the lowest lying resonance.  To proceed, we will conduct a similar analysis including higher
energies by introducing a second scattering channel, a \( \pi\Delta \) channel.

Previously we were able to describe the scattering data to 1350 MeV with a \( \pi N \) scattering
channel.  To describe the data up to 1650 MeV, we require the next \( \pi\Delta \) scattering
channel to account for inelasticity beyond the \( \pi\Delta \) threshold.  To describe the
interactions in the \( \pi\Delta \) channel, we use the same functional form as \eref{eq:g} to
describe the \( \Delta^{(0)} \rightarrow \pi\Delta \) interaction, and the same potential as
\eref{eq:V} to describe the \( \pi\Delta \rightarrow \pi\Delta \) and \( \pi\Delta \rightarrow \pi
N \) interactions.
This introduces three new fit parameters for the system: \( g_{\pi\Delta}^{\Delta}\,, \) \( v_{\pi N\pi \Delta}\,, \) and \( v_{\pi\Delta\pi\Delta}\,. \)

In fitting the experimental data, we must reproduce both the \( \pi N \) phase shift \( \delta_{\pi
  N}\,, \) and the inelasticity \( \eta\,. \) Unlike the single-channel case, the increased number
of parameters and the inclusion of the inelasticity make it difficult to fit the data for larger values of \( \Lambda \).
Further difficulty is introduced by the absence of phase shift data in the \( \pi\Delta \) channel.
Variation in the parameters is more unstable, and good fits to the scattering data above approximately \( \Lambda = 1.2 \) GeV are elusive.

\begin{table}
  \centering
  \caption{Two-channel fit parameters constrained to \( \pi N \) scattering data up to 1650 MeV. Here, \( \Lambda \) is fixed for each of the two fits while all other parameters are allowed to vary.}
  \begin{center}
    \begin{ruledtabular}
      \begin{tabular}{cccc}
        Parameter & Fit V & Fit VI \\
        \noalign{\smallskip}
        \hline
        \noalign{\smallskip}
        \( m_{\Delta}^{(0)} / \text{ GeV} \) & 1.3837 & 1.4405  \\
        \( g_{\pi N}^{\Delta} \)            & 0.1286  & 0.1041 \\
        \( g_{\pi\Delta}^{\Delta} \)         & 0.1324  & 0.0171 \\
        \( v_{\pi N,\pi N} \)               & -0.0103 & -0.0233 \\
        \( v_{\pi N,\pi\Delta} \)            & -0.0811 & -0.0220 \\
        \( v_{\pi\Delta,\pi\Delta} \)         & -0.0015 & -0.0645 \\
        \( \Lambda / \text{ GeV} \)       & 0.8000  & 1.2000 \\
        \noalign{\smallskip}
        \hline
        \noalign{\smallskip}
        DOF & 27 & 27 \\
        \( \chi^2 \) & 304.29 & 377.67 \\
        \( \chi^2 / \text{ DOF} \) & 11.27 & 13.99 \\
        \noalign{\smallskip}
        \hline
        \noalign{\smallskip}
        \( \alpha_{2} / \text{ GeV}^{-1} \) & 0.893 & 0.636 \\
        \( \alpha_{4} / \text{ GeV}^{-3} \) & -0.481 & -0.089 \\
        \noalign{\smallskip}
        \hline
        \noalign{\smallskip}
        Pole 1 / GeV & \( 1.210 - 0.049i \) & \( 1.211 - 0.049i \)\\
        Pole 2 / GeV & \( 1.434 - 0.207i \) & \( 1.449 - 0.053i \)\\
      \end{tabular}
    \end{ruledtabular}
  \end{center}
  \label{tab:2c}
\end{table}

For the SAID data up to 1650 MeV, we are able to obtain fits between \( \Lambda = 0.8 \) GeV and \(
\Lambda = 1.2 \) GeV with a significantly reduced \( \chi^2 \) compared to the single-channel
case. The fit parameters at these two limits for $\Lambda$ are presented in \tref{tab:2c}.

Again, our fits provide an excellent description of the experimental scattering data.  Following
Refs.~\cite{Fettes:1998ud} and \cite{Meissner:1999vr} and assigning a 3\% or 5\% uncertainty to the
scattering data, we find the introduction of 3\% uncertainties provides a $\chi^2$/DOF of 0.26 for
fits to 1.65 GeV with 50 DOF.  Similarly 5\% uncertainties provide a $\chi^2$/DOF of 0.10.  This
contrasts Refs.~\cite{Fettes:1998ud} and \cite{Meissner:1999vr} where they report $\chi^2$/DOF
values exceeding 0.77 for fits constrained within 1.3 GeV.

While both fits generate a pole at approximately the PDG pole position of \( 1.210 - 0.050i\, \)
GeV, they are not able to describe the scattering data equally well.  It is also worth noting that
a second pole was found for each of the two sets of fit parameters, though with significantly
different values for the imaginary components of each pole.  These values are somewhat comparable
to the PDG pole position for the \( \Delta \)(1600) at \( (1.510\pm0.050) - (0.135\pm0.035)i\, \)
GeV \rcite{10.1093/ptep/ptaa104}, though the imaginary component of the second pole for Fit VI is
considerably smaller.

\begin{figure*}
  \centering
  \includegraphics[width=0.46\textwidth]{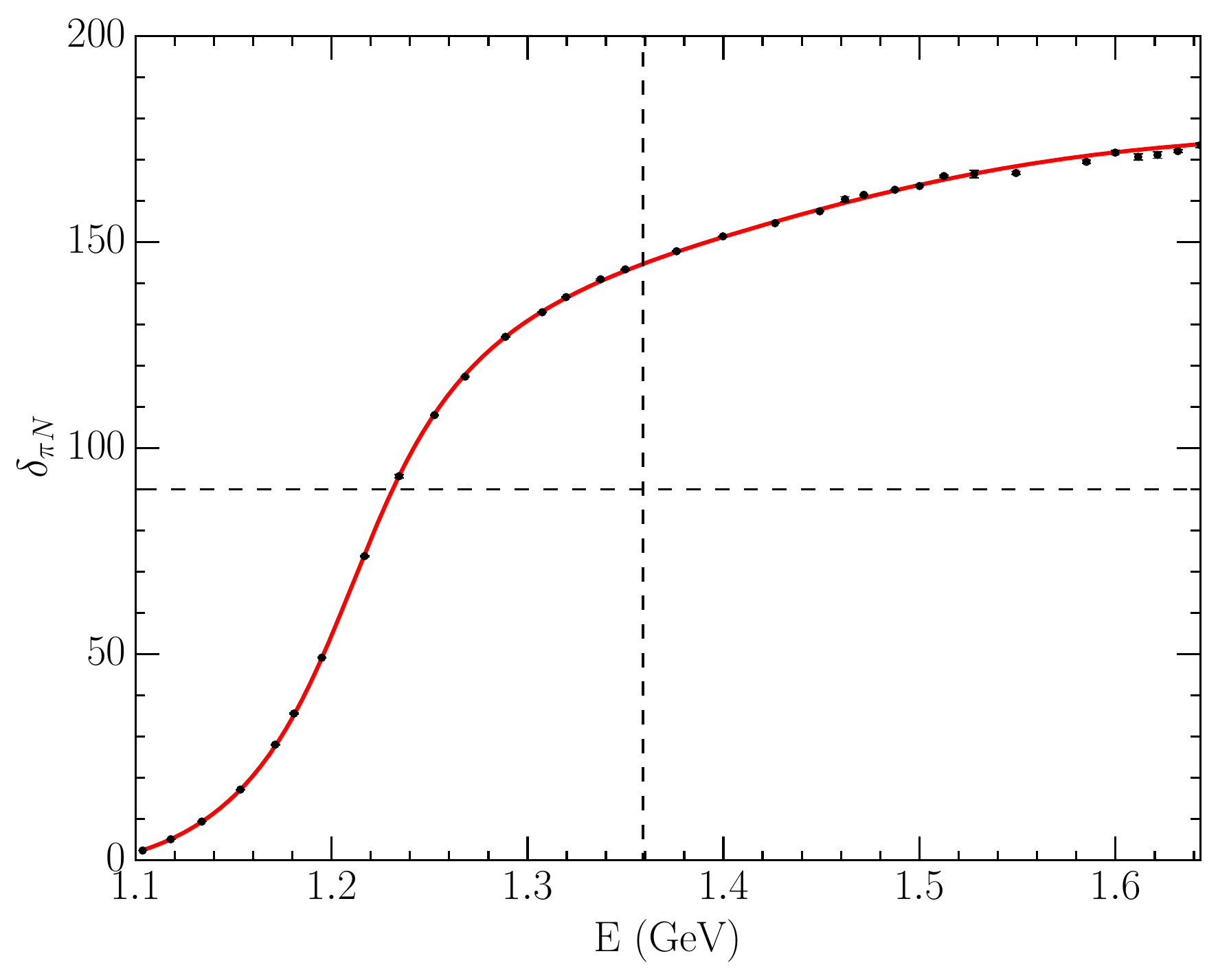}
  \includegraphics[width=0.46\textwidth]{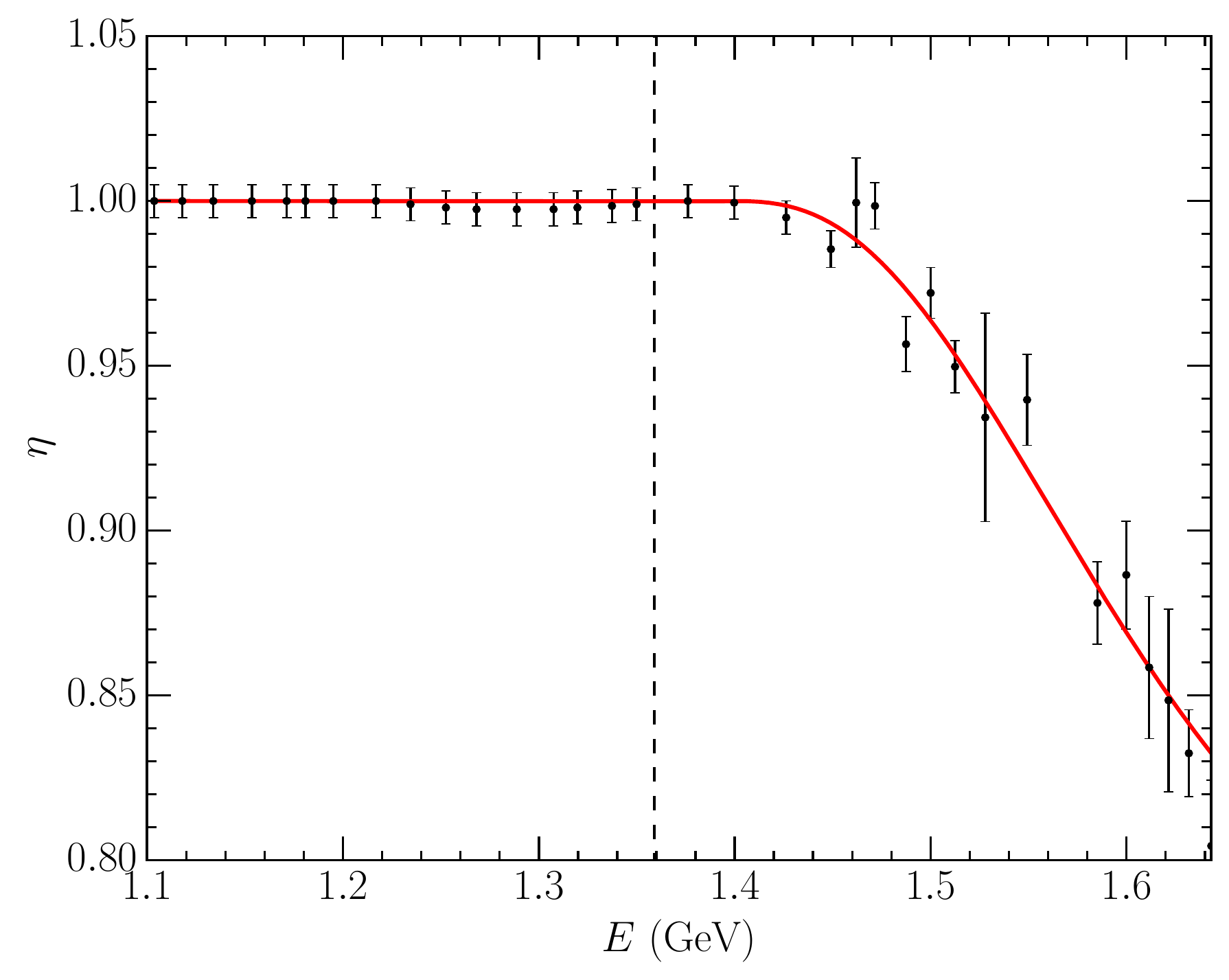}~\\
  \includegraphics[width=0.46\textwidth]{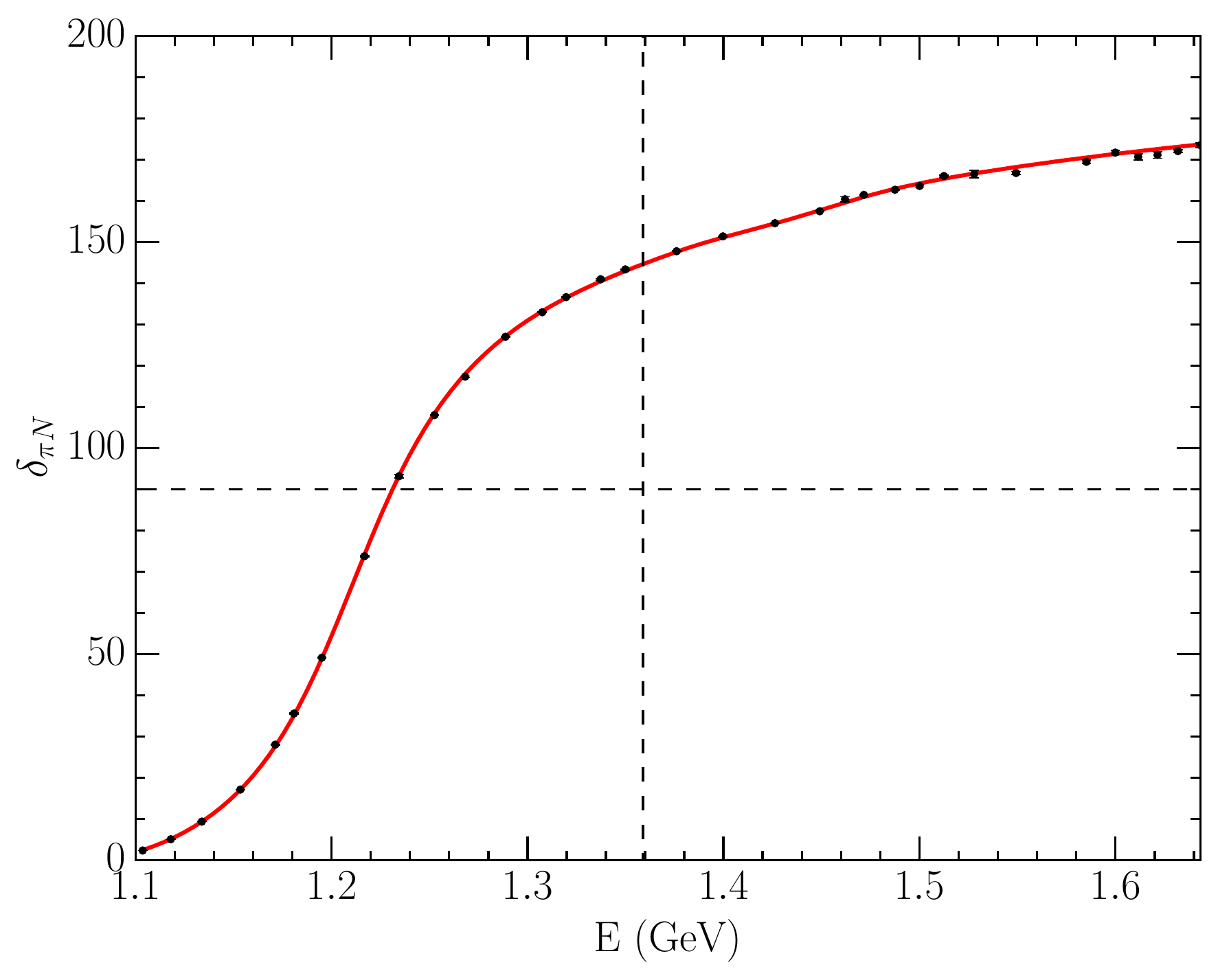}
  \includegraphics[width=0.46\textwidth]{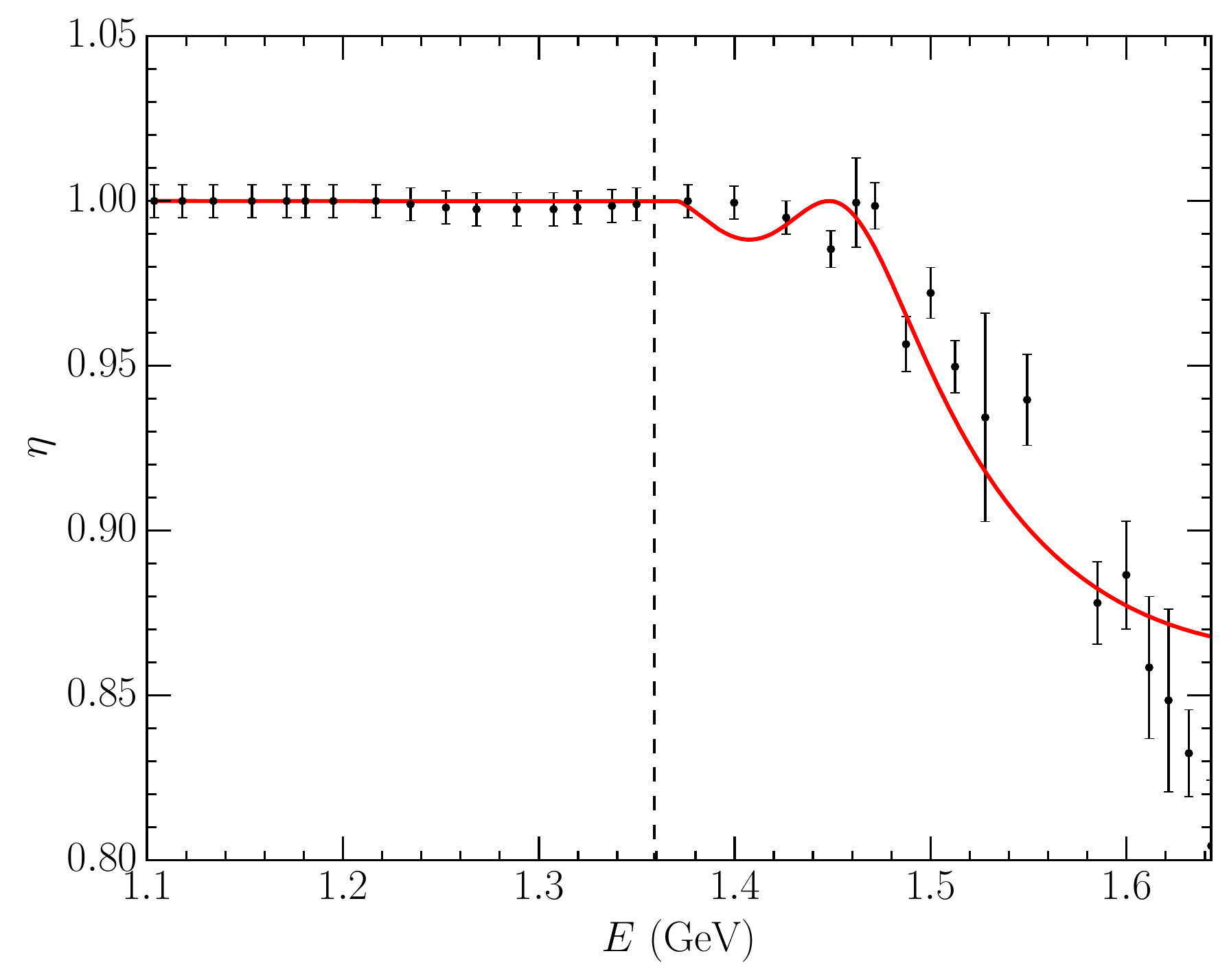}~\\
  \caption{\(P\)-wave \( \pi N \) phase shifts and inelasticities.  The solid points are
    experimental data obtained from \refref{site:SAID,Workman:2012hx}. The solid (red) curve is the fit of HEFT
    to the scattering data. The horizontal dashed line highlights a phase shift of 90 degrees,
    while the vertical dashed line illustrates the position of the \( \pi\Delta \) threshold. The
    upper plots illustrate the best fit when the regulator parameter \( \Lambda = 0.8 \) GeV (Fit
    V), while the lower plots illustrate the best fit when the regulator parameter \( \Lambda =
    1.2 \) GeV (Fit VI). Only phenomenologically motivated values associated with the induced pseudoscalar
    form factor of a baryon are able to give a good description of the inelasticity.}
  \label{fig:1b2c_fits}
\end{figure*}

As illustrated in \fref{fig:1b2c_fits}, both fits are able to describe the scattering phase shifts
up to 1650 MeV; however, only the smaller value of \( \Lambda = 0.8 \) GeV is able to give a good
description of the inelasticity.

In order to observe the effect these differing fits have on the
corresponding lattice energy levels, we explore the lattice volume and \( \Lambda \) dependence of
the finite-volume eigenmodes in the next two sections.

\subsection{Finite-Volume Dependence}

\begin{figure}
  \centering
  \includegraphics[width=0.45\textwidth]{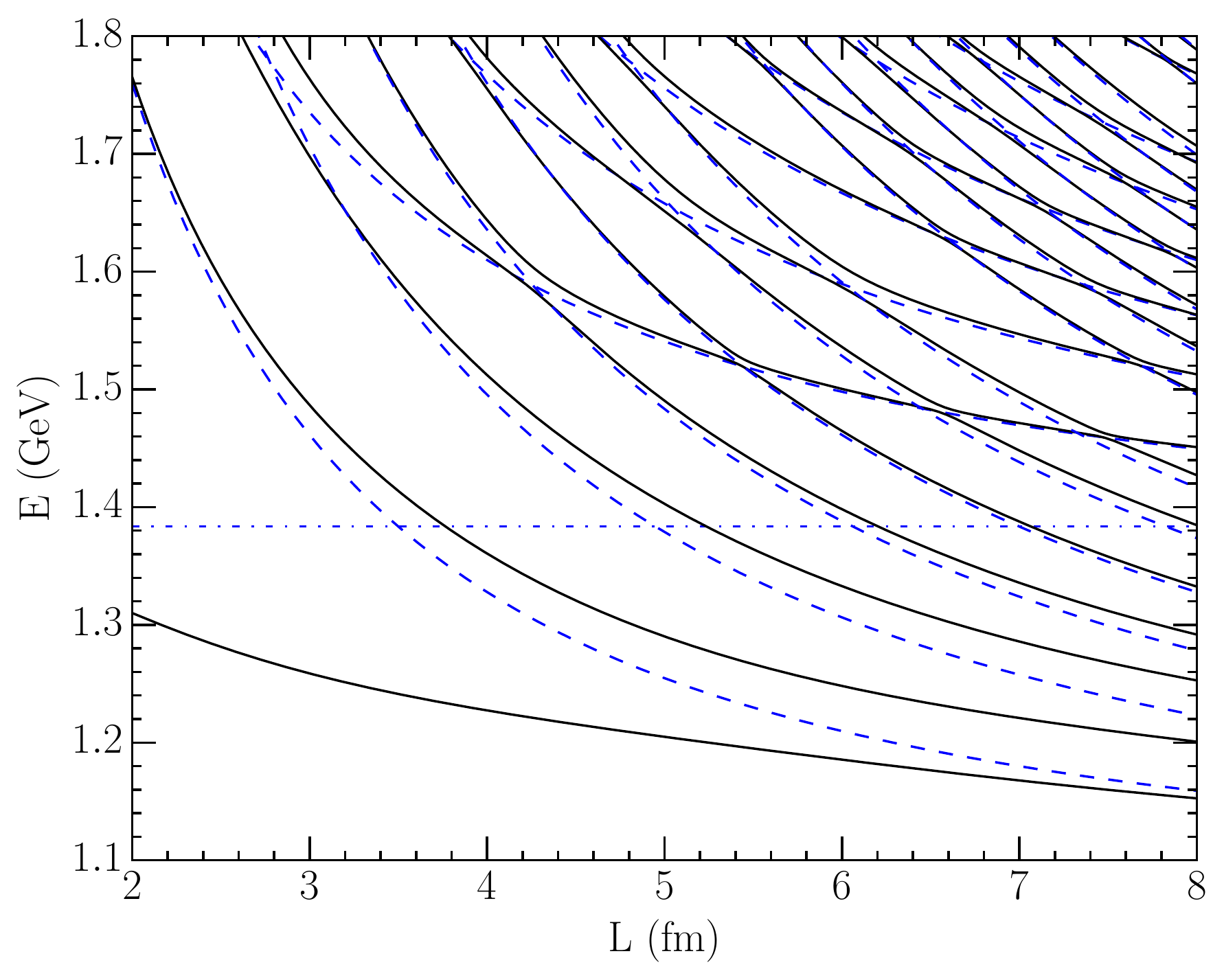}
  \caption{Two-channel lattice volume dependence of the energy eigenvalues of the Hamiltonian
    for the fit to experimental data with $\Lambda = 0.8$ GeV. The solid lines
    represent the energy eigenvalues.  The horizontal dot-dashed line is the bare mass and the
    curved dashed lines are the \( \pi N \) and \( \pi\Delta \) scattering states at $k = ( n_x^2
    + n_y^2 + n_z^2 )^{1/2}\, 2\pi/L$.}
  \label{fig:1b2c_EvL_800MeV}
\end{figure}

\begin{figure}
  \centering
  \includegraphics[width=0.45\textwidth]{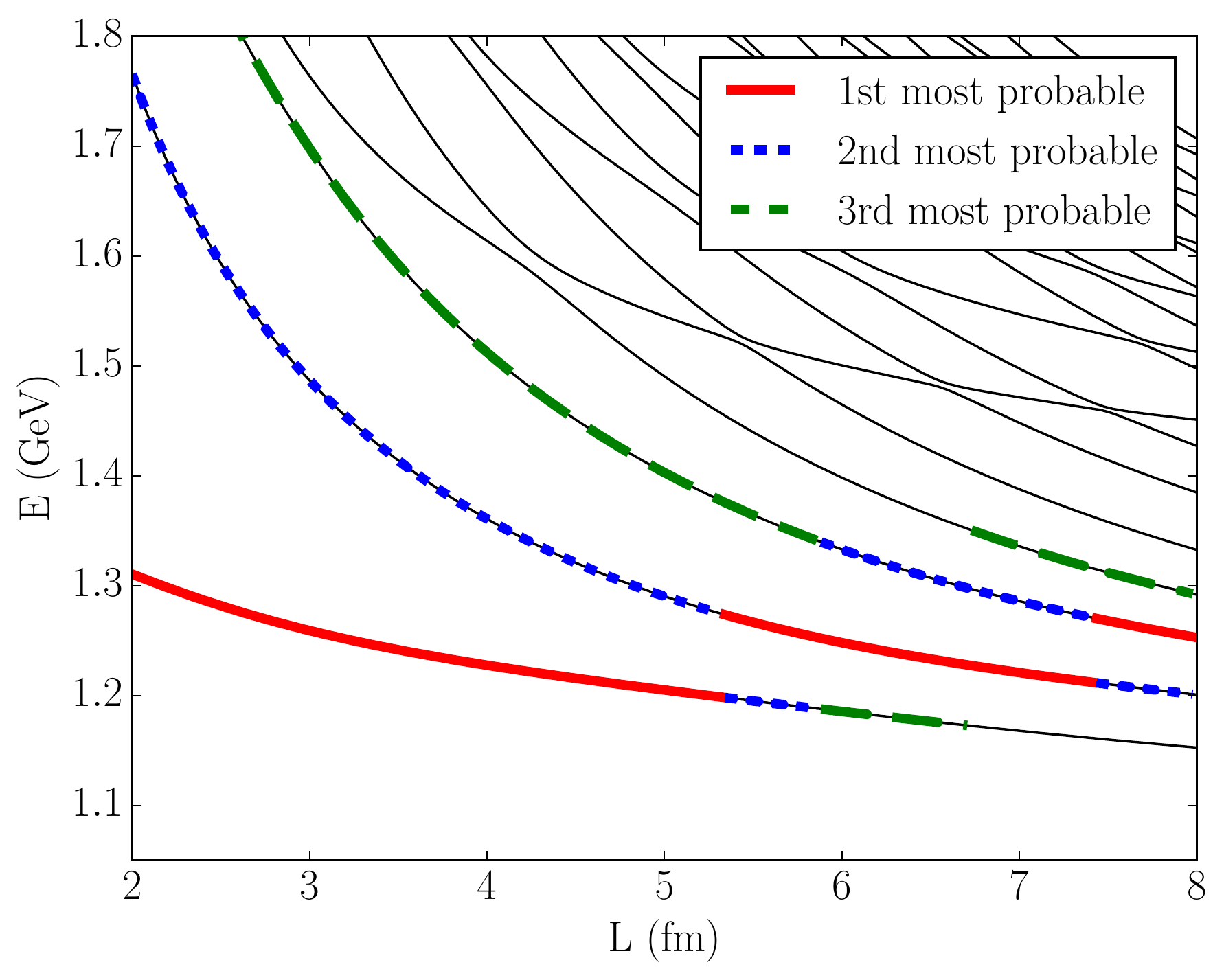}
  \caption{Two-channel lattice volume dependence of the energy eigenvalues of the Hamiltonian
    for the fit to experimental data with $\Lambda = 0.8$ GeV. The solid (red),
    short-dashed (blue) and long-dashed (green) highlights on the energy eigenvalues correspond to the states
    with the largest, second-largest and third-largest contribution from the bare \( \Delta \)
    basis state respectively.}
  \label{fig:1b2c_EvL_Lam800MeV_bare}
\end{figure}

Adding an additional channel to the Hamiltonian matrix only requires an additional row/column for
each channel at each momentum \( k_n\,. \) Therefore the free Hamiltonian takes the form
\begin{eqnarray}
&&H_0 =\\
&&\,\text{diag}\left( m_{\Delta}^{(0)}, \,\omega_{\pi N}(k_1), \,\omega_{\pi\Delta}(k_1),
                                            \,\omega_{\pi N}(k_2), \,\omega_{\pi\Delta}(k_2),
  \,\ldots \right)\,. \nonumber
\end{eqnarray}
Similarly, we can write the interaction Hamiltonian as
\begin{equation}
  H_I =
  \begin{pmatrix}
    0 & \bar{G}_{\pi N}^{\Delta}(k_1) & \bar{G}_{\pi\Delta}^{\Delta}(k_1) & \hdots \\
    \bar{G}_{\pi N}^{\Delta}(k_1) & \bar{V}_{\pi N,\pi N}(k_1,k_1) & \bar{V}_{\pi N,\pi\Delta}(k_1,k_1) & \hdots \\
    \bar{G}_{\pi\Delta}^{\Delta}(k_1) & \bar{V}_{\pi\Delta,\pi N}(k_1,k_1) & \bar{V}_{\pi\Delta,\pi\Delta}(k_1,k_1) & \hdots \\
    \vdots & \vdots & \ddots & \vdots
  \end{pmatrix} \, .
\end{equation}
Taking the Hamiltonian and solving the eigenvalue equation for varying lattice lengths, \( L\,, \)
we can generate the finite-volume spectra seen in \fref{fig:1b2c_EvL_800MeV} and
\fref{fig:1b2c_EvL_Lam800MeV_bare} for $\Lambda = 0.8$ GeV.

Below the \( \pi\Delta \) threshold of approximately 1350 MeV, the two-channel finite-volume
spectrum has the same form as the single-channel spectrum, which is to be expected as both the
single-channel fit and the two-channel fit perform equally well below the \( \pi\Delta \)
threshold.  Above this threshold, however, mixing between \( \pi N \) and \( \pi\Delta \)
basis states results in avoided level crossings.
These are apparent in the energy eigenvalues above the $\pi \Delta$ threshold of 1350 MeV.  For the
smaller volumes where the states are forced to interact in the finite volume, the energy gaps are
significant, and can be used to infer scattering observables from the finite-volume spectrum.

\subsection{Regulator Parameter Dependence}

As in the single-channel analysis reported in \sref{sec:1c_reg_dep}, we proceed to understand how
this finite-volume spectrum depends on the choice of the regulator parameter \( \Lambda \, , \)
albeit over a smaller range of \( \Lambda\,. \) The range of \( \Lambda \) considered isn't a
problem however, as using our single-channel results to guide us, we expect that optimal results
will be obtained in the physically-motivated region around \( \Lambda = 0.8 \) GeV
for a dipole form factor.

\begin{figure}
  \centering
  \includegraphics[width=0.46\textwidth]{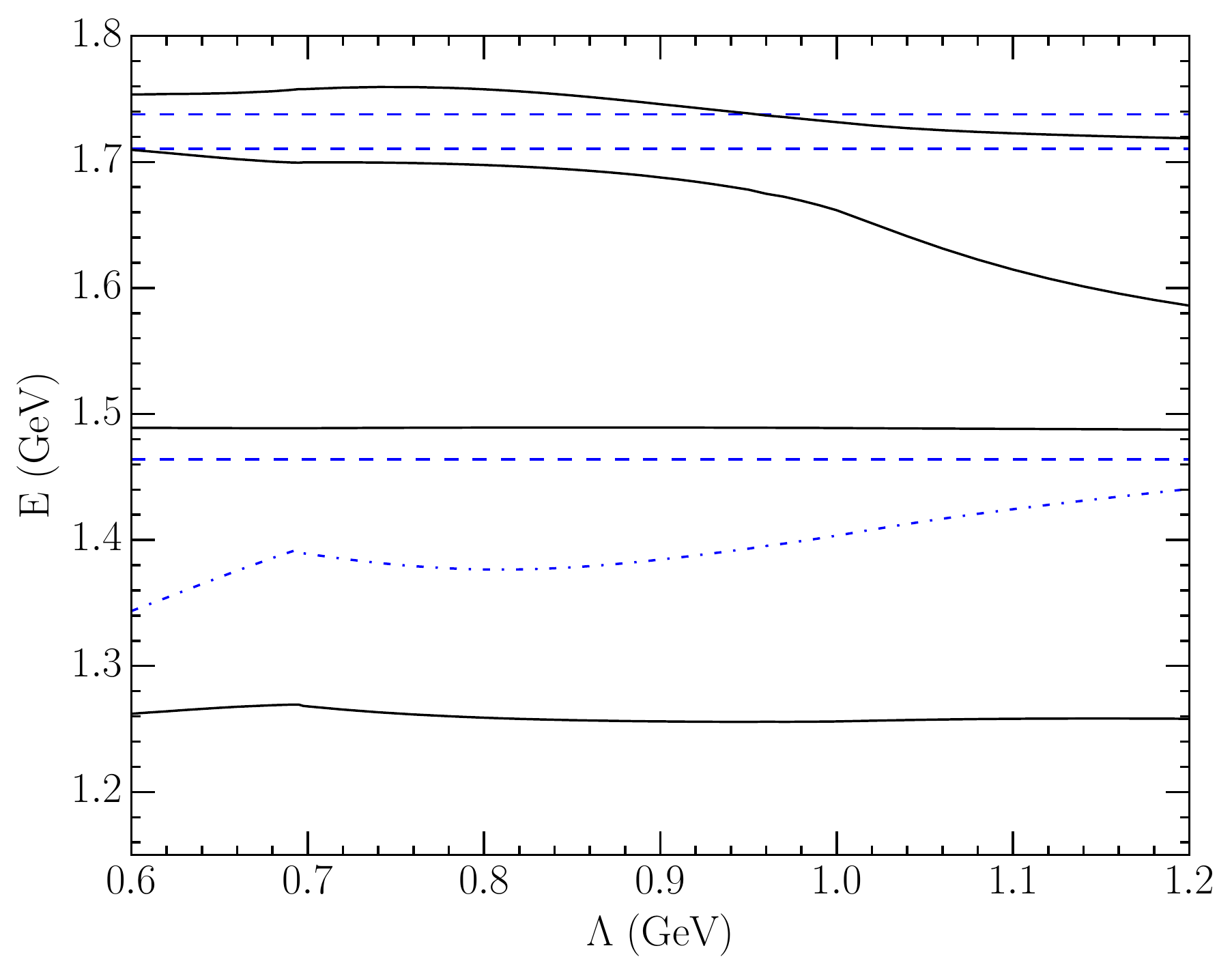}~\\
  \includegraphics[width=0.46\textwidth]{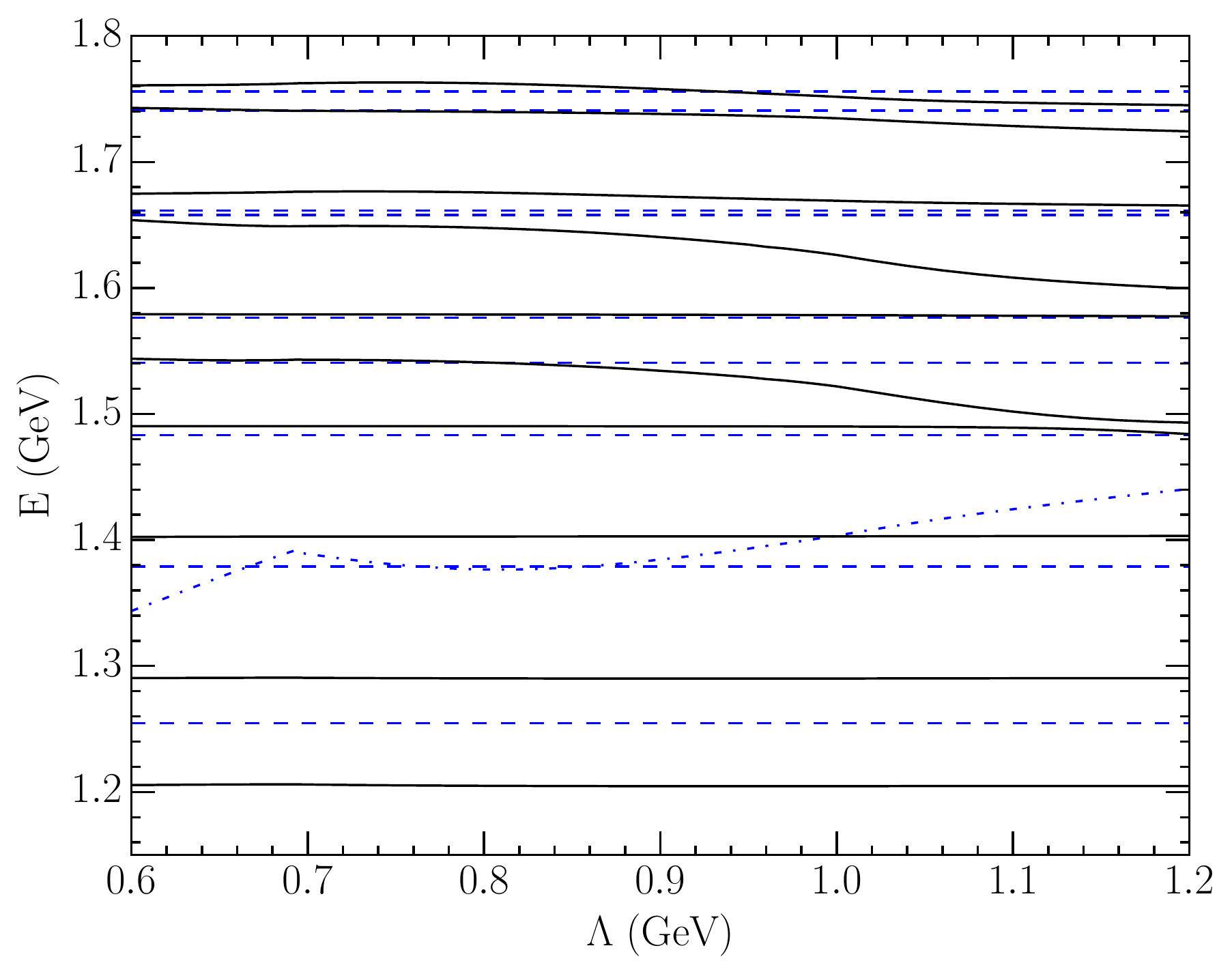}
  \caption{Regulator parameter, \( \Lambda \), dependence of the energy eigenvalues of the two
    scattering-channel HEFT finite-volume energy levels for the 2.99 fm lattice (top) and the 5.0
    fm lattice (bottom).  The solid (black) lines illustrate the energy eigenvalues, the thin
    horizontal dashed (blue) lines are the non-interacting energy levels of the \( \pi N \) and \(
    \pi\Delta \) basis states, and the thin dash-dot (blue) curve is the mass of the bare \( \Delta \)
    basis state.}
  \label{fig:1b2c_EvLambda}
\end{figure}

The results obtained by solving for the eigenvalues of the Hamiltonian with \( \Lambda \) varying
from 0.8 GeV to 1.2 GeV are presented in \fref{fig:1b2c_EvLambda}.  Here, for both \( L=2.99 \) fm
and \( L=5.0 \) fm, we see that, unlike the single-channel case, we observe a \( \Lambda \) dependence
of the energy eigenvalues within the range of energies considered in fitting the scattering data,
this time to 1650 MeV.

On the 3 fm lattice, the third state drops below 1650 MeV for large $\Lambda$.  On the 5 fm
lattice, both the fifth and seventh states display a strong $\Lambda$ dependence.  Thus, high
quality lattice QCD simulations covering several low-lying states hold the potential to constrain
$\Lambda$ and its associated parameter set.  Thus it is possible in principle to predict the
inelasticity from lattice QCD simulations.

Indeed, it is the lack of a complete experimental data set that prevents HEFT from maintaining
model independence via the L\"uscher formalism.  The presence of multiple open, coupled channels
constrained by data from only one scattering channel leaves the Hamiltonian model unconstrained.
This time, variation in $\Lambda$ leads to different roles for the scattering channels in describing
the $\pi N$ scattering data.  As we will see, lattice QCD can provide the additional information
required to constrain the Hamiltonian.

\begin{figure*}
  \centering
  \includegraphics[width=0.46\textwidth]{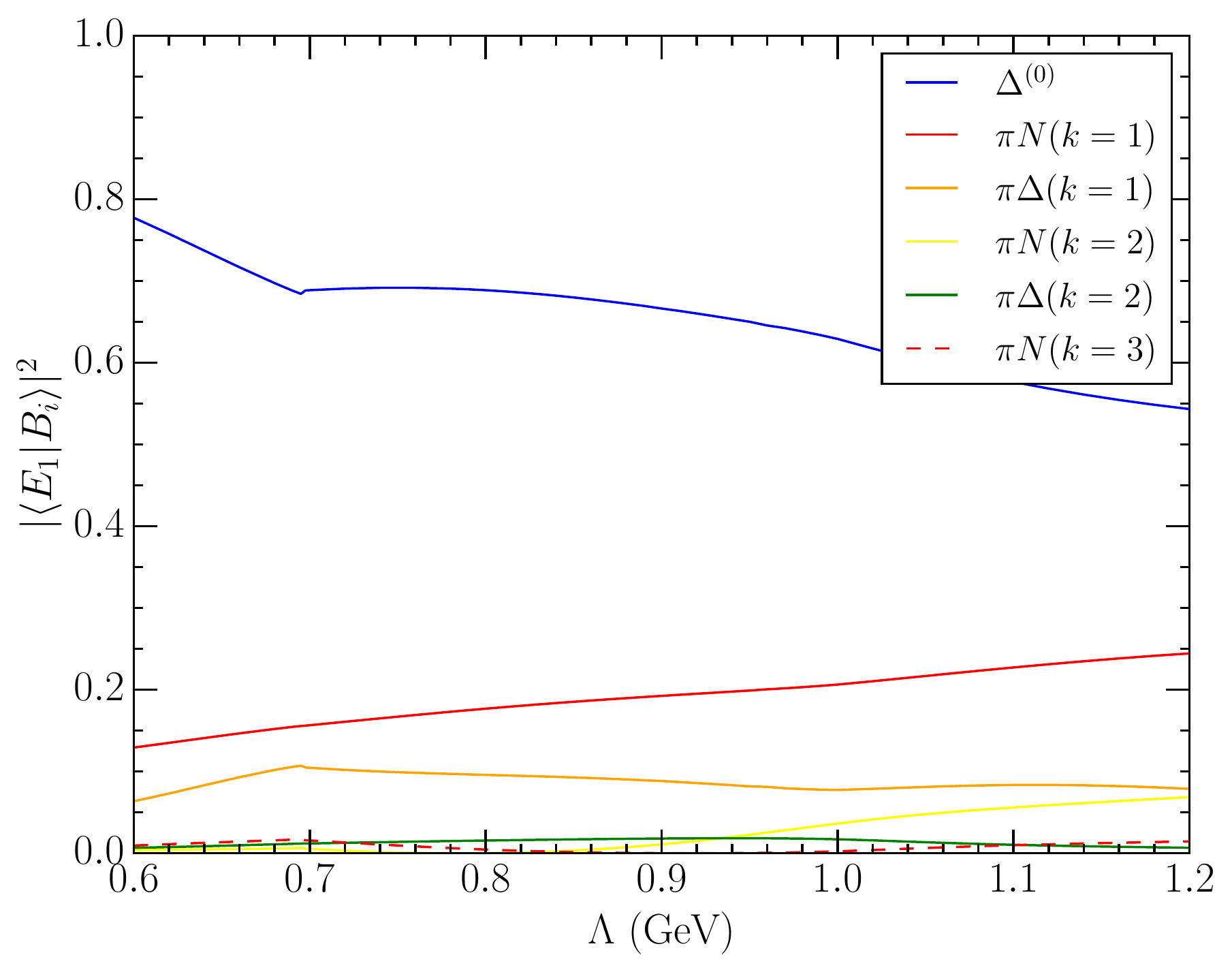}
  \includegraphics[width=0.46\textwidth]{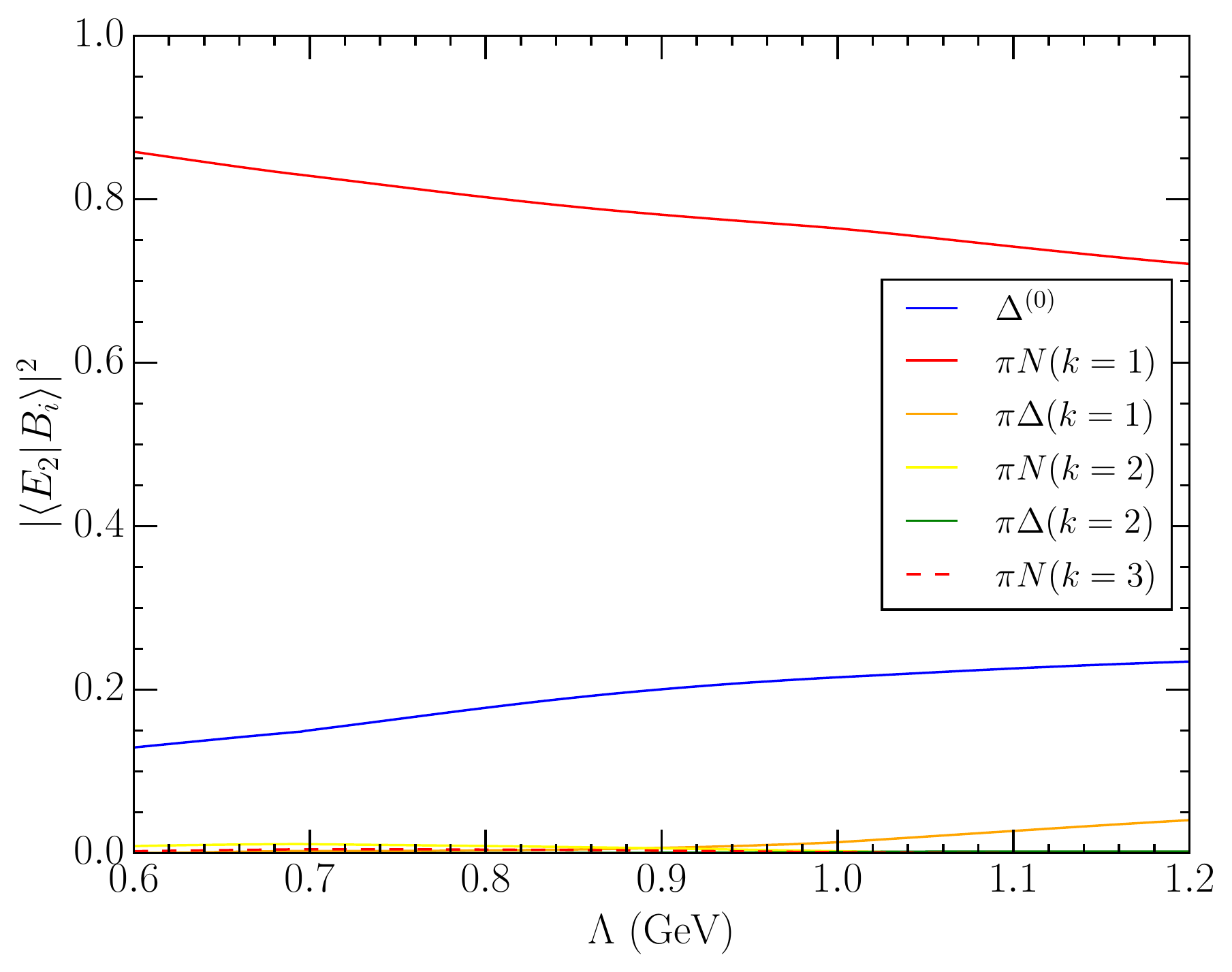}~\\
  \includegraphics[width=0.46\textwidth]{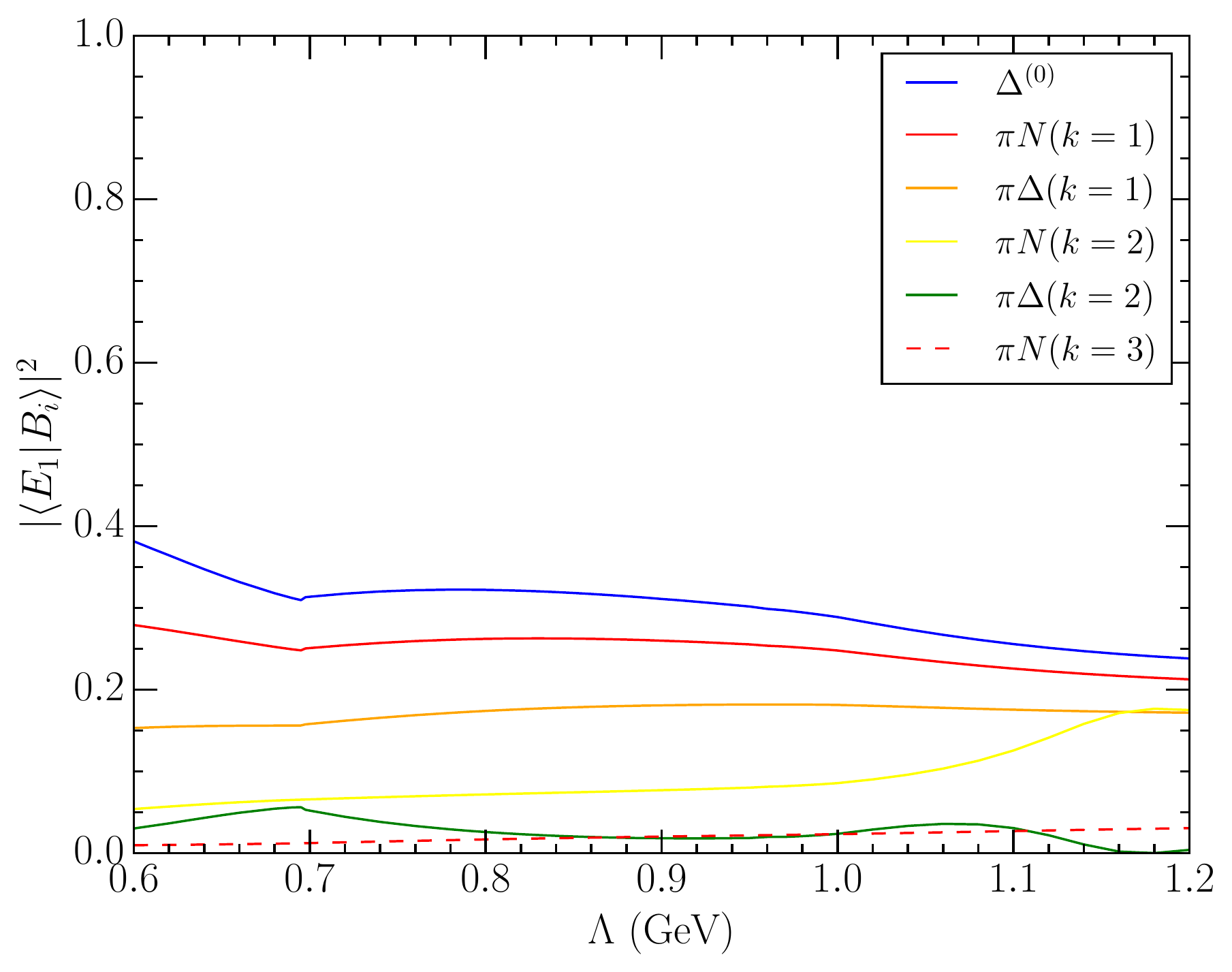}
  \includegraphics[width=0.46\textwidth]{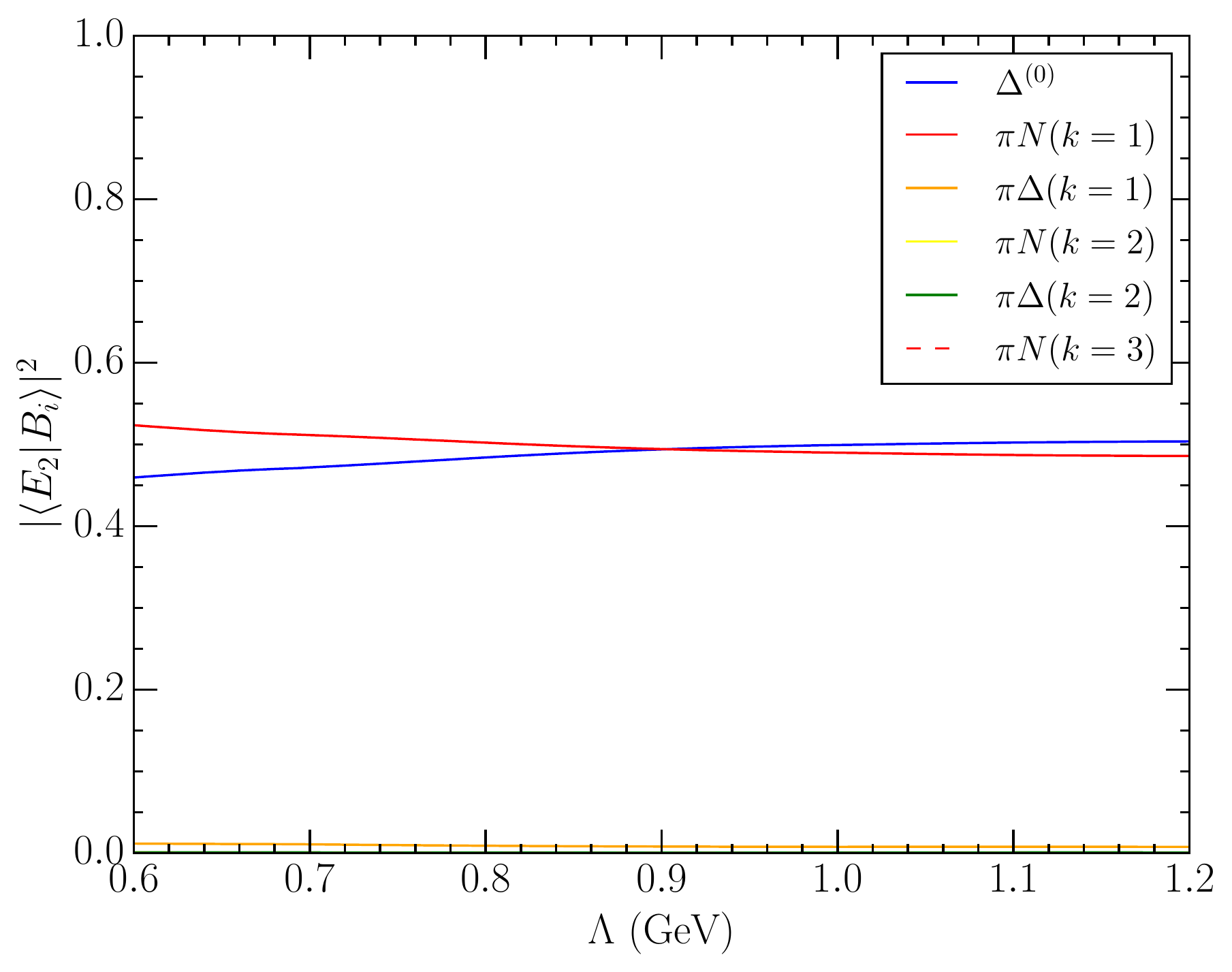}
  \caption{Regulator parameter, $\Lambda$, dependence of the eigenvectors describing the
    composition of the two lowest lying energy eigenstates in the two-channel case with \( \pi N \)
    and \( \pi\Delta \) scattering channels. The top two plots present results for a lattice volume of 2.99 fm,
    whereas the bottom two panels are for $L$ = 5.0 fm. The left plot shows the eigenvectors for the
    lowest-lying state, while the right plot shows the next energy eigenstate.  The bare basis
    states contribute most strongly to the lowest energy eigenstate at 2.99 fm.  At 5 fm, the
    lowest state is only just dominated by the bare basis state, with a strong contribution from many scattering states, while the first excited eigenstate is an approximately equal mixing of the bare state and the \( \pi N(k=1) \) scattering state.}
  \label{fig:1b2c_Eigenvectors}
\end{figure*}

As in the single-channel case, it is of interest to analyse how the eigenvectors describing the
composition of the energy eigenstates vary with \( \Lambda\,. \) Although we have an additional
channel, the overall behaviour of the eigenvectors is similar to that observed in the single
channel case.  \fref{fig:1b2c_Eigenvectors} illustrates the $\Lambda$ dependence of the
composition.  As \( \Lambda \) increases, enhanced short-distance mixing between the basis states
replaces contributions from the bare state.

However, given that the experimental inelasticities can only be described with $\Lambda \sim 0.8$
GeV -- in accord with phenomenologically motivated values associated with the induced pseudoscalar form
factor of a baryon -- we turn our attention to variation of the composition over the range $0.6 \le
\Lambda \le 1.0$ GeV.  Over this range, the composition of the energy eigenstates shows very little
dependence on the regulator parameter. Thus, physical insight into the meson-baryon rescattering
contributions to the energy eigenstates can be extracted.

\subsection{Pion Mass Dependence of the Finite-Volume Spectrum}
\label{sec:2c_LQCD}

In the single-channel system, we were able to demonstrate that for $\Lambda \alt 4$ GeV where a bare basis state
is included, the reproduction of the \lqcd results was largely independent of choice of \(
\Lambda\,. \) In the two-channel case, we have already found that increasing \( \Lambda \) to only
1.2 GeV results in a poorer description of the experimental inelasticity.  Still, it is desirable
to explore how variation of the regulator parameter and the ability to reproduce the inelasticity
manifests in the pion mass dependence of the spectrum.

\begin{figure}
  \centering
  \includegraphics[width=0.46\textwidth]{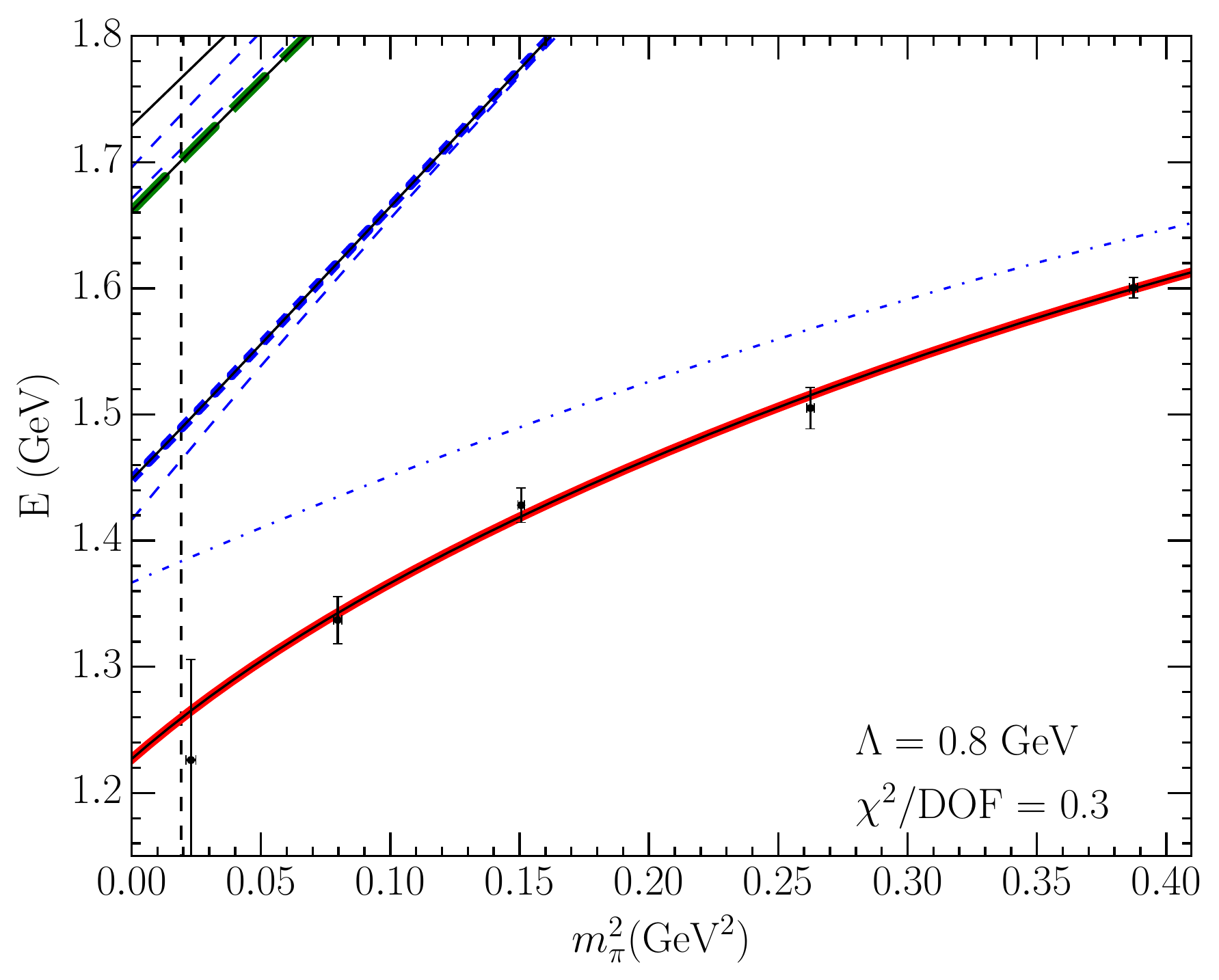}~\\
  \includegraphics[width=0.46\textwidth]{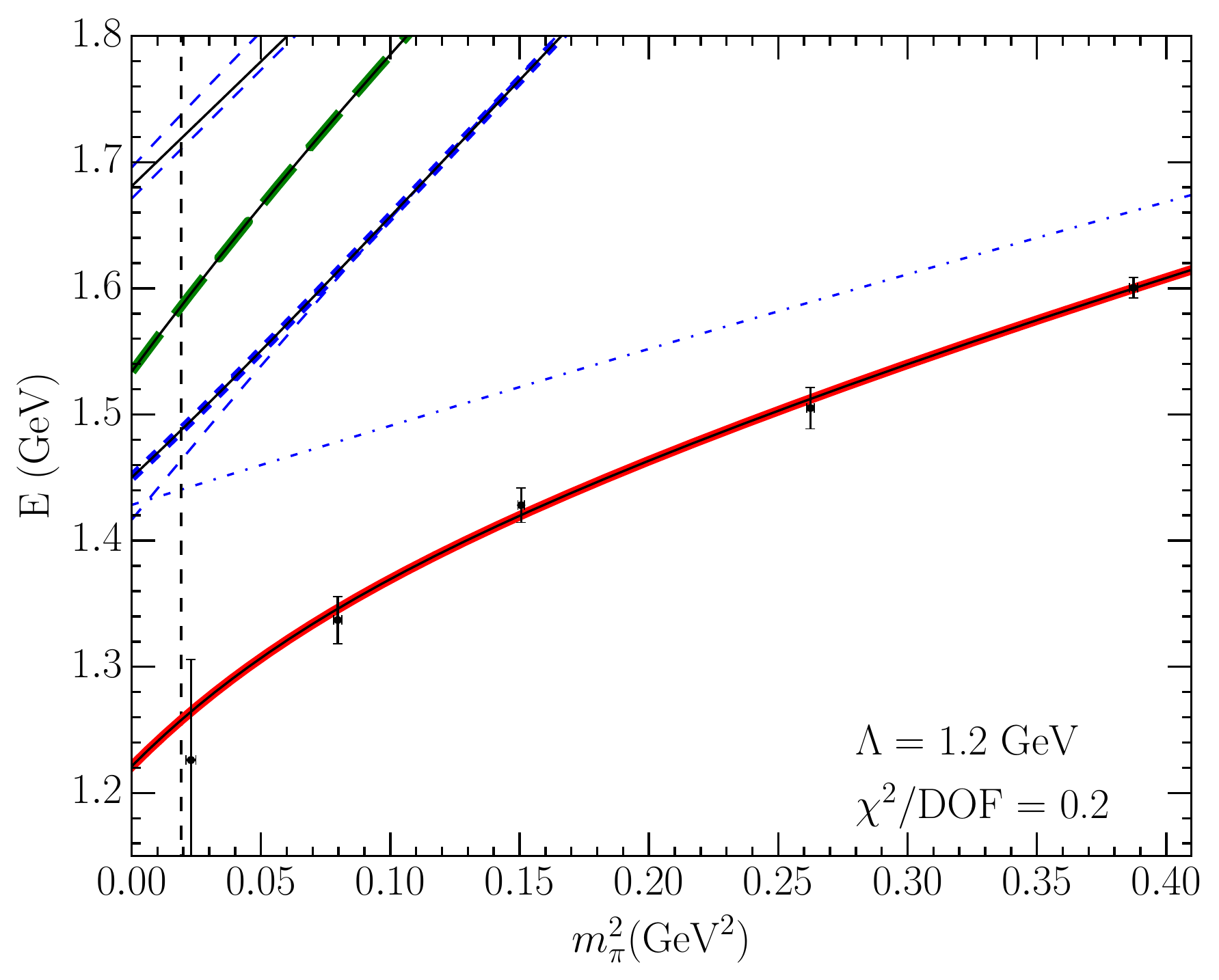}
  \caption{Pion mass dependence of the finite-volume HEFT energy eigenvalues for \( \Lambda = 0.8
    \) GeV (top) and \( \Lambda = 1.2 \) GeV (bottom) for the PACS-CS lattice length $L = 2.99$
    fm. The vertical dashed (black) line denotes the physical pion mass.  The thin diagonal (blue)
    dashed lines show the two-particle non-interacting energies and the thin more-horizontal (blue)
    dot-dash line
    illustrates the bare basis state mass. The solid (black) points are the lowest lying \( \Delta
    \) energies from \lqcd \cite{PACS-CS:2008bkb}.
    The solid black curves illustrate the finite-volume
    energy levels predicted by HEFT from fits to experimental phase shifts and inelasticities.
    These lines are dressed by solid (red), short-dashed (blue) and long-dashed (green) highlights
    indicating states with the largest, second-largest and third-largest contribution from the bare
    basis state $\ket{\Delta_0}$ respectively.  As the PACS-CS results follow from local
    three-quark operators, they are expected to lie on a solid (red) energy eigenstate.
    The quoted $\chi^2$/DOF are for the lowest-lying energy eigenvalue with respect to the PACS-CS data points.}
  \label{fig:1b2c_Evmpi_3fm}
\end{figure}

Results for two values of $\Lambda$ are presented in \fref{fig:1b2c_Evmpi_3fm}.  As in the
single-channel system, both values of \( \Lambda \) are able to produce the correct pion-mass
extrapolation for the lowest-lying state, despite the difficulties in fitting the inelasticity.

Thus to resolve a dependence relevant to the inelasticity, one must look to higher states in the
spectrum, more sensitive to the opening of $\pi\Delta$ channel.
The second excited state shows a large degree of variance with respect to the parameter set used.
As the scattering phase shifts are well-reproduced for \( E \leq 1650 \) MeV, this parameter
dependence is associated with the varying success in describing the inelasticity as presented
in \fref{fig:1b2c_fits}.  As such, the consideration of the first three energy levels in lattice
QCD for a lattice size of $\sim 3$ fm and a pion mass near the physical regime should
be sufficient to constrain the parameters to predict
both the \( \pi N \) phase shift and the inelasticity.  On larger
volumes $\sim 5$ fm, it was the fifth and seventh states that showed a sensitivity to the
inelasticity.

\begin{figure}
  \centering
  \includegraphics[width=0.46\textwidth]{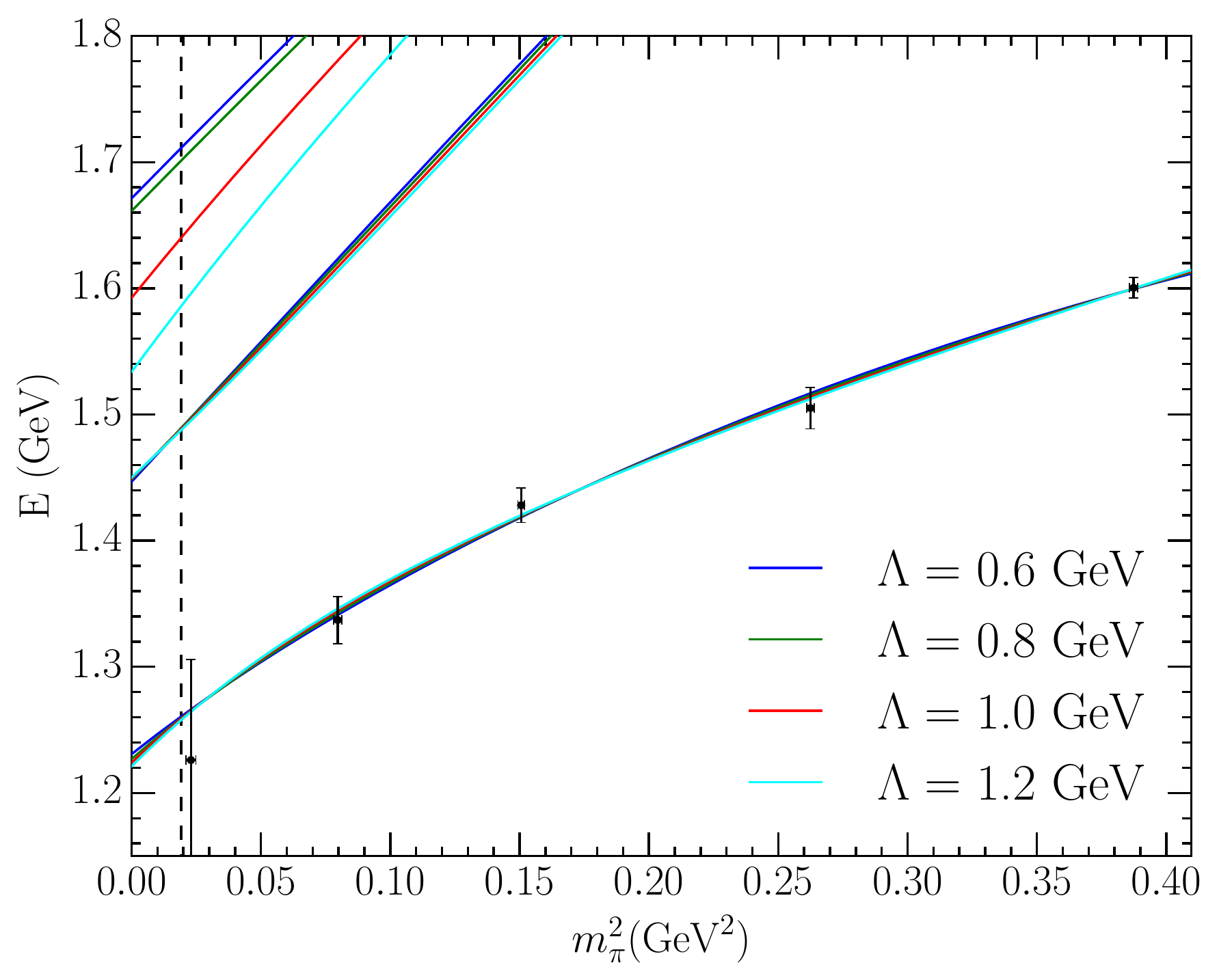}
  \caption{Two-channel Pion mass dependence of the three lowest-lying finite-volume HEFT
    eigenvalues at \( L = 2.99 \) fm using a dipole regulator, where the data points are the
    PACS-CS data.  Four parameter sets corresponding to each value of \( \Lambda \) are
    displayed, each with a corresponding bare mass expansion fit to the PACS-CS data.}
  \label{fig:1b2c_reg_comparison}
\end{figure}

Figure \ref{fig:1b2c_reg_comparison} serves to demonstrate more generally the spectral-dependence
arising from the choice of \( \Lambda\,, \) and its associated parameter set for a 3 fm lattice.
With sufficient high-quality lattice QCD results, the $\Lambda$ variation will be
constrained such that there is a unique Hamiltonian and a unique set of Hamiltonian eigenvectors
describing the eigenstate composition.

\subsection{Comparison with Contemporary \lqcd Results}

\begin{figure}
  \includegraphics[width=0.24\textwidth]{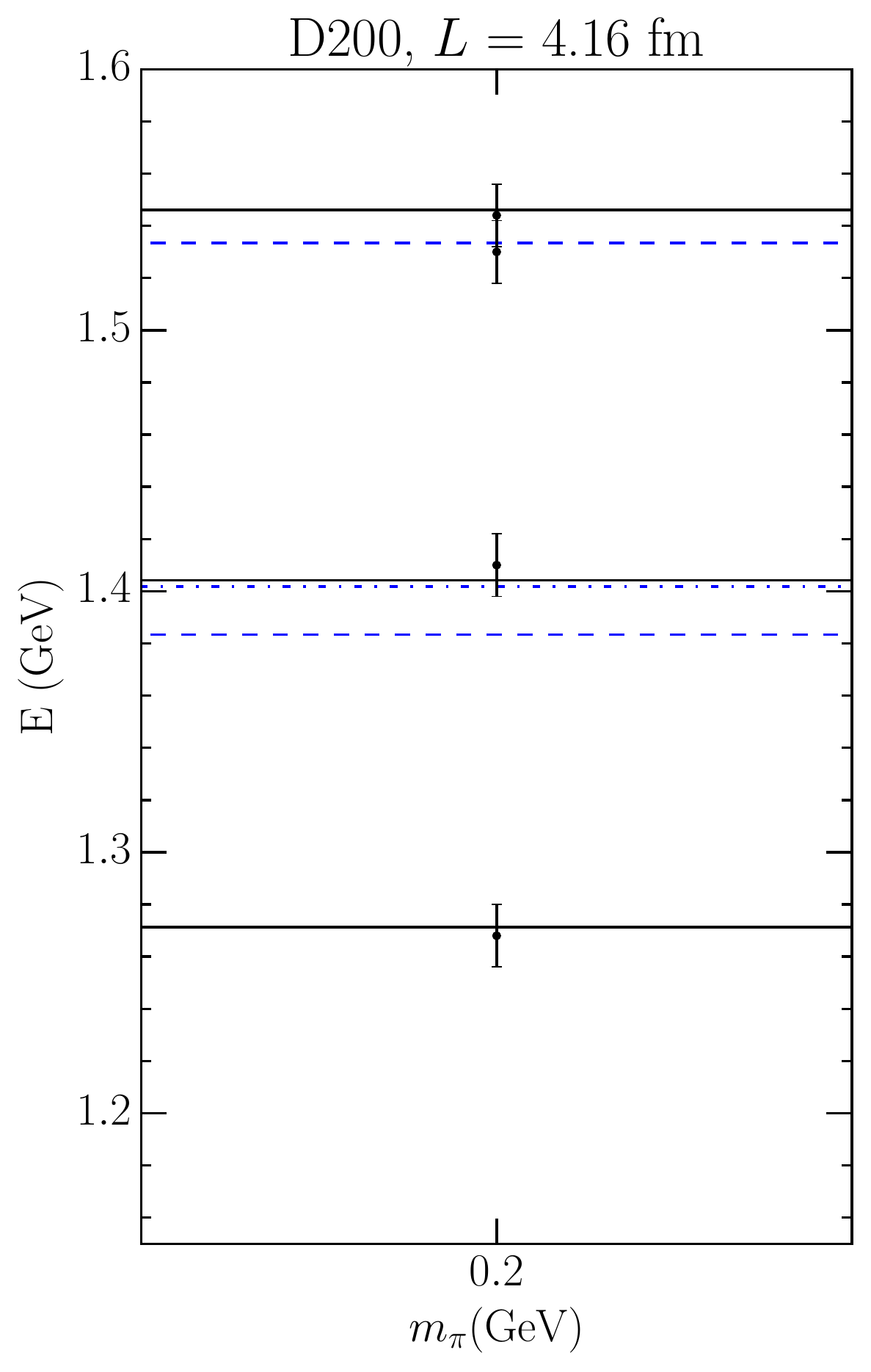}~
  \includegraphics[width=0.24\textwidth]{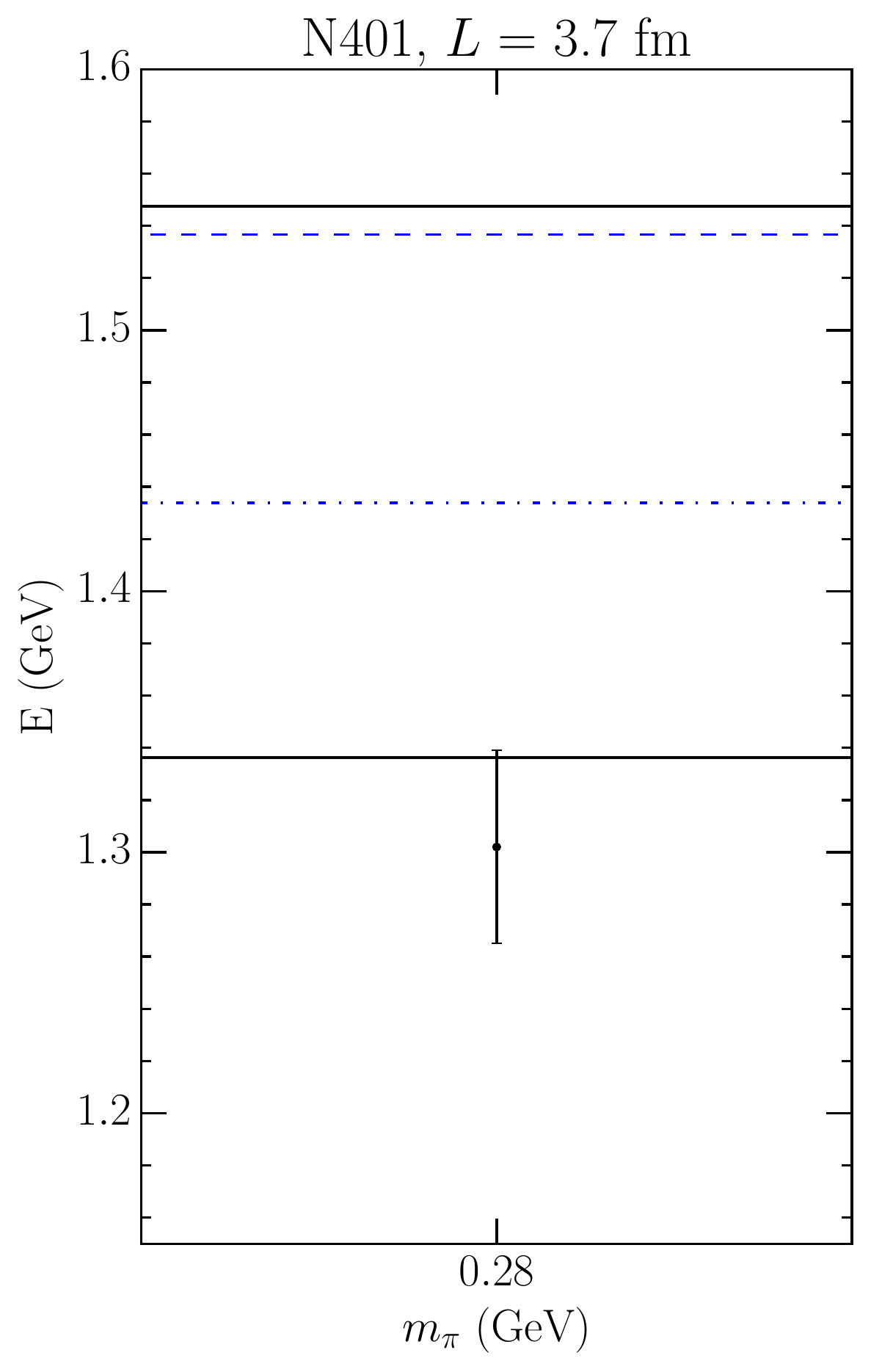}
  \caption{Comparison between the energy eigenvalues calculated in HEFT
    constrained by scattering data and the PACS-CS results for the quark-mass
      dependence (solid black lines) and \lqcd data from the CLS consortium (data points) for
    ensembles D200 (left) \rref{Morningstar:2021ewk}, and N401 (right) \rref{Andersen:2017una}.  The
    dashed blue lines denote non-interacting basis states, while the dot-dashed blue line is the
    mass of the bare basis state.}
  \label{fig:lqcd_CLS}
\end{figure}

In recent years, advances in \lqcd have allowed for several new studies of the \( \Delta(1232)\,
. \) Modern analyses include two-particle momentum-projected interpolating fields designed to more
directly access the two-particle scattering states.  Utilising our two-channel fit with \( \Lambda
= 0.8 \) GeV (Fit V of \tref{tab:2c}), comparison can be made between the finite-volume energy
eigenvalues calculated in HEFT and contemporary \lqcd results.  Figure \ref{fig:lqcd_CLS} provides
a comparison of the HEFT predictions developed herein with lattice QCD results from the CLS
consortium \rref{Andersen:2017una,Morningstar:2021ewk}.

From the ensemble labelled D200, with \( m_\pi = 0.2 \) GeV and \( L = 4.16 \) fm, points
corresponding with the three lowest-lying states are provided.  HEFT is able to predict them within
one standard deviation.
Moreover, HEFT correctly predicts the large increase in the scattering state energy relative to the non-interacting state for the first excitation.
Considering the eigenvectors, the lowest-lying state is dominated by the
bare basis state, while the second state is dominated by the \( \pi N(k=1) \) basis state with smaller mixing of other nearby basis states. This composition agrees with the \lqcd results where the first excitation couples strongly to momentum projected two-particle \( \pi N \) interpolators. We also note that only one of the two states reported in Hg for the second excitation is associated with \( J = 3/2 \) \rref{Morningstar:2022}.  The lowest-lying state from HEFT also agrees with the one
state available from the N401 ensemble with \( m_\pi = 0.28 \) GeV and \( L = 3.7 \) fm.


\section{Conclusion}
\label{sec:con}
We have examined the process of renormalisation in nonperturbative Hamiltonian Effective Field
Theory (HEFT).  As a nonperturbative
extension of effective field theory
incorporating the L\"uscher formalism,
HEFT provides a bridge between the
infinite-volume scattering data of experiment and the finite-volume spectrum of energy eigenstates
in \lqcd.

HEFT brings the insight of experimental data to the finite-volume of the lattice through the
parametrisation of a Hamiltonian built on a basis of non-interacting multiparticle states.
Through a process of constraining Hamiltonian parameters to scattering data, and then solving for
the eigenmodes of a finite-volume matrix Hamiltonian, one obtains finite-volume energy eigenvalues
and eigenvectors describing the composition of the finite volume states.  A key question is to
ascertain the regularisation-scheme dependence of these eigenvectors.

Using the FRR scheme, an expression for the \( S \)-matrix was obtained by
solving the coupled-channel, Bethe-Salpeter equations, and the phase shifts and inelasticities for
the system were then extracted.  These quantities were fit to the SAID Partial-Wave Analysis
Facility experimental scattering data \cite{site:SAID,Workman:2012hx} by adjusting the parameters of the
Hamiltonian.  These optimised parameters then serve as inputs for a Hamiltonian matrix model.  By
solving the eigenvalue equation for this Hamiltonian the finite-volume energy eigenvalues and
eigenvectors describing the composition of the finite volume states were resolved.

We considered the \( P \)-wave, \( I(J^P) = \frac{3}{2}(\frac{3}{2}^{+}) \) \( \Delta \) resonance
channel.  A simple description of the \( \Delta \) is to consider only the mixing of a bare \(
\Delta \) with a two-particle \( \pi N \) state.  Considering a basic system such as this allowed
for the development of intuition into the results of HEFT.  By using a dipole
  regulator and considering values for the regulator parameter ranging from \( \Lambda = 0.8 \)
GeV to \( \Lambda = 8.0 \) GeV, we were able to fit the scattering data with varying degrees of
success.  These fits produced a pole in agreement with the value listed by the Particle Data Group.
By solving the Hamiltonian matrix for this system at varying lattice sizes \( L\,, \) a
finite-volume energy spectrum was found.

We recall that the L\"uscher method of relating phase shifts to eigenvalues measured in lattice QCD is model independent.
Furthermore, it has been proven that the relationship between phase shifts and energy levels produced by HEFT and the L\"uscher method are identical up to corrections of order $\exp(-m_{\pi}L)$.
Thus the energy levels produced by the HEFT method are model independent within the energy range over which the experimental phase shifts are reproduced.
This has been demonstrated in the single-channel analysis of \sref{sec:1c}.

While the eigenvalues of single-channel
matrix Hamiltonian constrained to experimental scattering data do not
depend on the regulator parameter \( \Lambda \), the eigenvectors of the Hamiltonian do show a
significant dependence on $\Lambda$.  As \( \Lambda \) increases and short-distance interactions
are allowed between the effective fields, the bare basis state contribution to low-lying
finite-volume states decreases, implying that a bare basis state may not be necessary for a
description of the \( \Delta \) resonance.  To resolve the validity of this conjecture, the pion
mass dependence of the HEFT was examined.

Extending HEFT beyond the physical pion mass allows for a direct comparison to \lqcd calculations
available at many pion masses.  In performing this comparison, it was revealed that by utilising a
quadratic form for the bare mass extrapolation, any value of $\Lambda \alt 4$ GeV is able to
reproduce the \lqcd data well.
In addition, independence on the choice between Gaussian and dipole regulators is observed.
However, without a bare state, although the scattering data was
able to be reproduced in the resonance region, the \lqcd data was not reproduced away from the
physical point.

Having gained intuition into the structure of the \( \Delta \) in the simple single-channel
case, the more complicated two-channel system was considered, allowing the scattering data to be
described at higher energies and introducing an inelasticity for consideration.
While the scattering phase shifts could be described for various values of $\Lambda$, an accurate
description of the inelasticity required a value of $\Lambda$ consistent with phenomenologically
motivated values associated with the induced pseudoscalar form factor of a baryon.

Thus consideration of the two-channel case has made it clear that nonperturbative HEFT does not
enjoy the same freedom in selecting a regulator as in $\chi$PT.  Whereas any value of $\Lambda$ is
admissible in the power counting regime of $\chi$PT (which is usually regarded to include the
physical pion mass), in the two-channel HEFT case the regulator and its associated parameter set
contains the physics necessary to describe the inelasticity. $\Lambda$ must take physically
motivated values to enable a description of the inelasticity.

By considering the pion mass dependence of the spectra generated by different regulator parameters,
and comparing them to \lqcd results, it becomes clear that future high-precision lattice QCD results
for excited states near the opening of the $\pi\Delta$ channel will be able to constrain the
Hamiltonian and make predictions for both the \( \pi N \) phase shift and the inelasticity.
Finally, comparison was made with contemporary \lqcd results from the CLS consortium for the \(
\Delta \) spectrum.  Agreement was observed for all the eigenstates available.

Future work should consider the effects of moving frames on the \( \Delta \) system in
HEFT~\cite{Li:2021mob} in order to compare with a larger range of recent \lqcd
results~\rref{Andersen:2017una,Morningstar:2021ewk}.  Additionally, the understanding of the roles
of the bare mass and regulator parameters will prove useful for the study of new areas of interest
such as the odd-parity nucleon resonances, where recent research~\cite{Stokes:2019zdd} has
indicated that two bare states may be required to adequately describe the system.

\section*{Acknowledgements}
D.B.L. thanks Albert Kong for discussions in the early phase of this research
and the organisers of the July 2019 CERN Workshop ``Advances in Lattice Gauge Theory'' where many
of the questions addressed herein were first posed.
The authors would also like to thank Colin Morningstar for helpful discussions on the nature of the states observed in lattice QCD.
This research was supported by the Australian Government Research Training Program Scholarship, and with supercomputing resources provided by the Phoenix HPC service at the University of Adelaide.
This research was undertaken with the assistance of resources from the National Computational
Infrastructure (NCI), provided through the National Computational Merit Allocation Scheme, and
supported by the Australian Government through Grant No.~LE190100021 and the University of Adelaide Partner Share.
This research was supported by the Australian Research Council through ARC Discovery Project Grants Nos. DP180100497 (A.W.T.) and DP190102215 and DP210103706 (D.B.L.).
J.-J. Wu was supported
by the Fundamental Research Funds for the Central Universities,
by the National Key R$\&$D Program of China under
Contract No. 2020YFA0406400,
and by the Key Research Program of the Chinese Academy of
Sciences, Grant NO. XDPB15.

\newpage
\bibliography{DeltaRenormReferences}

\begin{thebibliography}{10}

\bibitem{Luscher:1985dn}
M.~L{\"u}scher.
\newblock Volume dependence of the energy spectrum in massive quantum field
  theories {{I}}. {{Stable}} particle states.
\newblock {\em Communications in Mathematical Physics}, 104(2):177--206, June
  1986.

\bibitem{Luscher:1986pf}
M.~L{\"u}scher.
\newblock Volume dependence of the energy spectrum in massive quantum field
  theories {{II}}. {{Scattering}} states.
\newblock {\em Communications in Mathematical Physics}, 105(2):153--188, June
  1986.

\bibitem{Luscher:1990ux}
Martin L{\"u}scher.
\newblock Two-particle states on a torus and their relation to the scattering
  matrix.
\newblock {\em Nuclear Physics B}, 354(2):531--578, May 1991.

\bibitem{He:2005ey}
Song He, Xu~Feng, and Chuan Liu.
\newblock Two particle states and the {{S}}-matrix elements in multi-channel
  scattering.
\newblock {\em JHEP}, 07:011, 2005.

\bibitem{Lage:2009zv}
Michael Lage, Ulf-G. Meissner, and Akaki Rusetsky.
\newblock A {{Method}} to measure the antikaon-nucleon scattering length in
  lattice {{QCD}}.
\newblock {\em Physics Letters}, B681:439--443, 2009.

\bibitem{Bernard:2010fp}
V.~Bernard, M.~Lage, U.~G. Meissner, and A.~Rusetsky.
\newblock Scalar mesons in a finite volume.
\newblock {\em JHEP}, 01:019, 2011.

\bibitem{Guo:2012hv}
Peng Guo, Jozef Dudek, Robert Edwards, and Adam~P. Szczepaniak.
\newblock Coupled-channel scattering on a torus.
\newblock {\em Physical Review}, D88(1):014501, 2013.

\bibitem{Hu:2016shf}
B.~Hu, R.~Molina, M.~D{\"o}ring, and A.~Alexandru.
\newblock Two-flavor {{Simulations}} of the \$\textbackslash{}rho(770)\$ and
  the {{Role}} of the \${{K}}\textbackslash{}bar {{K}}\$ {{Channel}}.
\newblock {\em Physical Review Letters}, 117(12):122001, September 2016.

\bibitem{Li:2012bi}
Ning Li and Chuan Liu.
\newblock Generalized {{L}}\textbackslash{}"uscher {{Formula}} in
  {{Multi}}-channel {{Baryon}}-{{Meson Scattering}}.
\newblock {\em Physical Review D}, 87(1):014502, January 2013.

\bibitem{Hansen:2012bj}
Maxwell~T. Hansen and Stephen~R. Sharpe.
\newblock {Multiple-channel generalization of Lellouch-Lüscher formula}.
\newblock {\em PoS}, LATTICE2012:127, 2012.

\bibitem{Doring:2018xxx}
M.~D{\"o}ring, H.-W. Hammer, M.~Mai, J.-Y. Pang, A.~Rusetsky, and J.~Wu.
\newblock Three-body spectrum in a finite volume: The role of cubic symmetry.
\newblock {\em Physical Review D}, 97(11):114508, June 2018.

\bibitem{Hansen:2019nir}
Maxwell~T. Hansen and Stephen~R. Sharpe.
\newblock Lattice {{QCD}} and {{Three}}-particle {{Decays}} of {{Resonances}}.
\newblock {\em Annual Review of Nuclear and Particle Science}, 69(1):65--107,
  October 2019.

\bibitem{Blanton:2019vdk}
Tyler~D. Blanton, Fernando {Romero-L{\'o}pez}, and Stephen~R. Sharpe.
\newblock \${{I}} = 3\$ three-pion scattering amplitude from lattice {{QCD}}.
\newblock {\em Physical Review Letters}, 124(3):032001, January 2020.

\bibitem{PACS-CS:2008bkb}
S.~Aoki et~al.
\newblock {2+1 Flavor Lattice QCD toward the Physical Point}.
\newblock {\em Phys. Rev. D}, 79:034503, 2009.

\bibitem{Andersen:2017una}
Christian~Walther Andersen, John Bulava, Ben Hörz, and Colin Morningstar.
\newblock {Elastic $I=3/2$ $p$-wave nucleon-pion scattering amplitude and the
  $\Delta$(1232) resonance from N$_f$=2+1 lattice QCD}.
\newblock {\em Phys. Rev.}, D97(1):014506, 2018.

\bibitem{Morningstar:2021ewk}
Colin Morningstar, John Bulava, Andrew~D. Hanlon, Ben H\"orz, Daniel Mohler,
  Amy Nicholson, Sarah Skinner, and Andr\'e Walker-Loud.
\newblock {Progress on Meson-Baryon Scattering}.
\newblock {\em PoS}, LATTICE2021:170, 2022.

\bibitem{Silvi:2021uya}
Giorgio Silvi et~al.
\newblock {$P$-wave nucleon-pion scattering amplitude in the $\Delta$(1232)
  channel from lattice QCD}.
\newblock {\em Phys. Rev. D}, 103(9):094508, 2021.

\bibitem{Wu:2017qve}
Jia-jun Wu, Derek~B. Leinweber, Zhan-wei Liu, and Anthony~W. Thomas.
\newblock {Structure of the Roper Resonance from Lattice QCD Constraints}.
\newblock {\em Phys. Rev. D}, 97(9):094509, 2018.

\bibitem{Hall:2014uca}
Jonathan M.~M. Hall, Waseem Kamleh, Derek~B. Leinweber, Benjamin~J. Menadue,
  Benjamin~J. Owen, Anthony~W. Thomas, and Ross~D. Young.
\newblock {Lattice QCD Evidence that the \ensuremath{\Lambda}(1405) Resonance
  is an Antikaon-Nucleon Molecule}.
\newblock {\em Phys. Rev. Lett.}, 114(13):132002, 2015.

\bibitem{Hall:2013qba}
J.~M.~M. Hall, A.~C.~P. Hsu, D.~B. Leinweber, A.~W. Thomas, and R.~D. Young.
\newblock {Finite-volume matrix Hamiltonian model for a $\Delta \to N\pi$
  system}.
\newblock {\em Phys. Rev. D}, 87(9):094510, 2013.

\bibitem{Wu:2014vma}
Jia-Jun Wu, T.-S.~H. Lee, A.~W. Thomas, and R.~D. Young.
\newblock Finite-volume {{Hamiltonian}} method for coupled channel interactions
  in lattice {{QCD}}.
\newblock {\em Physical Review C}, 90(5):055206, November 2014.

\bibitem{Young:2002ib}
Ross~Daniel Young, Derek~Bruce Leinweber, and Anthony~William Thomas.
\newblock {Convergence of chiral effective field theory}.
\newblock {\em Prog. Part. Nucl. Phys.}, 50:399--417, 2003.

\bibitem{Leinweber:2003dg}
Derek~Bruce Leinweber, Anthony~William Thomas, and Ross~Daniel Young.
\newblock {Physical nucleon properties from lattice QCD}.
\newblock {\em Phys. Rev. Lett.}, 92:242002, 2004.

\bibitem{Leinweber:2005cm}
Derek~Bruce Leinweber, Anthony~William Thomas, and Ross~Daniel Young.
\newblock {Power counting regime of chiral extrapolation and beyond}.
\newblock {\em PoS}, LAT2005:048, 2006.

\bibitem{Gell-Mann:1968hlm}
Murray Gell-Mann, R.~J. Oakes, and B.~Renner.
\newblock {Behavior of current divergences under SU(3) x SU(3)}.
\newblock {\em Phys. Rev.}, 175:2195--2199, 1968.

\bibitem{Leinweber:1999ig}
Derek~Bruce Leinweber, Anthony~William Thomas, Kazuo Tsushima, and
  Stewart~Victor Wright.
\newblock {Baryon masses from lattice QCD: Beyond the perturbative chiral
  regime}.
\newblock {\em Phys. Rev. D}, 61:074502, 2000.

\bibitem{Hall:2010ai}
J.~M.~M. Hall, D.~B. Leinweber, and R.~D. Young.
\newblock {Power Counting Regime of Chiral Effective Field Theory and Beyond}.
\newblock {\em Phys. Rev. D}, 82:034010, 2010.

\bibitem{Hall:2011en}
J.M.M. Hall, F.X. Lee, D.B. Leinweber, K.F. Liu, N.~Mathur, R.D. Young, and
  J.B. Zhang.
\newblock {Chiral extrapolation beyond the power-counting regime}.
\newblock {\em Phys. Rev. D}, 84:114011, 2011.

\bibitem{Hall:2012pk}
J.M.M. Hall, D.B. Leinweber, and R.D. Young.
\newblock {Chiral extrapolations for nucleon magnetic moments}.
\newblock {\em Phys. Rev. D}, 85:094502, 2012.

\bibitem{Hall:2013oga}
J.M.M. Hall, D.B. Leinweber, and R.D. Young.
\newblock {Chiral extrapolations for nucleon electric charge radii}.
\newblock {\em Phys. Rev. D}, 88(1):014504, 2013.

\bibitem{Guichon:1982zk}
Pierre A.~M. Guichon, Gerald~A. Miller, and Anthony~William Thomas.
\newblock {The Axial Form-factor of the Nucleon and the Pion - Nucleon Vertex
  Function}.
\newblock {\em Phys. Lett. B}, 124:109--112, 1983.

\bibitem{Thomas:1982kv}
Anthony~William Thomas.
\newblock {Chiral Symmetry and the Bag Model: A New Starting Point for Nuclear
  Physics}.
\newblock {\em Adv. Nucl. Phys.}, 13:1--137, 1984.

\bibitem{Miller:1980hp}
Gerald~A. Miller, Anthony~William Thomas, and S.~Theberge.
\newblock {Pionic Corrections in the MIT Bag Model}.
\newblock {\em Comments Nucl. Part. Phys.}, 10(3):101--108, 1981.

\bibitem{Li:2019qvh}
Yan Li, Jia-Jun Wu, Curtis~D. Abell, Derek~B. Leinweber, and Anthony~W. Thomas.
\newblock {Partial Wave Mixing in Hamiltonian Effective Field Theory}.
\newblock {\em Phys. Rev. D}, 101(11):114501, 2020.

\bibitem{Liu:2015ktc}
Zhan-Wei Liu, Waseem Kamleh, Derek~B. Leinweber, Finn~M. Stokes, Anthony~W.
  Thomas, and Jia-Jun Wu.
\newblock Hamiltonian effective field theory study of the
  \$\textbackslash{}mathbf\{\vphantom\}{{N}}\^*(1535)\vphantom\{\}\$ resonance
  in lattice {{QCD}}.
\newblock {\em Physical Review Letters}, 116(8):082004, February 2016.

\bibitem{Arndt:1985vj}
Richard~A. Arndt, John~M. Ford, and L.~David Roper.
\newblock {PION - NUCLEON PARTIAL WAVE ANALYSIS TO 1100-MeV}.
\newblock {\em Phys. Rev. D}, 32:1085, 1985.

\bibitem{Workman:2012hx}
R.~L. Workman, R.~A. Arndt, W.~J. Briscoe, M.~W. Paris, and I.~I. Strakovsky.
\newblock {Parameterization dependence of T matrix poles and eigenphases from a
  fit to $\pi$N elastic scattering data}.
\newblock {\em Phys. Rev. C}, 86:035202, 2012.

\bibitem{site:SAID}
{INS Data Analysis Center}.
\newblock \url{http://gwdac.phys.gwu.edu/}.
\newblock Online, Solution W108.

\bibitem{Fettes:1998ud}
Nadia Fettes, Ulf-G. Meissner, and Sven Steininger.
\newblock {Pion - nucleon scattering in chiral perturbation theory. 1. Isospin
  symmetric case}.
\newblock {\em Nucl. Phys. A}, 640:199--234, 1998.

\bibitem{Meissner:1999vr}
Ulf-G. Meissner and J.~A. Oller.
\newblock {Chiral unitary meson baryon dynamics in the presence of resonances:
  Elastic pion nucleon scattering}.
\newblock {\em Nucl. Phys. A}, 673:311--334, 2000.

\bibitem{Juli_D_az_2007}
B.~Juli{\'{a} }-D{\'{\i}}az, T.-S.~H. Lee, A.~Matsuyama, and T.~Sato.
\newblock Dynamical coupled-channel model of $\pi n$ scattering in the w $\leq$
  2 gev nucleon resonance region.
\newblock {\em Physical Review C}, 76(6), dec 2007.

\bibitem{Sato:2009de}
T.~Sato and T.~S.~H. Lee.
\newblock {Dynamical Models of the Excitations of Nucleon Resonances}.
\newblock {\em J. Phys. G}, 36:073001, 2009.

\bibitem{10.1093/ptep/ptaa104}
Particle~Data Group.
\newblock {Review of Particle Physics*}.
\newblock {\em Progress of Theoretical and Experimental Physics}, 2020(8), 08
  2020.
\newblock 083C01.

\bibitem{Miller:1979kg}
Gerald~A. Miller, Anthony~William Thomas, and S.~Theberge.
\newblock {Pion - Nucleon Scattering in the Cloudy Bag Model}.
\newblock {\em Phys. Lett. B}, 91:192--195, 1980.

\bibitem{Theberge:1980ye}
S.~Theberge, Anthony~William Thomas, and Gerald~A. Miller.
\newblock {The Cloudy Bag Model. 1. The (3,3) Resonance}.
\newblock {\em Phys. Rev. D}, 22:2838, 1980.
\newblock [Erratum: Phys.Rev.D 23, 2106 (1981)].

\bibitem{McGovern:1998tm}
Judith~A. McGovern and Michael~C. Birse.
\newblock {On the absence of fifth order contributions to the nucleon mass in
  heavy baryon chiral perturbation theory}.
\newblock {\em Phys. Lett. B}, 446:300--305, 1999.

\bibitem{Ren_2020}
X.-L. Ren, E.~Epelbaum, J.~Gegelia, and Ulf-G. Mei{\ss}ner.
\newblock Meson-baryon scattering in resummed baryon chiral perturbation theory
  using time-ordered perturbation theory.
\newblock {\em The European Physical Journal C}, 80(5), may 2020.

\bibitem{Morningstar:2022}
C.~Morningstar.
\newblock private communication, June 2022.

\bibitem{Li:2021mob}
Yan Li, Jia-jun Wu, Derek~B. Leinweber, and Anthony~W. Thomas.
\newblock {Hamiltonian effective field theory in elongated or moving finite
  volume}.
\newblock {\em Phys. Rev. D}, 103(9):094518, 2021.

\bibitem{Stokes:2019zdd}
Finn~M. Stokes, Waseem Kamleh, and Derek~B. Leinweber.
\newblock {Elastic Form Factors of Nucleon Excitations in Lattice QCD}.
\newblock {\em Phys. Rev. D}, 102(1):014507, 2020.

\end{thebibliography}
\bibliographystyle{unsrt}
\end{document}